\documentclass[12pt,a4paper]{article}
\usepackage{a4wide,cite}
\usepackage{epsfig}
\newcommand{\be}{\begin{equation}}
\newcommand{\ee}{\end{equation}}
\newcommand{\bea}{\begin{eqnarray}}
\newcommand{\eea}{\end{eqnarray}}

\newcommand{\ep}{\varepsilon}
\newcommand{\nn}{\nonumber}

\newcommand{\pslash}{p\hspace{-2mm}/}
\newcommand{\half}{{\textstyle\frac{1}{2}}}
\newcommand{\quarter}{{\textstyle\frac{1}{4}}}

\newcommand{\Cl}[2]{{\mbox{Cl}}_{#1}\left(#2\right)}

\catcode`\@=11
\def\Im{\mathop{\rm Im}\nolimits}

\begin{document}

 \thispagestyle{empty}
 \begin{flushright}
 {MZ-TH/99-63} \\[3mm]
 {hep-ph/0008171} \\[3mm]
 {August 2000}
 \end{flushright}
 \vspace*{2.0cm}
 \begin{center}
 {\large \bf
 Quark-gluon vertex in arbitrary gauge and dimension}
 \end{center}
 \vspace{1cm}
 \begin{center}
 A.I.~Davydychev$^{a,}$\footnote{On leave from 
 Institute for Nuclear Physics, Moscow State University, 
 119899, Moscow, Russia. Email address:
 davyd@thep.physik.uni-mainz.de}, \ \
 P.~Osland$^{b,}$\footnote{Email address: Per.Osland@fi.uib.no} \ \
 \ \ and \ \
 L.~Saks$^{b,}$\footnote{Email address: Leo.Saks@fi.uib.no}
\\
 \vspace{1cm}
$^{a}${\em
 Department of Physics,
 University of Mainz, \\
 Staudingerweg 7,
 D-55099 Mainz, Germany}
\\
\vspace{.3cm}
$^{b}${\em
 Department of Physics,
 University of Bergen, \\
 All\'egaten 55,
 N-5007 Bergen, Norway}
\\
\end{center}
 \hspace{3in}
 \begin{abstract}
One-loop off-shell contributions to the quark-gluon vertex are calculated,
in an arbitrary covariant gauge and in arbitrary space-time dimension,
including quark-mass effects.
It is shown how one can get results for all on-shell limits of interest
directly from the off-shell expressions.
In order to demonstrate that the Ward--Slavnov--Taylor
identity for the quark-gluon vertex is satisfied,
we have also calculated the corresponding one-loop contribution
involving the quark-quark-ghost-ghost vertex.
 \end{abstract}

%%%%%%%%%%%%%%%%%%%%%%%%%%%%%%%%%%%%%%%%%%%%%%%%%%%%%%%%%%%%%%%%%%%%%%%
\newpage

%%%%%%%%%%%%%%%%%%%%%%%%%%%%%%%%%%%%%%%%%%%%%%%%%%%%%%%%%%%%%%%%%%%%%%%
\section{Introduction}
\setcounter{equation}{0}
%%%%%%%%%%%%%%%%%%%%%%%%%%%%%%%%%%%%%%%%%%%%%%%%%%%%%%%%%%%%%%%%%%%%%%%
Tests of perturbative Quantum Chromodynamics \cite{QCD,MarPag}
are steadily reaching new levels of precision \cite{QCD-overview},
necessitating information on higher-loop results for a wide range 
of processes.
Higher-order QCD effects are also frequently required for 
background estimates in searches for signs of new physics.

In spite of its fundamental role, the quark-gluon vertex 
has not been explored in detail, even at one loop.
It is of course very much related to the electron-photon vertex
of Quantum Electrodynamics, a basic and nontrivial aspect of which 
is the anomalous
magnetic moment, or $g-2$, which provides a powerful test of the
whole concept of Quantum Field Theories.
The quark-gluon vertex differs from the electron-photon vertex
already at one loop, by the contributions of an additional Feynman 
diagram, involving the three-gluon vertex.
In fact, apart from introducing additional colour structure, this
non-Abelian diagram introduces at the one-loop level 
a kinematical structure which is absent in the QED vertex.

For special kinematical configurations, and special gauges,
several results are available.
Already around 1980, the symmetric off-shell case was considered
by Pascual and Tarrach (PT) in \cite{PT} (see also \cite{CG})
in an arbitrary covariant gauge for massless 
quarks\footnote{Numerical analysis of two-loop QCD vertices
in this limit is given in a recent paper~\cite{ChS}.}.
The emphasis was on comparing the $\overline{\mbox{MS}}$ and 
Weinberg's renormalization schemes.
The symmetric off-shell case was also considered in \cite{DTP} by
Dung, Tarasov and Phuoc (DTP), for massive quarks,
restricted to the scalar function multiplying $\gamma_{\mu}$. 
For massless quarks, some on-shell results are available,
mainly in the Feynman gauge, presented
by Nowak, Prasza{\l}owicz and S{\l}omi{\'n}ski (NPS) \cite{NPS}.

The situation when one (gluon or quark) momentum vanishes
has been studied in more detail. Technically, in this case
all three-point functions effectively reduce to 
two-point integrals. For massless quarks, 
some one-loop results in arbitrary covariant gauge  
have been obtained by Braaten and Leveille \cite{BL}.
In the Feynman gauge,
also two-loop corrections have been presented in \cite{BL}.
Moreover, in a recent paper by Chetyrkin and R{\'e}tey
\cite{ChR}, renormalized expressions for three-loop-order
QCD vertices have been obtained for such zero-momentum
configurations, in an arbitrary covariant gauge.

The QED contribution, proportional to the ``Abelian''
contribution to the quark-gluon vertex, has been studied more 
systematically. An early paper by Ball and Chiu (BC) \cite{BC1} presented
a systematic kinematical decomposition of the vertex, and gave
off-shell results for the one-loop QED vertex in Feynman gauge.
Their work was extended to arbitrary covariant gauge by
K{\i}z{\i}lers{\"u}, Reenders and Pennington (KRP) \cite{KRP}.
All above-mentioned papers deal with the (dimensionally-regulated
\cite{dimreg}) four-dimensional case. Results for the three-dimensional 
QED contribution are also available (for massless fermions), 
due to Bashir, K{\i}z{\i}lers{\"u} and Pennington (BKP) \cite{BKP}.  
A summary of all these one-loop results
is given in Table~1.
In addition to this table, we note that
another special gauge which has been investigated
is the Fried--Yennie gauge \cite{ALV}.

%================================================================
\begin{table}[ht]
\begin{center}
\begin{minipage}{160mm}
{\renewcommand{\arraystretch}{1.5}
\begin{tabular}{||l||l|l|l|l|l||}
\hline
          &\multicolumn{3}{|c|}{\parbox[c]{63mm}{all momenta
                                        off-shell\phantom{${}_I$}}}
          &\multicolumn{2}{|c||}{\parbox[c]{45mm}{some momenta
                                        on-shell\phantom{${}_I$}}} \\
           \cline{2-6}
\phantom{$P^2$} &\multicolumn{2}{|c|}{\parbox[c]{38mm}{general case}}
                &$p_1^2=p_2^2=p_3^2$
&$p_3=0$ & \parbox[c]{25mm}{\raggedright $p_1^2\!=\!p_2^2\!=\!0$~or 
                             $p_2^2\!=\!p_3^2\!=\!0$} \\
           \cline{2-3}
\phantom{$P^2$} &\parbox[c]{23mm}{QED}
                &\parbox[c]{17mm}{QCD} & & & \\
\hline \hline
\parbox[c]{17mm}{Feynman \\ gauge}
              &\parbox[c]{23mm}{\raggedright BC \cite{BC1}}
              &\parbox[c]{17mm}{}
              &\parbox[c]{25mm}{\raggedright special~case~of \\ 
                                PT~\cite{PT},~$m\!=\!0$}
              &\parbox[c]{23mm}{\raggedright special~case~of \\
                                BL~\cite{BL},~$m\!=\!0$}
              &\parbox[c]{26mm}{\raggedright NPS~\cite{NPS},~$m\!=\!0$} \\
\hline
\parbox[c]{17mm}{Arbitrary \\ covariant \\ gauge}
              &\parbox[c]{23mm}{\raggedright KRP
                                \cite{KRP};\phantom{\Large I}\\
                                             BKP~\cite{BKP},~$3d$ }
              &\parbox[c]{17mm}{}
              &\parbox[c]{25mm}{\raggedright PT~\cite{PT},~$m\!=\!0$; \\
                                DTP \cite{DTP}, \\
                                $\gamma_{\mu}$ part}
              &\parbox[c]{23mm}{\raggedright BL~\cite{BL},~$m\!=\!0$}
              &\parbox[c]{25mm}{\raggedright } \\
\hline
\end{tabular}
} %end arraystretch
\end{minipage}
\end{center}
\caption{Kinematics and gauges considered in other one-loop 
studies.
None of these results is valid for arbitrary dimension $n$.}
\end{table}
%================================================================

Among non-covariant gauges, we would like to mention
the Coulomb gauge. In some sense, it is more 
``physical'', but technically rather challenging \cite{Leib}.
A rather different approach to QCD vertex functions is provided
by lattice calculations
\cite{lattice}.
The quark-gluon vertex functions may also serve as a basis for modeling
the photon-nucleon vertices \cite{nuclear} and the quark-Reggeon
vertex~\cite{FFQ}.

{} From Table~1, one can see that, even if we consider the
results in (or around) four dimensions, there are still several
``white spots''.
The aim of the present paper is to cover {\em all} such remaining spots.
Moreover, we present results which are valid
for an {\em arbitrary} value of the space-time dimension.
Apart from the quark-gluon vertex itself, we also consider the
related two-point functions, and the quark-quark-ghost-ghost vertex
function, in order to be able to check
that the obtained quark-gluon vertex function obeys 
the Ward--Slavnov--Taylor (WST) identity \cite{WST}.

At the one-loop level, the simple and well-known Dirac-matrix structure
of the lowest-order quark-gluon vertex gets modified. 
In the general case, twelve
structures are needed to decompose it \cite{BC1}.
Thus, twelve scalar functions multiplying these tensor structures
are to be calculated.
These scalar functions depend on the gauge parameter,
the space-time dimension, quark mass(es), and the kinematical invariants
($p_1^2$, $p_2^2$, $p_3^2$). 
Four of them (the ``longitudinal'' ones) are involved 
in the WST identity, whereas the remaining eight are unconstrained.

There are several reasons why the one-loop results calculated in
arbitrary gauge and dimension $n$ are of special interest: \\
(i) knowing the results in arbitrary gauge, one can explicitly
keep track of gauge invariance for physical quantities; \\
(ii) if one is interested in the two-loop calculation of the quark-gluon
coupling, one should know one-loop contributions in more detail; \\
(iii) results in arbitrary dimension make it possible to consider
all on-shell limits {\em directly} from these expressions (see Section~4), 
this is impossible if one only has 
the results valid around four dimensions; \\
(iv) QCD is also a theory of interest in three and two dimensions
(see, e.g., \cite{23d}); \\
(v) as we shall see, the results for arbitrary dimension are not much
more cumbersome than those considered around four dimensions (in some
respects, they are even more transparent and instructive).

The paper is organized as follows.
In Section~2, we introduce the notation for the two- and three-point
functions to be considered, and discuss their decomposition in terms of
scalar functions as well as the corresponding
Ward--Slavnov--Taylor identity.
In Section~3, we present the most general off-shell results
for the quark-gluon vertex. Section~4 contains the corresponding
expressions for special limits of interest. In Section~5,
we conclude with a summary and a discussion of the results.
Then, we have several appendices where some further results
and technical details are presented, such as the formulae used
to decompose the quark-gluon vertex (Appendix~A), relevant results for
the scalar integrals (Appendices~B and C), results for
the one-loop contribution involved in checking
the WST identity (Appendix~D), and general results for the transverse part
of the quark-gluon vertex (Appendix~E).

%%%%%%%%%%%%%%%%%%%%%%%%%%%%%%%%%%%%%%%%%%%%%%%%%%%%%%%%%%%%%%%%%%%%%%%
\section{Preliminaries}
\setcounter{equation}{0}
%%%%%%%%%%%%%%%%%%%%%%%%%%%%%%%%%%%%%%%%%%%%%%%%%%%%%%%%%%%%%%%%%%%%%%%
We shall here establish some notation, and discuss the 
functions involved in the Ward--Slavnov--Taylor identity
for the quark-gluon vertex.

\subsection{Notation}

A graphical representation of the quark-gluon vertex is 
given in Fig.~1.\footnote{To produce the figures, 
the {\sf AXODRAW} package \cite{axodraw} was used.}
%%%%%%%%%%%%%%%%%%%%%%%%%%%%%%%%%%%%%%%%%%%%%%%%%%%%%%%%%%%%%%%%%%%%%%%%
\begin{figure}[htb]
\refstepcounter{figure}
\label{Fig:1}
\addtocounter{figure}{-1}
\begin{center}
\setlength{\unitlength}{1cm}
\begin{picture}(5.0,4.5)
\put(-3,0){
%\mbox{\epsfysize=5cm\epsffile{fig1.eps}}}
\mbox{\epsfysize=5cm\epsffile{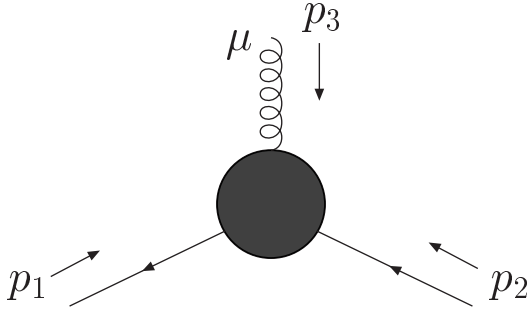}}}
\end{picture}
\vspace*{-4mm}
\caption{Kinematics of the quark-gluon vertex.}
\end{center}
\end{figure}
%%%%%%%%%%%%%%%%%%%%%%%%%%%%%%%%%%%%%%%%%%%%%%%%%%%%%%%%%%%%%%%%%%%%%%
The momentum of the outgoing quark is denoted by $p_1$, 
$p_2$ is the momentum of the incoming quark, whereas $p_3$ 
and $\mu$ are the momentum and the Lorentz index of the gluon,
respectively.
All momenta are ingoing, $p_1+p_2+p_3=0$.
The lowest-order quark-gluon vertex is
\begin{equation}
\label{qg_vert} 
g(T^{a})_{ji}\left[\gamma_{\mu}\right]_{\beta \alpha} ,
\end{equation}
where $T^{a}$ are colour matrices corresponding to the
fundamental representation of the gauge group.
As a rule, it will be implied that the SU($N$) group
is considered, with $N$ being the number of colours
(we can put $N=3$ in the end).

When one calculates radiative corrections to the quark-gluon
vertex, other Dirac matrix structures arise, in addition
to $\gamma_{\mu}$, Eq.~(\ref{qg_vert}).
The total number of independent structures is 12
(see, e.g., in \cite{PT,BC1}, and also in Appendix~A of this paper). 
Extracting the over-all colour structure, we can present the 
one-particle irreducible quark-gluon vertex as
\begin{equation}
\label{qqg}
\Gamma_{\mu}^{a}(p_1,p_2,p_3)
= g T^{a}\, \Gamma_{\mu}(p_1,p_2,p_3) \; ,
\end{equation}
where matrix notation in both colour and Dirac matrices is 
understood.

At the one-loop level, we have two contributions to the
quark-gluon vertex which are shown in Fig.~2. 
Their colour factors are 
proportional to $(C_F-\half C_A)$ and
$C_A$, respectively, where $C_F$ and $C_A$ denote
eigenvalues of the quadratic Casimir operator in
the fundamental and adjoint representations, respectively.
For the SU($N$) gauge group,
\begin{equation}
C_A=N, \hspace{10mm}
C_F=\frac{N^2-1}{2N} \; .
\end{equation}
%%%%%%%%%%%%%%%%%%%%%%%%%%%%%%%%%%%%%%%%%%%%%%%%%%%%%%%%%%%%%%%%%%%%%%%%
\begin{figure}[htb]
\refstepcounter{figure}
\label{Fig:2}
\addtocounter{figure}{-1}
\begin{center}
\setlength{\unitlength}{1cm}
\begin{picture}(6.0,5)
\put(-6,-1.5){
%\mbox{\epsfysize=8cm\epsffile{qqg-loops.eps}}}
\mbox{\epsfysize=8cm\epsffile{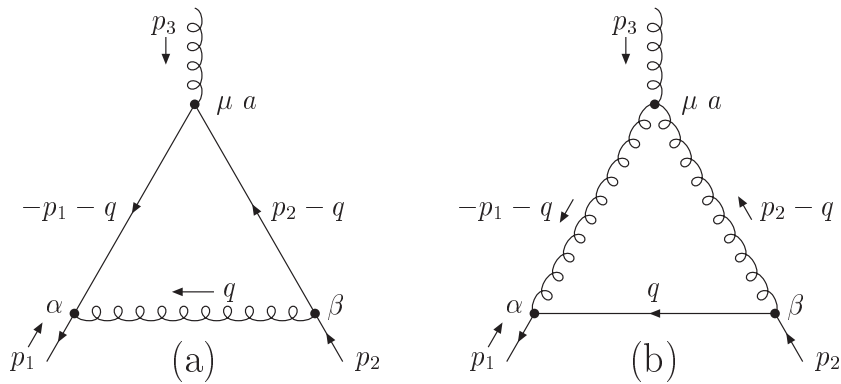}}}
\end{picture}
\caption{The two one-loop diagrams.}
\end{center}
\end{figure}
%%%%%%%%%%%%%%%%%%%%%%%%%%%%%%%%%%%%%%%%%%%%%%%%%%%%%%%%%%%%%%%%%%%%%%

The first, ``Abelian'' contribution is completely similar 
to the one-loop correction to the fermion-photon vertex
in QED. The difference is only in the over-all 
factor. Formally, we can get the one-loop QED vertex from
the considered QCD vertex by putting $C_A=0$.
The second diagram in Fig.~2 is essentially non-Abelian
and appears due to the self interaction of gluons.

If quarks are massive, it is clear that diagrams $a$ and $b$ 
in Fig.~2 involve ``triangle'' integrals with two and one massive 
lines, respectively:
\begin{eqnarray}
\label{defJ2}
J_2(\nu_1,\nu_2,\nu_3) &\equiv& \int
 \frac{\mbox{d}^n q}{ \left[(p_2 -q )^2-m^2\right]^{\nu_1}
                      \left[(p_1 +q )^2-m^2\right]^{\nu_2}  
      (q^2)^{\nu_3} } ,
\\ 
\label{defJ1}
J_1(\nu_1,\nu_2,\nu_3) &\equiv& \int
 \frac{\mbox{d}^n q}{ \left[(p_2 -q )^2\right]^{\nu_1}  
                      \left[(p_1 +q )^2\right]^{\nu_2}
      \left[q^2-m^2\right]^{\nu_3} } ,
\end{eqnarray}
where $n=4-2\ep$ is the space-time dimension in the 
framework of dimensional regularization \cite{dimreg}.
Understanding the subscript of $J$ as the number of massive
propagators (cf.\ in \cite{BD-TMF}), we can extend this 
notation to the massless integrals $J(\nu_1,\nu_2,\nu_3)$ 
considered in \cite{DOT1}, via
\begin{equation}
J(\nu_1,\nu_2,\nu_3)\leftrightarrow J_0(\nu_1,\nu_2,\nu_3) \; .
\end{equation}
Integrals with Lorentz indices can be reduced to the scalar
ones using the standard techniques \cite{BF+PV}
(see also in \cite{PLB'91,Tar}).
Using the integration-by-parts technique \cite{ibp}
(see also in \cite{JPA}), all integrals with higher integer powers 
of propagators can be algebraically reduced to integrals
with the powers equal to one or zero (for details, see
Appendix~B).

As in \cite{DOT1}, we shall extract from the 
expressions for one-loop integrals a factor
\begin{equation}
\label{eta}
\eta \equiv
\frac{\Gamma^2(\frac{n}{2}-1)}{\Gamma(n-3)} \;
     \Gamma(3-{\textstyle{n\over2}}) =
\frac{\Gamma^2(1-\varepsilon)}{\Gamma(1-2\varepsilon)} \;
\Gamma(1+\varepsilon) .
\end{equation}
A natural extension of the notation used in \cite{DOT1} 
is to introduce the functions $\varphi_i$ ($i=1,2$) such that
\begin{equation}
J_i(1,1,1)={\mbox{i}}\; \pi^{n/2}\; \eta\; \varphi_i(p_1^2,p_2^2,p_3^2;m).
\end{equation}
In this sense, the function $\varphi$, Eq.~(2.14) of \cite{DOT1},
would correspond to $\varphi_0$, which also represents the massless
limit of $\varphi_i$ ($i=1,2$).
Moreover, we can reserve the notations $J_3$ and $\varphi_3$ for the
triangle integral with all three massive lines, which occurs
in the three-gluon vertex (the massive quark loop contribution).

Then, for the two-point integrals we introduce functions
\begin{equation}
\kappa_i(p_l^2;m)\equiv \kappa_{i,l} ,
\end{equation}
where $p_l$ ($l=1,2,3$) is the external momentum of the two-point
function, whereas the subscript ($i=0,1,2$) shows how many
of the two internal propagators are massive.
In this way, $\kappa_0$ (coinciding with the $\kappa$ defined in
Eq.~(2.15) of \cite{DOT1})
corresponds to the two-point function with massless lines,
\begin{equation}
\label{J1_110}
J_1(1,1,0)=J_0(1,1,0)={\mbox{i}}\; \pi^{n/2}\; \eta\; \kappa_{0,3} \; ,
\end{equation}
and analogously for $J_0(0,1,1)$ and $J_0(1,0,1)$, with $\kappa_{0,1}$
and $\kappa_{0,2}$, respectively. Then,
$\kappa_1$ corresponds to the two-point function with one massive
and one massless line,
\begin{eqnarray}
\label{j1_011}
J_1(0,1,1)=J_2(0,1,1)={\mbox{i}}\; \pi^{n/2}\; \eta\; \kappa_{1,1} , \\
\label{j1_101}
J_1(1,0,1)=J_2(1,0,1)={\mbox{i}}\; \pi^{n/2}\; \eta\; \kappa_{1,2} .
\end{eqnarray}
Finally, $\kappa_2$ corresponds to the two-point function with two
massive lines,
\begin{equation}
\label{J2_110}
J_2(1,1,0)=J_3(1,1,0)={\mbox{i}}\; \pi^{n/2}\; \eta\; \kappa_{2,3} ,
\end{equation}
and similarly for $J_3(1,0,1)$ and $J_3(0,1,1)$ (which would involve
$\kappa_{2,2}$ and $\kappa_{2,1}$, respectively).
The massless two-point functions introduced in Eq.~(2.15) of \cite{DOT1} 
can be identified as
$\kappa_i\leftrightarrow \kappa_{0,i}$.

The new feature, as compared to the massless case, is the
appearance of the ``tadpole'' integral
\begin{equation}
\label{J1_001}
J_1(0,0,1)=J_2(1,0,0)=J_2(0,1,0)={\mbox{i}}\pi^{n/2}\;
\frac{\Gamma(1\!+\!\ep)}{\ep(1\!-\!\ep)} (m^2)^{1-\ep}
= {\mbox{i}}\pi^{n/2}\; \eta m^2 \widetilde{\kappa} ,
\hspace{4mm}
\end{equation}
with 
\begin{equation}
\label{kappa_tilde}
\widetilde{\kappa} \equiv\widetilde{\kappa}(m^2)
\equiv \frac{\Gamma(1-2\ep)}{\Gamma^2(1-\ep)}\;
\frac{1}{\ep(1-\ep)}\; (m^2)^{-\ep} .
\end{equation}
Let us recall that the massless tadpoles vanish in
the framework of dimensional regularization \cite{dimreg},
\begin{equation}
J_1(1,0,0)=J_1(0,1,0)=J_2(0,0,1)=0.
\end{equation}

We shall also introduce some notation to keep track of the various
orders in the perturbative expansion.
For a quantity $X$ (e.g. any of the scalar functions contributing to the
propagators or the vertices), we shall denote the zero-loop-order
contribution as $X^{(0)}$, and the one-loop-order contribution as
$X^{(1)}$, so that the perturbative expansion looks like
\begin{equation}
\label{X01}
X = X^{(0)} + X^{(1)} + \ldots
\end{equation}

\subsection{Two-point functions}

The lowest-order gluon propagator is
\begin{equation}
\label{gl_prop}
-\mbox{i} \delta^{a_1 a_2} \; \frac{1}{p^2}
\left( g_{\mu_1 \mu_2} - \xi \; \frac{p_{\mu_1} p_{\mu_2}}{p^2}
\right) ,
\end{equation}
where $\xi$ is the gauge parameter corresponding to a general
covariant gauge, defined such that $\xi=0$ is the Feynman gauge.
Here and henceforth, a causal prescription is understood,
$1/p^2 \leftrightarrow 1/(p^2 +\mbox{i}0)$.
For the present purposes, loop corrections to the gluon propagator 
(see, e.g., Eqs.~(2.7) and (C.1) of \cite{DOT1}) are not required.

We shall denote the quark propagator as $S(p)$.
The two scalar functions 
$\alpha(p^2)$ and $\beta(p^2)$ in the 
inverse quark propagator are defined via
\begin{equation}
\mbox{i} S^{-1}(p) \equiv \alpha(p^2)\pslash + \beta(p^2) I \; , 
\end{equation}
where $\pslash\equiv p^{\mu}\gamma_{\mu}$, whereas $I$ is the unit matrix
in the space of Dirac matrices.
At the lowest order, $\alpha^{(0)}=1$ and $\beta^{(0)}=-m$.
For the next-to-leading order, one needs to calculate 
the one-loop diagram shown in Fig.~3, which yields
(for $n$ near 4, see \cite{Muta})
%%%%%%%%%%%%%%%%%%%%%%%%%%%%%%%%%%%%%%%%%%%%%%%%%%%%%%%%%%%%%%%%%%%%%%%%
\begin{figure}[htb]
\refstepcounter{figure}
\label{Fig:3}
\addtocounter{figure}{-1}
\begin{center}
\setlength{\unitlength}{1cm}
\begin{picture}(5.0,3.5)
\put(-1.7,0){
%\mbox{\epsfysize=4cm\epsffile{fig_q.eps}}}
\mbox{\epsfysize=4cm\epsffile{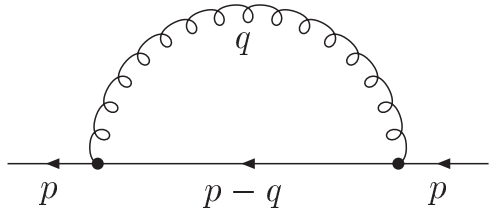}}}
\end{picture}
\caption{Quark self-energy diagram.}
\end{center}
\end{figure}
%%%%%%%%%%%%%%%%%%%%%%%%%%%%%%%%%%%%%%%%%%%%%%%%%%%%%%%%%%%%%%%%%%%%%%
\begin{eqnarray}
\alpha^{(1)}(p^2) &=& 
\frac{g^2\; \eta}{(4\pi)^{n/2}}\; \frac{C_F}{2 p^2}\;
(n-2)(1-\xi) \left[ (p^2+m^2) \kappa_1(p^2;m) 
- m^2 \widetilde{\kappa}(m^2)
\right] \; ,
\\
\beta^{(1)}(p^2) &=& 
-\frac{g^2\; \eta}{(4\pi)^{n/2}}\; C_F \; m (n-\xi) \kappa_1(p^2;m) 
\; .
\end{eqnarray}
The ghost propagator is
\begin{equation}
\label{gh_se}
\widetilde{D}^{a_1 a_2}(p^2)
= {\mbox{i}}\, \delta^{a_1 a_2} \frac{G(p^2)}{p^2} \; .
\end{equation}
The lowest-order result is $G^{(0)}=1$, whereas the one-loop contribution reads
\begin{equation}
\label{G(1)}
G^{(1)}(p^2) = \frac{g^2 \; \eta}{(4\pi)^{n/2}} \;
 \frac{C_A}{4}
\left[ 2 + (n-3)\xi \right] \; \kappa_0(p^2)  .
\end{equation}
Note that in the Fried--Yennie gauge \cite{FY} (see also in \cite{AbrSol}), 
$\xi=-2$, $G^{(1)}(p^2)$ is finite as $n\to 4$.
Moreover, if one chooses $\xi=-2/(n-3)$ as the $n$-dimensional 
generalization of this gauge
\cite{dipl,Adkins}, then the r.h.s.\ of Eq.~(\ref{G(1)}) vanishes. 

\subsection{Ward--Slavnov--Taylor identity}

The WST identity \cite{WST}
for the quark-gluon vertex $\Gamma_{\mu}(p_1,p_2,p_3)$
reads (see, e.g., in \cite{PT-QCD,MarPag})
\begin{equation}
\label{WST_qqg}
p_3^{\mu}\; \Gamma_{\mu}(p_1,p_2,p_3)
=G(p_3^2)\left[ S^{-1}(-p_1)\; H(p_1,p_2,p_3)
-\overline{H}(p_2,p_1,p_3)\; S^{-1}(p_2) \right],
\end{equation}
where $G(p^2)$ (see Eq.~(\ref{gh_se})) is the scalar function associated 
with the ghost propagator.

The function $H$ (and the ``conjugated'' function $\overline{H}$)
involves the complete four-point quark-quark-ghost-ghost vertex, as shown in 
Fig.~4. 
%%%%%%%%%%%%%%%%%%%%%%%%%%%%%%%%%%%%%%%%%%%%%%%%%%%%%%%%%%%%%%%%%%%%%%%%
\begin{figure}[htb]
\refstepcounter{figure}
\label{Fig:4}
\addtocounter{figure}{-1}
\begin{center}
\setlength{\unitlength}{1cm}
\begin{picture}(6.0,4.5)
\put(-4.3,-0.5){
%\mbox{\epsfysize=5.5cm\epsffile{fig-wst4.eps}}}
\mbox{\epsfysize=5.5cm\epsffile{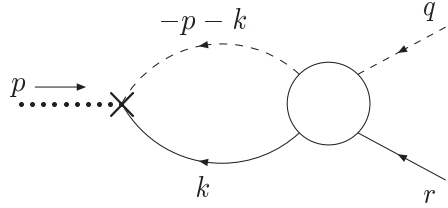}}}
\end{picture}
\caption{Graphical representation of the $H$ function.}
\end{center}
\end{figure}
%%%%%%%%%%%%%%%%%%%%%%%%%%%%%%%%%%%%%%%%%%%%%%%%%%%%%%%%%%%%%%%%%%%%%%
%%%%%%%%%%%%%%%%%%%%%%%%%%%%%%%%%%%%%%%%%%%%%%%%%%%%%%%%%%%%%%%%%%%%%%%%
\begin{figure}[htb]
\refstepcounter{figure}
\label{Fig:5}
\addtocounter{figure}{-1}
\begin{center}
\setlength{\unitlength}{1cm}
\begin{picture}(6.0,3.7)
\put(-4.2,0.){
%\mbox{\epsfysize=4.5cm\epsffile{fig-wst5.eps}}}
\mbox{\epsfysize=4.5cm\epsffile{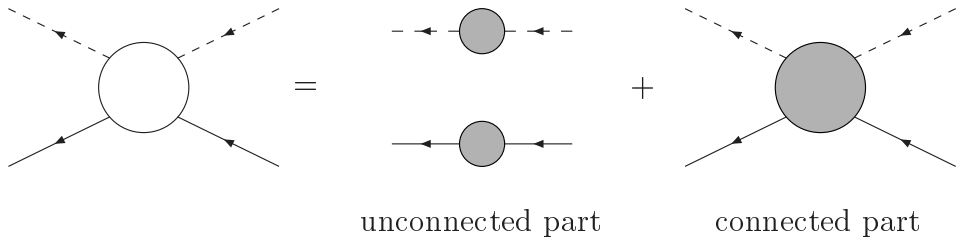}}}
\end{picture}
\caption{Unconnected and connected parts of the 
quark-quark-ghost-ghost amplitude.}
\end{center}
\end{figure}
%%%%%%%%%%%%%%%%%%%%%%%%%%%%%%%%%%%%%%%%%%%%%%%%%%%%%%%%%%%%%%%%%%%%%%
To get the $H$ function, we need to ``join'' the out-quark
and out-ghost lines in a non-standard vertex (denoted by a cross)
and integrate over the resulting loop momentum.
It should be noted that the complete quark-quark-ghost-ghost vertex
involved in the WST identity can be decomposed into a connected
and an unconnected piece, as shown in Fig.~5. Moreover,
the connected part can be further split in terms of
the proper (one-particle irreducible) vertices, see Fig.~6.
%%%%%%%%%%%%%%%%%%%%%%%%%%%%%%%%%%%%%%%%%%%%%%%%%%%%%%%%%%%%%%%%%%%%%%%%
\begin{figure}[htb]
\refstepcounter{figure}
\label{Fig:6}
\addtocounter{figure}{-1}
\begin{center}
\setlength{\unitlength}{1cm}
\begin{picture}(6.0,4.5)
\put(-5.5,0.){
%\mbox{\epsfysize=4.5cm\epsffile{fig-wst6.eps}}}
\mbox{\epsfysize=4.5cm\epsffile{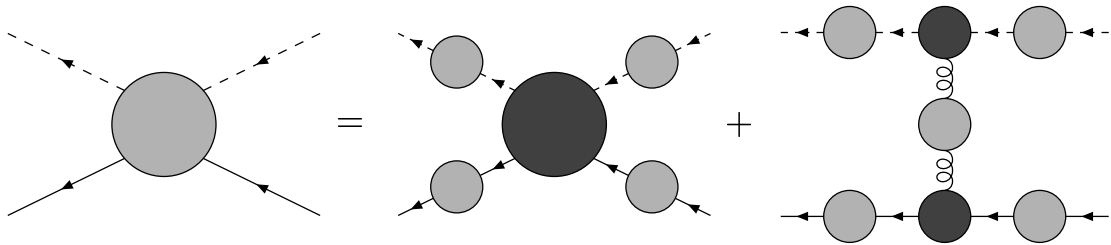}}}
\end{picture}
\caption{Connected part of the quark-quark-ghost-ghost 
amplitude in terms of proper (dark) vertices.}
\end{center}
\end{figure}
%%%%%%%%%%%%%%%%%%%%%%%%%%%%%%%%%%%%%%%%%%%%%%%%%%%%%%%%%%%%%%%%%%%%%%

We note that the first diagram on the r.h.s. of the equation shown 
in Fig.~6 (the diagram involving the proper four-point function) 
does not not have a zero-loop (tree) contribution. Its perturbative 
expansion starts from the one-loop boxes, corresponding to 
the exchange by two gluons.
Since the $H$ function involves an extra loop integration
(see Fig.~4), this proper four-point function does not
contribute to the one-loop-order $H$ function, $H^{(1)}$,
which is shown in Fig.~7.
%%%%%%%%%%%%%%%%%%%%%%%%%%%%%%%%%%%%%%%%%%%%%%%%%%%%%%%%%%%%%%%%%%%%%%%%
\begin{figure}[htb]
\refstepcounter{figure}
\label{Fig:h1}
\addtocounter{figure}{-1}
\begin{center}  
\setlength{\unitlength}{1cm}
\begin{picture}(6.0,4.5)
\put(-6.5,-0.5){
%\mbox{\epsfysize=7cm\epsffile{h1.eps}}}
\mbox{\epsfysize=7cm\epsffile{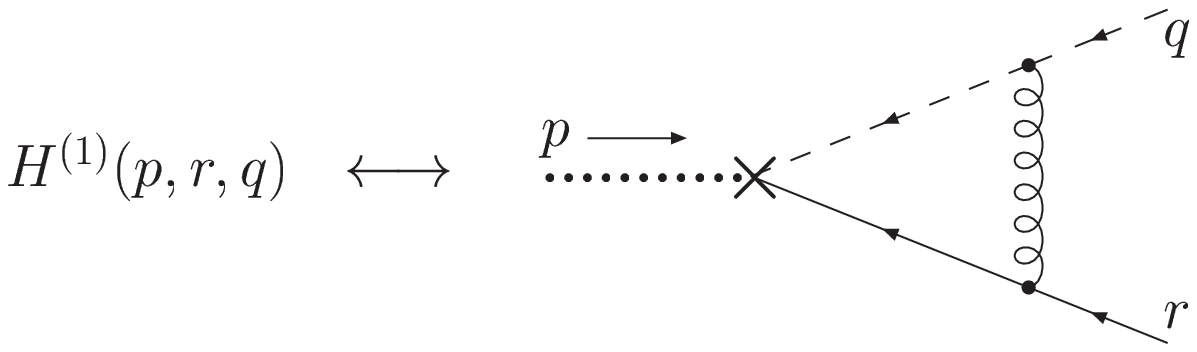}}}
\end{picture}
\caption{The one-loop order function $H^{(1)}$.}
\vspace*{-4mm}
\end{center}
\end{figure}
%%%%%%%%%%%%%%%%%%%%%%%%%%%%%%%%%%%%%%%%%%%%%%%%%%%%%%%%%%%%%%%%%%%%%%  

The $H$ function can be decomposed in terms of scalar functions
(``form factors'') as
\begin{equation}
\label{H}
H(p_1,p_2,p_3)=\chi_0(p_1^2,p_2^2,p_3^2) I
+\chi_1(p_1^2,p_2^2,p_3^2) \pslash_1
+\chi_2(p_1^2,p_2^2,p_3^2) \pslash_2
+\chi_3(p_1^2,p_2^2,p_3^2) \sigma_{\mu\nu}p_1^{\mu}p_2^{\nu} ,
\end{equation}
with
\begin{equation}
\sigma_{\mu \nu}=\half \left(\gamma_{\mu} \gamma_{\nu} 
                            - \gamma_{\nu} \gamma_{\mu} \right) \; .
\end{equation}
The ``conjugated'' function $\overline{H}$ can be written in terms 
of the same scalar functions,
\begin{equation}
\label{Hbar}
\overline{H}(p_2,p_1,p_3)=\chi_0(p_2^2,p_1^2,p_3^2) I
-\chi_2(p_2^2,p_1^2,p_3^2) \pslash_1
-\chi_1(p_2^2,p_1^2,p_3^2) \pslash_2
+\chi_3(p_2^2,p_1^2,p_3^2) \sigma_{\mu\nu}p_1^{\mu}p_2^{\nu} .
\end{equation}
At the lowest order, $\chi_0^{(0)}=1$ and $\chi_i^{(0)}=0$ ($i=1,2,3$).
The one-loop results for the $\chi_i$ functions (valid for
arbitrary values of $n$ and $\xi$) are presented in Appendix~D. 

At the one-loop level, it is convenient to ``split'' the 
WST identity into two separate identities, corresponding
to the contributions of the two diagrams shown in Fig.~2.
To do this, we need to rewrite the one-loop contribution
to the r.h.s.\ of (\ref{WST_qqg}) in terms of colour
coefficients $(C_F-\half C_A)$ and $C_A$, in analogy with
the two contributions to the l.h.s. On the r.h.s., all 
one-loop contributions are proportional to $C_A$, 
except for the quark self energies, which
contain $C_F$. Therefore, all we need to do is to
represent this $C_F$ as $(C_F-\half C_A)+\half C_A$.
In this way, we get two separate WST identities for
the contributions of diagrams $a$ and $b$ in Fig.~2,
\begin{eqnarray}
\label{WST1}
p_3^{\mu} \Gamma_{\mu}^{(1a)}(p_1,p_2,p_3)
&=& \left(C_F-\half C_A\right)\; C_F^{-1} \;
\left[ S^{-1}(-p_1) - S^{-1}(p_2) \right]^{(1)} ,
\\
\label{WST2}
p_3^{\mu} \Gamma_{\mu}^{(1b)}(p_1,p_2,p_3)
&=& \left[ S^{-1}(-p_1) \right]^{(0)}\; H^{(1)}(p_1,p_2,p_3)
- {\overline{H}}^{(1)}(p_2,p_1,p_3)\; 
     \left[ S^{-1}(p_2) \right]^{(0)}
\nn \\
&& + \half C_A \; C_F^{-1} \;
\left[ S^{-1}(-p_1) - S^{-1}(p_2) \right]^{(1)}\; H^{(0)}
\nn \\
&& + 2 G^{(1)}(p_3^2)\; 
\left[ S^{-1}(-p_1) - S^{-1}(p_2) \right]^{(0)}\; H^{(0)} ,
\end{eqnarray}
where, following the convention of (\ref{X01}),
the superscripts ``(0)'' and ``(1)'' correspond to
the zero-loop and one-loop contributions, respectively.

The first identity, Eq.~(\ref{WST1}), has, up to a factor, the same form 
as the Abelian (QED) identity, also known as the Ward--Fradkin--Takahashi
identity \cite{WFT}. The second identity, Eq.~(\ref{WST2}),
is the non-Abelian one.

\subsection{Decomposition of the quark-gluon vertex}

Keeping in mind the WST identity (\ref{WST_qqg}), it is useful to split
the quark-gluon vertex into a longitudinal part and a transverse part,
\begin{equation}
\label{LT}
\Gamma_{\mu}(p_1,p_2,p_3)
= \Gamma_{\mu}^{({\rm L})}(p_1,p_2,p_3) 
+ \Gamma_{\mu}^{({\rm T})}(p_1,p_2,p_3) ,
\end{equation}
where 
\begin{equation}
p_3^{\mu}\Gamma_{\mu}^{({\rm T})}(p_1,p_2,p_3)=0
\end{equation}
and, therefore, $\Gamma_{\mu}^{({\rm T})}$ does not contribute
to the l.h.s. of Eq.~(\ref{WST_qqg}).

In general, we shall just extend the decomposition of the QED vertex
suggested in \cite{BC1} (see also in Ref.~\cite{KRP}) to the 
QCD case. The longitudinal part of the vertex
can be represented as 
\begin{equation}
\label{GammaL}
\Gamma_{\mu}^{({\rm L})}(p_1,p_2,p_3)
= \sum\limits_{i=1}^4 \lambda_i(p_1^2,p_2^2,p_3^2)\;
L_{i,\mu}(p_1,p_2) ,
\end{equation}
with
\begin{eqnarray}
\label{L_i}
L_{1,\mu} &=& \gamma_{\mu},
\nn \\
L_{2,\mu} &=& (\pslash_1-\pslash_2)(p_1-p_2)_{\mu},
\nn \\
L_{3,\mu} &=& (p_1-p_2)_{\mu},
\nn \\
L_{4,\mu} &=& \sigma_{\mu \nu}\, (p_1-p_2)^{\nu} .
\end{eqnarray}
Using Eq.~(\ref{WST_qqg}), the functions $\lambda_i$ can be related to
the functions $\alpha$, $\beta$, $G$ and $\chi_i$.
For instance, in the ``Abelian'' case (i.e., when we consider only
the first, QED-like diagram in Fig.~2) the functions
$\lambda_i$ ($i=1,2,3,4$) would be equal to (up to a colour factor)
\begin{equation}
\label{lambdas_QED}
\half\left[ \alpha(p_1^2)+\alpha(p_2^2) \right], \hspace{5mm}
\frac{\alpha(p_1^2)-\alpha(p_2^2)}{2(p_1^2-p_2^2)}, \hspace{5mm}
-\frac{\beta(p_1^2)-\beta(p_2^2)}{p_1^2-p_2^2},  \hspace{5mm}
\mbox{and} \hspace{5mm} 0 ,
\end{equation}
respectively (see in \cite{BC1,KRP})\footnote{Since in the QED case
the longitudinal functions have such simple representations
(\ref{lambdas_QED}) in terms of $\alpha(p_{1,2}^2)$ and
$\beta(p_{1,2}^2)$, there was no need in Refs.~\cite{BC1,KRP}
to introduce a special notation for these functions.
In the presence of the non-Abelian contribution, the situation
becomes more complicated. This is why we have introduced
longitudinal functions $\lambda_i$ in Eq.~(\ref{GammaL}).}.

The transverse part of the vertex, which does not contribute to the
WST  identity (\ref{WST_qqg}), can be presented as \cite{BC1}
\begin{eqnarray}
\label{GammaT}
\Gamma_{\mu}^{({\rm T})}(p_1,p_2,p_3)=\sum\limits_{i=1}^8 
\tau_{i}(p_1^2,p_2^2,p_3^2) \; T_{i,\mu}(p_1,p_2),
\end{eqnarray}
where the transverse tensors are the following:
\begin{eqnarray}
\label{T_i}
T_{1,\mu}
&=& p_{1\mu}(p_2 p_3)-p_{2\mu}(p_1 p_3),
\nn \\
T_{2,\mu}&=&
-\left[p_{1\mu}(p_2 p_3)-p_{2\mu}(p_1 p_3)\right](\pslash_1-\pslash_2),
\nn \\
T_{3,\mu}&=&p_3^2\gamma_{\mu}-p_{3\mu}\pslash_3,
\nn \\
T_{4,\mu}&=&\left[p_{1\mu}(p_2 p_3)-p_{2\mu}(p_1 p_3)\right]
\sigma_{\nu \lambda}\, p_1^{\nu}p_2^{\lambda},
\nn \\
T_{5,\mu}&=& \sigma_{\mu \nu}\, p_3^\nu,
\nn \\
T_{6,\mu}&=&\gamma_{\mu} (p_1^2-p_2^2)+(p_1-p_2)_{\mu}\pslash_3,
\nn \\
T_{7,\mu}&=&-\half
(p_1^2-p_2^2)\left[\gamma_{\mu}(\pslash_1-\pslash_2)-(p_1-p_2)_{\mu}\right]
-(p_1-p_2)_{\mu}\sigma_{\nu\lambda}\, p_1^{\nu}p_2^{\lambda},
\nn \\
T_{8,\mu}&=&
-\gamma_{\mu}\sigma_{\nu\lambda}\, p_1^{\nu}p_2^{\lambda}
+p_{1\mu}\pslash_2-p_{2\mu}\pslash_1.
\end{eqnarray}
The connection of $\lambda$'s and $\tau$'s with the naive decomposition
basis is discussed in Appendix~A.

Applying charge conjugation to the quark-gluon vertex, 
i.e., interchanging quark and anti-quark, 
the following relation is obtained (see, e.g., in Ref.~\cite{KRP}):
\begin{eqnarray}
\label{c_vertex_c_inv}
{\rm C}\;\Gamma_{\mu}(p_1,p_2,p_3)\;{\rm C}^{-1}
=-\Gamma_{\mu}^{\rm T}(p_2,p_1,p_3).
\end{eqnarray}
Interchanging the quark momenta ($p_1 \leftrightarrow p_2$)
and using the fact that
\begin{eqnarray}
{\rm C}\;\gamma_{\mu}\;{\rm C}^{-1}=-\gamma_{\mu}^{\rm T},
\end{eqnarray}
one finds that all $L_\mu$ and $T_\mu$ 
are odd, except for $L_{4,\mu}$ and $T_{6,\mu}$:
\begin{eqnarray}
L_{i,\mu}(p_1,p_2)&=&-L_{i,\mu}^{\rm T}(p_2,p_1), 
\qquad i=1,2,3,
\nn \\
L_{4,\mu}(p_1,p_2)&=& L_{4,\mu}^{\rm T}(p_2,p_1), 
\nn \\
T_{i,\mu}(p_1,p_2)&=&-T_{i,\mu}^{\rm T}(p_2,p_1), 
\qquad i=1,2,3,4,5,7,8,
\nn \\
T_{6,\mu}(p_1,p_2)&=& T_{6,\mu}^{\rm T}(p_2,p_1).
\end{eqnarray}
To satisfy Eq.~(\ref{c_vertex_c_inv}), all $\lambda$'s and $\tau$'s
must be symmetric under the interchange of $p_1^2$ and $p_2^2$, except 
$\lambda_4$ and $\tau_6$, which are odd:
\begin{eqnarray}
\label{symmetry_of_taus}
\lambda_i(p_1^2,p_2^2,p_3^2) &=&
\lambda_i(p_2^2,p_1^2,p_3^2), \qquad i=1,2,3, 
\nn \\
\lambda_4(p_1^2,p_2^2,p_3^2) &=&
-\lambda_4(p_2^2,p_1^2,p_3^2) ,
\nn \\
\tau_i(p_1^2,p_2^2,p_3^2)
&=&\tau_i(p_2^2,p_1^2,p_3^2), \qquad i=1,2,3,4,5,7,8,
\nn \\
\tau_6(p_1^2,p_2^2,p_3^2)
&=&- \tau_6(p_2^2,p_1^2,p_3^2).
\end{eqnarray}
An important corollary of these relations is that in the
case $p_1^2=p_2^2\equiv p^2$ the $\lambda_4$ and $\tau_6$
functions must vanish,
\begin{equation}
\label{l4t6}
\lambda_4(p^2,p^2,p_3^2) = 0, \qquad
\tau_6(p^2,p^2,p_3^2) = 0 \; .
\end{equation}

Furthermore, in Ref.~\cite{KRP} a modification of the basis 
(\ref{GammaT})--(\ref{T_i}) has been proposed, which has an advantage
in dealing with kinematical singularities\footnote{In Ref.~\cite{KRP}, the
notation $\sigma_i$ and $S_i$ was used for what we call 
$\widetilde{\tau}_i$ and $\widetilde{T}_i$.}. 
Namely, the transverse part is represented as
\begin{eqnarray}
\label{GammaTtilde}
\Gamma_{\mu}^{({\rm T})}(p_1,p_2,p_3)=\sum\limits_{i=1}^8
\widetilde{\tau}_i(p_1^2,p_2^2,p_3^2) \; \widetilde{T}_{i,\mu}(p_1,p_2),
\end{eqnarray}
where
\begin{eqnarray}
\widetilde{T}_{4,\mu}
&=& \frac{2}{p_2^2-p_1^2}\left[ 2T_{4,\mu} - p_3^2 T_{7,\mu} \right] 
   = p_3^2[p_{1\mu} - p_{2\mu} - \gamma_\mu(\pslash_1 - \pslash_2)]
     -2(p_1 + p_2)_\mu \sigma_{\nu\lambda}\, p_1^\nu p_2^\lambda, \nn \\
\widetilde{T}_{i,\mu}&=&T_{i,\mu}, \hspace{10mm} (i\neq 4), \nn \\
\end{eqnarray}
with
\begin{equation}
\widetilde{\tau}_4=\quarter(p_2^2-p_1^2) \tau_4, \qquad
\label{tau7t} 
\widetilde{\tau}_7=\tau_7+\half p_3^2 \tau_4 .
\end{equation}
Moreover, as we shall see below, in the on-shell limit the following
modifications of $\lambda_2$ and $\lambda_3$ turn out to be useful:
\begin{equation}
\label{lambda23t}
\widetilde{\lambda}_2 \equiv \lambda_2 +\half p_3^2\tau_2,
\hspace{10mm}
\widetilde{\lambda}_3 \equiv \lambda_3 -\half p_3^2\tau_1 \; .
\end{equation}

%%%%%%%%%%%%%%%%%%%%%%%%%%%%%%%%%%%%%%%%%%%%%%%%%%%%%%%%%%%%%%%%%%%%%%%
\section{Off-shell results}
\setcounter{equation}{0}
%%%%%%%%%%%%%%%%%%%%%%%%%%%%%%%%%%%%%%%%%%%%%%%%%%%%%%%%%%%%%%%%%%%%%%%

Before presenting results for the $\lambda_i$ and $\tau_i$
functions, let us 
introduce the following notation for the Gram determinants
occurring in the denominators:
\begin{eqnarray}
\label{Eq:calK}
{\cal{K}} &\equiv& p_1^2 p_2^2 - (p_1 p_2)^2 \; ,
\\
\label{calM1}
{\cal{M}}_1 &\equiv&(p_1^2-m^2)(p_2^2-m^2)+m^2 p_3^2 \;, 
\\
\label{calM2}
{\cal{M}}_2 &\equiv&
(p_1^2-m^2)(p_2^2-m^2)p_3^2+m^2 (p_1^2-p_2^2)^2 \; .
\end{eqnarray}
In fact, ${\cal K}$ is a symmetric function of 
$p_1^2$, $p_2^2$ and $p_3^2$. It can be 
rewritten as $-\quarter\lambda(p_1^2,p_2^2,p_3^2)$,
where $\lambda(x,y,z)$ is the K\"all\'{e}n function
(cf.\ Eq.~(3.2) of \cite{DOT1}).
Note that ${\cal{M}}_1$ can also be represented as
\[
{\cal{M}}_1 
= p_1^2 p_2^2 +2 (p_1 p_2) m^2+m^4
= {\cal{K}} + \left[ (p_1 p_2)+m^2 \right]^2 \; .   
\]

To distinguish between the contributions of the two one-loop 
diagrams in Fig.~2, we shall use the letters $a$ and $b$:
\begin{equation}
\lambda_i^{(1)} = \lambda_i^{(1a)} + \lambda_i^{(1b)} \; ,
\hspace{10mm}
\tau_i^{(1)} = \tau_i^{(1a)} + \tau_i^{(1b)} \; .
\end{equation}
For the calculation we used the algebraic programming system 
{\sf REDUCE} \cite{reduce}.
Further technical details can be found in Appendices~A and B.

\subsection{Results for the longitudinal part of the vertex}

The general results for the longitudinal functions of the vertex are
reasonably compact, even in a general covariant gauge and arbitrary
dimension, and are given below for the two diagrams.

\subsubsection{Diagram $a$}

\begin{eqnarray}
\label{lam1(1a)}
\lambda_1^{(1a)}(p_1^2,p_2^2,p_3^2) &=&
\frac{g^2\eta \left(C_F-\half C_A\right)}{(4\pi)^{n/2}} 
      \frac{(n-2)(1-\xi)}{4 p_1^2 p_2^2}
\nn \\ 
&& \times
\left[ p_2^2 (p_1^2+m^2) \kappa_{1,1}+p_1^2 (p_2^2+m^2) \kappa_{1,2}
         -(p_1^2+p_2^2) m^2 \widetilde{\kappa} \right] \; ,
\\
\lambda_2^{(1a)}(p_1^2,p_2^2,p_3^2) &=&
\frac{g^2\eta \left(C_F-\half C_A\right)}{(4\pi)^{n/2}}\; 
\frac{(n-2)(1-\xi)}{4 p_1^2 p_2^2 (p_1^2-p_2^2)}
\nn \\
&& \times
\left[p_2^2 (p_1^2+m^2) \kappa_{1,1} -p_1^2 (p_2^2+m^2) \kappa_{1,2} 
         +(p_1^2-p_2^2) m^2 \widetilde{\kappa} \right] \; ,
\\
\lambda_3^{(1a)}(p_1^2,p_2^2,p_3^2) &=&
\frac{g^2\eta \left(C_F-\half C_A\right)}{(4\pi)^{n/2}}\;
\frac{(n-\xi) m}{p_1^2-p_2^2} 
\left( \kappa_{1,1}-\kappa_{1,2} \right) \; ,
\\
\lambda_4^{(1a)}(p_1^2,p_2^2,p_3^2) &=& 0 \; .
\end{eqnarray}
Note that diagram $a$ does not contribute to $\lambda_4^{(1)}$.

\subsubsection{Diagram $b$}

\begin{eqnarray}
\lambda_1^{(1b)}(p_1^2,p_2^2,p_3^2) &=&
- \frac{g^2\eta C_A}{(4\pi)^{n/2}} \frac{1}{16{\cal K}}
\bigg\{ 
(2\!-\!\xi) {\cal K}
\left[ 2 (p_1^2\!+\!p_2^2\!-\!2 m^2) \varphi_1
          -n \kappa_{1,1} -n \kappa_{1,2}-4 \kappa_{0,3} \right]
\nn \\ && 
+ \big[2+(n-3)\xi\big] (p_1^2-p_2^2)^2
\left[ (p_1 p_2) \varphi_1 + m^2 \varphi_1 + \kappa_{0,3} \right]
\nn \\ &&
+ \big[2+(n-3)\xi\big] (p_1^2-p_2^2)
\left[ p_2^2 \kappa_{1,2} - p_1^2 \kappa_{1,1} 
       +(p_1 p_2) \big( \kappa_{1,1} - \kappa_{1,2} \big) \right]
\nn \\ && 
-(n-2) (2-\xi){\cal{K}}\,m^2
\bigg[ \frac{\kappa_{1,1}}{p_1^2}
      +\frac{\kappa_{1,2}}{p_2^2}
      -\frac{p_1^2+p_2^2}{p_1^2 p_2^2} \widetilde{\kappa} \bigg]
\bigg\} \; ,
\\
\lambda_2^{(1b)}(p_1^2,p_2^2,p_3^2) &=&
\frac{g^2\eta C_A}{(4\pi)^{n/2}} \frac{1}{16{\cal K}(p_1^2\!-\!p_2^2)} 
\bigg\{ 
\big[ 2\!+\!(n\!-\!3)\xi \big] p_3^2 (p_1^2\!-\!p_2^2)
\left[ (p_1 p_2)\varphi_1 + m^2 \varphi_1 + \kappa_{0,3} \right]
\nn \\ &&
+ \big[2+(n-3)\xi\big]p_3^2 
\left[ p_2^2 \kappa_{1,2} - p_1^2 \kappa_{1,1}
       +(p_1 p_2) \big( \kappa_{1,1} - \kappa_{1,2} \big) \right]
\nn \\ &&
+(n-2) (2-\xi) {\cal K} 
\bigg[ \frac{p_1^2+m^2}{p_1^2} \kappa_{1,1} 
      -\frac{p_2^2+m^2}{p_2^2} \kappa_{1,2}
         +\frac{p_1^2-p_2^2}{p_1^2 p_2^2} m^2 
       \widetilde{\kappa} \bigg] \bigg\} \; ,
\\
\lambda_3^{(1b)}(p_1^2,p_2^2,p_3^2) &=&
\frac{g^2\eta C_A}{(4\pi)^{n/2}}\; \frac{m}{8 {\cal M}_1}
\bigg\{ 
(n-4)\xi p_3^2 \big[ (p_1 p_2) +m^2 \big] \varphi_1
\nn \\ && 
+(n-3) \xi \left[ (p_2^2-p_3^2-m^2) \kappa_{1,1}
+(p_1^2-p_3^2-m^2) \kappa_{1,2} + p_3^2  \kappa_{0,3} \right]
\nn \\ && 
-(n-2)\xi \frac{{\cal M}_1 (p_1 p_2)}{p_1^2-p_2^2}
\left(\frac{\kappa_{1,1}}{p_1^2}-\frac{\kappa_{1,2}}{p_2^2}\right)
  + \big[ 4(n-1) - n \xi \big] {\cal M}_1  
\frac{\kappa_{1,1}-\kappa_{1,2}}{p_1^2-p_2^2}
\nn \\ && 
+(n-2)\xi \left[ 1- 
\frac{2 (p_1 p_2)+m^2}{p_1^2 p_2^2} (p_1 p_2) 
\right] m^2 \widetilde{\kappa}
\bigg\} \; ,
\\
\label{lam4(1b)}
\lambda_4^{(1b)}(p_1^2,p_2^2,p_3^2) &=& 
\frac{g^2\eta C_A}{(4\pi)^{n/2}}\; 
\frac{\xi m (p_1^2-p_2^2)}{16 {\cal K} {\cal M}_1}
\bigg\{ (n\!-\!3) p_3^2 \big[(p_1 p_2)\!+\!m^2\big]^2 \varphi_1
+ (n-3){\cal{K}}(\kappa_{1,1}+\kappa_{1,2})
\nn \\ &&  
+ {\cal K} p_3^2 \varphi_1
+(n\!-\!3)\big[(p_1 p_2)\!+\!m^2\big]
\left[ (p_1 p_3) \kappa_{1,1} \!+\! (p_2 p_3) \kappa_{1,2} \!+\!
p_3^2 \kappa_{0,3} \right]
\nn \\ && 
+(n-2) {\cal K} \left[ 
\frac{2 (p_1 p_2)+m^2}{p_1^2 p_2^2} m^2 \widetilde{\kappa}
 + \frac{{\cal M}_1}{p_1^2-p_2^2} 
\left(\frac{\kappa_{1,1}}{p_1^2}-\frac{\kappa_{1,2}}{p_2^2}\right)
\right] 
\bigg\} \; .
\end{eqnarray}

We have checked that these expressions, together with the results for the
two-point functions and for the
$\chi_i$ functions (given in Appendix~D), 
satisfy the WST identities (\ref{WST1})--(\ref{WST2}) 
at the one-loop level, for arbitrary values of $n$ and $\xi$.

Results for the $\lambda_i$ functions 
in the Feynman gauge ($\xi=0$) can be easily obtained from the
expressions presented above. 
We just note that 
\begin{eqnarray}
\left.\lambda_3^{(1b)}(p_1^2,p_2^2,p_3^2)\right|_{\xi=0} &=&
\frac{g^2\eta C_A}{(4\pi)^{n/2}}
\frac{(n-1) m}{2(p_1^2 - p_2^2)}
\left( \kappa_{1,1}-\kappa_{1,2} \right) \; ,
\\
\left.\lambda_4^{(1b)}(p_1^2,p_2^2,p_3^2)\right|_{\xi=0} &=& 0.
\end{eqnarray}

\subsection{Results for the transverse part of the vertex}

In an arbitrary covariant gauge, the results for most of 
the $\tau_i$ functions are rather cumbersome. 
The expressions can be made more compact 
by introducing certain linear combinations of the $\kappa_{i,l}$ 
functions, namely
\begin{eqnarray}
\label{combs}
&& (p_1 p_3) \kappa_{1,1}+(p_2 p_3) \kappa_{1,2}+p_3^2 \kappa_{i,3},
\quad
\kappa_{1,1} + \kappa_{1,2}-2 \kappa_{i,3},
\nn \\ 
&& \kappa_{1,1} + \kappa_{1,2}-2 \widetilde{\kappa}, \quad
\kappa_{1,1} + \kappa_{1,2}, \quad
\kappa_{1,1} - \kappa_{1,2} \; ,
\end{eqnarray}
where we should take, for the subscript of $\kappa_{i,3}$,
$i=2$ for diagram $a$ and $i=0$ for diagram $b$.
Note that the basis (\ref{combs}) is over-complete: we have got
five combinations of the four $\kappa$'s. 
This degree of freedom has allowed us to present the
results in a more compact form. 

We have collected results in an arbitrary covariant gauge in Appendix~E.
Here we present results for the $\tau_i$ functions in the Feynman gauge.

\subsubsection{Transverse functions in Feynman gauge: diagram $a$}

\begin{eqnarray}
\label{tau1(1a)}
\left.\tau_1^{(1a)}(p_1^2,p_2^2,p_3^2)\right|_{\xi=0} &=&
\frac{g^2\eta \left(C_F-\half C_A\right)}{(4\pi)^{n/2}} \;
\frac{n m}{2{\cal{K}}} 
\Biggl\{
2\left[ m^2-(p_1 p_2)\right]\varphi_2
+\kappa_{1,1}+\kappa_{1,2}-2\kappa_{2,3} 
\nn \\
&& 
+ (p_1-p_2)^2 \; \frac{\kappa_{1,1}-\kappa_{1,2}}{p_1^2-p_2^2}
\Biggr\} \; , \label{tau1-a}
\\
\left.\tau_2^{(1a)}(p_1^2,p_2^2,p_3^2)\right|_{\xi=0} &=&
\frac{g^2\eta \left(C_F-\half C_A\right)}{(4\pi)^{n/2}} \;
\frac{1}{8{\cal{K}}}
\Biggl\{
2 (4m^2-p_3^2) \varphi_2 
- \frac{2(n-1)}{{\cal{K}}} \left[ (p_1 p_2)-m^2 \right]^2 p_3^2 \varphi_2
\nn \\
&& - \frac{2(n-1)}{{\cal{K}}} \left[ (p_1 p_2)-m^2 \right]
\left[ (p_1 p_3)\kappa_{1,1} + (p_2 p_3)\kappa_{1,2} 
+ p_3^2 \kappa_{2,3} \right]
\nn \\
&& -(n-2)\frac{m^2(p_1 p_2)}{p_1^2 p_2^2} 
\left( \kappa_{1,1}+\kappa_{1,2}-2{\widetilde{\kappa}}\right) 
- (n-4) \left(\kappa_{1,1}+\kappa_{1,2}\right)
\nn \\
&& +(n-2)\left[ \frac{m^2(p_1 p_2)(p_1^2+p_2^2)}{p_1^2 p_2^2}
-2m^2-(p_1-p_2)^2\right]
\frac{\kappa_{1,1}-\kappa_{1,2}}{p_1^2-p_2^2}
\Biggr\} \; ,
\\
\left.\tau_3^{(1a)}(p_1^2,p_2^2,p_3^2)\right|_{\xi=0} &=&
\frac{g^2\eta \left(C_F-\half C_A\right)}{(4\pi)^{n/2}}\;
\frac{1}{16 {\cal{K}}}
\Biggl\{ \frac{2(n-1)}{{\cal{K}}}
p_3^2 (p_1-p_2)^2 \left[(p_1 p_2)-m^2\right]^2 \varphi_2
\nn \\
&& +\frac{2(n-1)}{{\cal{K}}} (p_1-p_2)^2 \left[(p_1 p_2)-m^2\right]
\left[ (p_1 p_3)\kappa_{1,1} + (p_2 p_3)\kappa_{1,2}
+ p_3^2 \kappa_{2,3} \right]
\nn \\
&& -8(n-2) (p_1^2-m^2)(p_2^2-m^2)\varphi_2
-2 (p_1-p_2)^2 \left[ 4(n-1)m^2-p_3^2\right]\varphi_2
\nn \\
&& +4(n-2) \left[(p_1 p_2)-m^2\right] 
\left( \kappa_{1,1}+\kappa_{1,2}-2\kappa_{2,3} \right)
\nn \\
&& +(n-4) (p_1-p_2)^2 \left(\kappa_{1,1}+\kappa_{1,2}\right)
\nn \\
&& +(n-2) \left[ 1-\frac{m^2(p_1 p_2)}{p_1^2 p_2^2} \right]
 (p_1^2-p_2^2) \left(\kappa_{1,1}-\kappa_{1,2}\right)
\nn \\
&&
+(n-2) m^2 \left[ \frac{(p_1 p_2)(p_1^2+p_2^2)}{p_1^2 p_2^2}-2\right]
\left( \kappa_{1,1} + \kappa_{1,2}- 2{\widetilde{\kappa}} \right)
\Biggr\} \; ,
\\
\left.\tau_4^{(1a)}(p_1^2,p_2^2,p_3^2)\right|_{\xi=0} &=&0,
\\
\left.\tau_5^{(1a)}(p_1^2,p_2^2,p_3^2)\right|_{\xi=0} &=& 
-\frac{g^2\eta \left(C_F-\half C_A\right)}{(4\pi)^{n/2}}
(n-4) m \varphi_2,
\\
\left.\tau_6^{(1a)}(p_1^2,p_2^2,p_3^2)\right|_{\xi=0} &=&
\frac{g^2\eta \left(C_F-\half C_A\right)}{(4\pi)^{n/2}}\;
\frac{p_1^2-p_2^2}{16 {\cal{K}}} 
\Biggl\{
2(4m^2-p_3^2)\varphi_2 
- \frac{2(n-1)}{{\cal{K}}} p_3^2 \left[(p_1 p_2)-m^2\right]^2 \varphi_2
\nn \\
&& -\frac{2(n-1)}{{\cal{K}}} \left[(p_1 p_2)-m^2\right]
\left[ (p_1 p_3)\kappa_{1,1} + (p_2 p_3)\kappa_{1,2}
+ p_3^2 \kappa_{2,3} \right]
\nn \\
&& +(n-2)\left[ \frac{m^2 (p_1 p_2) (p_1^2+p_2^2)}{p_1^2 p_2^2}
-2m^2 - (p_1-p_2)^2 \right] \frac{\kappa_{1,1}-\kappa_{1,2}}{p_1^2-p_2^2}
\nn \\
&& -(n-2)\frac{m^2(p_1 p_2)}{p_1^2 p_2^2} 
\left( \kappa_{1,1} + \kappa_{1,2}- 2{\widetilde{\kappa}} \right)
- (n-4) \left(\kappa_{1,1}+\kappa_{1,2}\right)
\Biggr\} \; ,
\\
\left.\tau_7^{(1a)}(p_1^2,p_2^2,p_3^2)\right|_{\xi=0} &=& 0,
\\
\label{tau8(1a)}
\left.\tau_8^{(1a)}(p_1^2,p_2^2,p_3^2)\right|_{\xi=0} &=&
\frac{g^2\eta \left(C_F-\half C_A\right)}{(4\pi)^{n/2}}
\frac{(6 \!-\! n)}{2 {\cal{K}}}
\Bigl\{
p_3^2 [(p_1 p_2)-m^2] \varphi_2
\nn \\   
&& +(p_1 p_3) \kappa_{1,1}
+(p_2 p_3) \kappa_{1,2} + p_3^2 \kappa_{2,3}
\Bigr\}. \label{tau8-a}
\end{eqnarray}

The function $\tau_4^{(1a)}$ becomes non-zero when
$\xi\ne0$ (see Appendix~E). 
In fact, the results for $\tau_1^{(1a)}$ and $\tau_8^{(1a)}$
in an arbitrary covariant gauge can be obtained from 
Eqs.~(\ref{tau1-a}) and (\ref{tau8-a}) by changing
the overall factors $n$ and $(6-n)$ into $(n-\xi)$
and $\left[6-n+(n-4)\xi\right]$, respectively.

\subsubsection{Transverse functions in Feynman gauge: diagram $b$}

\begin{eqnarray}
\left.\tau_1^{(1b)}(p_1^2,p_2^2,p_3^2)\right|_{\xi=0} &=& 
- \frac{g^2\eta C_A}{(4\pi)^{n/2}}
\frac{(n - 1) m}{4 {\cal{K}}} 
\Biggl\{ 2 [(p_1 p_2)+m^2] \varphi_1
-\kappa_{1,1}-\kappa_{1,2} +2 \kappa_{0,3}
\nn \\
&& - (p_1-p_2)^2\; \frac{\kappa_{1,1}-\kappa_{1,2}}{p_1^2 - p_2^2}
\Biggr\} ,
\\
\left.\tau_2^{(1b)}(p_1^2,p_2^2,p_3^2)\right|_{\xi=0} &=&
\frac{g^2\eta C_A}{(4\pi)^{n/2}}\;
\frac{1}{16{\cal{K}}}
\Biggl\{ 
\frac{2(n-1)}{{\cal{K}}}p_3^2 \left[ (p_1 p_2)+m^2\right]^2 \varphi_1
+4(n-3)\left[ (p_1 p_2)+m^2\right] \varphi_1
\nn \\
&& +2 p_3^2 \varphi_1 
+ \frac{2(n-1)}{{\cal{K}}} \left[ (p_1 p_2)+m^2\right]
\left[ (p_1 p_3) \kappa_{1,1} + (p_2 p_3) \kappa_{1,2}
+ p_3^2 \kappa_{0,3} \right]
\nn \\
&& -2 (n-3) \left( \kappa_{1,1}+\kappa_{1,2}-2\kappa_{0,3}\right)
+(n-4) \left( \kappa_{1,1}+\kappa_{1,2} \right)
\nn \\
&& -(n-2)\frac{m^2 (p_1 p_2)}{p_1^2 p_2^2} 
\left( \kappa_{1,1}+\kappa_{1,2}-2{\widetilde{\kappa}} \right) 
-(n-4) (p_1-p_2)^2 \frac{\kappa_{1,1}-\kappa_{1,2}}{p_1^2-p_2^2}
\nn \\
&& +(n-2)m^2\left[ \frac{(p_1 p_2) (p_1^2+p_2^2)}{p_1^2 p_2^2} -2 \right]
\frac{\kappa_{1,1}-\kappa_{1,2}}{p_1^2-p_2^2}
\Biggr\} \; ,
\\
\left.\tau_3^{(1b)}(p_1^2,p_2^2,p_3^2)\right|_{\xi=0} &=&
-\frac{g^2\eta C_A}{(4\pi)^{n/2}}\;
\frac{1}{32{\cal{K}}}   
\Biggl\{
\frac{2(n-1)}{{\cal{K}}}p_3^2 (p_1-p_2)^2 
\left[ (p_1 p_2)+m^2\right]^2 \varphi_1
\nn \\
&& +\frac{2(n-1)}{{\cal{K}}} (p_1-p_2)^2 \left[ (p_1 p_2)+m^2\right]
\left[ (p_1 p_3) \kappa_{1,1} + (p_2 p_3) \kappa_{1,2}
+ p_3^2 \kappa_{0,3} \right]
\nn \\
&& -8(n-2)\left[ (p_1 p_2)+m^2\right]^2 \varphi_1
+(n-4) (p_1-p_2)^2 \left( \kappa_{1,1}+\kappa_{1,2} \right)
\nn \\
&& +2 p_3^2 (p_1-p_2)^2 \varphi_1
+4 (n-2) \left[ (p_1 p_2)+m^2\right] 
\left( \kappa_{1,1}+\kappa_{1,2} - 2\kappa_{0,3}\right)
\nn \\
&& +(n-2) \left[ \frac{m^2(p_1 p_2)}{p_1^2 p_2^2} +1 \right] 
(p_1^2-p_2^2) \left(\kappa_{1,1}-\kappa_{1,2}\right)
\nn \\
&& -m^2 (n-2) \left[ \frac{(p_1 p_2)(p_1^2+p_2^2)}{p_1^2 p_2^2}-2\right]
\left( \kappa_{1,1}+\kappa_{1,2}-2{\widetilde{\kappa}} \right)
\Biggr\} \; ,
\\
\left.\tau_4^{(1b)}(p_1^2,p_2^2,p_3^2)\right|_{\xi=0} &=& 0,
\\
\left.\tau_5^{(1b)}(p_1^2,p_2^2,p_3^2)\right|_{\xi=0} &=& -
\frac{g^2\eta C_A}{(4\pi)^{n/2}} \;
\frac{3}{2}m \varphi_1,
\\
\left.\tau_6^{(1b)}(p_1^2,p_2^2,p_3^2)\right|_{\xi=0} &=& 
\frac{g^2\eta C_A}{(4\pi)^{n/2}} \;
\frac{p_1^2-p_2^2}{32{\cal{K}}}
\Biggl\{ \frac{2(n-1)}{{\cal{K}}} p_3^2 
\left[ (p_1 p_2)+m^2 \right]^2 \varphi_1
+ 4 \left[ (p_1 p_2)+m^2 \right] \varphi_1
\nn \\
&& + \frac{2(n-1)}{{\cal{K}}} \left[ (p_1 p_2)+m^2 \right]
\left[ (p_1 p_3) \kappa_{1,1} + (p_2 p_3) \kappa_{1,2}
+ p_3^2 \kappa_{0,3} \right]
\nn \\
&& + 2 p_3^2 \varphi_1
+(n-4) \left( \kappa_{1,1}+\kappa_{1,2} \right)
- 2 \left( \kappa_{1,1}+\kappa_{1,2} - 2 \kappa_{0,3} \right)
\nn \\
&& -(n-2) \frac{m^2 (p_1 p_2)}{p_1^2 p_2^2}
\left( \kappa_{1,1}+\kappa_{1,2} - 2 {\widetilde{\kappa}} \right)
+(n-4) (p_1-p_2)^2 \frac{\kappa_{1,1}-\kappa_{1,2}}{p_1^2-p_2^2}
\nn \\
&& + (n-2)m^2 \left[\frac{(p_1 p_2) (p_1^2+p_2^2)}{p_1^2 p_2^2} -2 \right]
\frac{\kappa_{1,1}-\kappa_{1,2}}{p_1^2-p_2^2}
\Biggr\} \; ,
\\
\left.\tau_7^{(1b)}(p_1^2,p_2^2,p_3^2)\right|_{\xi=0} &=&0,
\\
\label{tau8(1b)}
\left.\tau_8^{(1b)}(p_1^2,p_2^2,p_3^2)\right|_{\xi=0} &=& 
\frac{g^2\eta C_A}{(4\pi)^{n/2}}
\frac{3}{4 {\cal{K}}} 
\Bigl\{ p_3^2 \left[ (p_1 p_2)+m^2 \right] \varphi_1 
+2 {\cal{K}} \varphi_1 
\nn \\
&& + (p_1 p_3) \kappa_{1,1}
+ (p_2 p_3) \kappa_{1,2}
+p_3^2 \kappa_{0,3} 
\Bigr\} .
\end{eqnarray}

In the limit of massless quarks, $m\to0$, the above results simplify
considerably. 
First of all, in this limit, 
$\tau_1^{(1)}=\tau_4^{(1)}=\tau_5^{(1)}=\tau_7^{(1)}=0$
(this holds also in an arbitrary covariant gauge).
Furthermore, $\varphi_1\to\varphi_0$, $\varphi_2\to\varphi_0$,
$\kappa_{1,i}\to\kappa_{0,i}$, $\kappa_{2,i}\to\kappa_{0,i}$
and $\widetilde{\kappa}\to0$. 

\subsection{Comparison with other papers}

As we have already mentioned, the contributions of the first 
diagram should coincide, up to an overall factor, with
the one-loop contribution to the fermion-photon vertex in QED.
Formally, to get the QED case from our expressions, we can put
$C_A \Rightarrow 0$,
$C_F \Rightarrow 1$,
$g \Rightarrow e$
(the absolute value of the charge of the electron).
Therefore, our expressions also provide
a one-loop correction to the QED vertex, in an arbitrary covariant gauge
and dimension.

To get the expressions for the dimensionally-regulated four-dimensional
case, we put $n=4-2\ep$ and expand our results (including the
scalar integrals, see Appendix~C) in $\ep$, keeping singular
($1/\ep$) and finite terms.
In the Feynman gauge ($\xi=0$) and in four dimensions,
we reproduce the well-known results from \cite{BC1}.
There are a few misprints in \cite{BC1}
which were pointed out in \cite{KRP} (p.~1252).
We agree with most of these corrections given in \cite{KRP},
except for the following:
Eqs.~(3.12), (3.14) and (A19) of \cite{BC1} are correct,
they should not be changed.

For an arbitrary value of $\xi$ and in four dimensions,
one-loop contributions to the QED vertex have been calculated
by K{\i}z{\i}lers{\"u}, Reenders and Pennington \cite{KRP}.  
We basically reproduce their results given by Eqs.~(60)--(64),
(67)--(74), and (87),
except for $\tau_3$ and $\tau_6$, Eqs.~(69) and (72).
Specifically, to get agreement with our results,
we had to change the following\footnote{There is also an
obvious misprint in their (A14):
the denominator $[(k-2)^2-m^2]$ should read $[(k-w)^2-m^2]$.
We are grateful to the authors
of \cite{KRP} for confirming all mentioned misprints.}:
in the sixth line of $\tau_3$,
\[ 
+ \quarter q^2 m^2 [ p^2 (p^2-m^2) L + k^2 (k^2-m^2) L' ]
\; \rightarrow \; 
- \quarter q^2 m^2 [ p^2 (p^2+m^2) L + k^2 (k^2+m^2) L' ],
\]
and in the eleventh line of $\tau_6$,
\[
-\frac{3m^2}{8\Delta^2} p^2 q^2 (p^4-k^4)  
[(m^2+k^2)L-(m^2+p^2)L']
\; \rightarrow \;
-\frac{3m^2}{8\Delta^2} q^2 (p^4-k^4)
[p^2(m^2+k^2) L+k^2(m^2+p^2) L'] .
\]

For massless three-dimensional QED, results are given 
by Bashir, K{\i}z{\i}lers{\"u} and Pennington 
in \cite{BKP}.
We reproduce their results for the longitudinal vertex, 
Eqs.~(50) and (51),
as well as the transverse parts,\footnote{However,
we note that some factors of $\pi$ are inconsistent.
The result for $J_0$ (Appendix, Eq.~(5)) should read
$-2/\sqrt{-k^2p^2q^2}$ (no $\pi$).
The right-hand sides of Eqs.~(40) and (55) should have an extra $\pi$.
Finally, $K_0$ and $K^{(0)}$ are related like $J_0$ and $J^{(0)}$.
We are grateful to the authors of \cite{BKP} for confirming
these misprints.}
Eqs.~(55), (57)--(60).

Comparison with some other papers is given in Section~4, where
the corresponding special limits are considered.

\subsection{Renormalization}

In the limit $n\to 4$ ($\ep\to 0$) the only function in the quark-gluon
vertex which has an ultraviolet (UV) singularity {\em at one loop}
is the $\lambda_1^{(1)}$ function.
In an arbitrary covariant gauge, the UV-singular part
of $\lambda_1^{(1)}$ reads
\begin{eqnarray}
\lambda_1^{(1,{\rm UV})}&=&\frac{g^2\eta}{(4\pi)^{2-\ep}}
\left[ (1-\xi)\left( C_F - \half C_A \right) 
+ {\textstyle{3\over4}} (2-\xi) C_A
\right] \left( \frac{1}{\ep} + \ldots \right)
\nn \\
&=& \frac{g^2\eta}{(4\pi)^{2-\ep}}
\left[ (1-\xi) C_F 
+ {\textstyle{1\over4}} (4-\xi) C_A \right]
\left( \frac{1}{\ep} + \ldots \right) .
\end{eqnarray}
In the first line, the contributions from the first and the second
diagram are explicitly separated.
This result coincides with Eqs.~(A.55)--(A.57) of
Ref.~\cite{Muta}\footnote{The contribution of the first diagram is also
in agreement with Eq.~(11.65) of \cite{Bailin+Love}, whereas for
the contribution of the second diagram there seems to be a misprint
in Eq.~(11.70) of \cite{Bailin+Love}. Namely, in {\em their} notation, the
factor $(1-\xi)$ in Eq.~(11.70) should read $(1+\xi)$.
Their $\xi$ and $\ep$ correspond to our
$(1-\xi)$ and $2\ep$, respectively.}.  

The divergent parts of the two-point functions are as follows:
\begin{eqnarray}
\alpha^{(1,{\rm UV})}&=&\frac{g^2\eta}{(4\pi)^{2-\ep}}
C_F (1-\xi) \left( \frac{1}{\ep} + \ldots \right) ,  
\\
\beta^{(1,{\rm UV})}&=&-\frac{g^2\eta}{(4\pi)^{2-\ep}}
m\, C_F (4-\xi) \left( \frac{1}{\ep} + \ldots \right) ,  
\\
G^{(1,{\rm UV})}&=&\frac{g^2\eta}{(4\pi)^{2-\ep}}
\frac{C_A}{4}\; (2+\xi) \left( \frac{1}{\ep} + \ldots \right) .
\end{eqnarray}
The results for $\alpha$ and $\beta$ agree with
Eq.~(2.5.138) of \cite{Muta} (our $\alpha$ and $\beta$
correspond to his $(-B)$ and $mA$, respectively),
and with Eq.~(11.55) of \cite{Bailin+Love}.
Among the $\chi_i$ functions, only $\chi_0$ has a UV-singularity
at one loop,
\begin{equation}
\chi_0^{(1,{\rm UV})} = \frac{g^2\eta}{(4\pi)^{2-\ep}}
\; \frac{C_A}{2}\; (1-\xi)  \left( \frac{1}{\ep} + \ldots \right) .
\end{equation}

Using our results we have checked that all other functions 
$\lambda_i^{(1)}$ ($i\ne 1$), $\chi_i^{(1)}$ ($i=1,2,3$) and
$\tau_i^{(1)}$ have no UV-singularities. This was one
of the important checks on self consistency of the calculation.

To renormalize the above expressions,
we need to subtract the $1/\ep$ poles, (possibly) getting some constant $R$
instead, depending on the renormalization scheme:
\begin{equation}
\left( \frac{1}{\ep} + \ldots \right) \rightarrow
\left( R + \ldots \right) .
\end{equation}
In the $\overline{\mbox{MS}}$ scheme $R=0$, because (see Eq.~(\ref{eta}))
\begin{equation}
\eta = e^{-\gamma\ep} \left[ 1 + {\cal{O}}(\ep^2) \right] ,
\end{equation}
so that $e^{-\gamma\ep}$ and $(4\pi)^{\ep}$ are absorbed by the
$\overline{\mbox{MS}}$ re-definition of the coupling constant $g^2$.

This procedure can be also re-formulated in the language of
the renormalization $Z$-factors, by analogy with
Section~VIII of \cite{DOT2}.

%%%%%%%%%%%%%%%%%%%%%%%%%%%%%%%%%%%%%%%%%%%%%%%%%%%%%%%%%%%%%%%%%%%%%%%
\section{Some special cases}
\setcounter{equation}{0}
%%%%%%%%%%%%%%%%%%%%%%%%%%%%%%%%%%%%%%%%%%%%%%%%%%%%%%%%%%%%%%%%%%%%%%%

A few limits are of special interest:
\begin{itemize}
\item 
the symmetric case, when $p_1^2=p_2^2=p_3^2$;
\item 
the on-shell limit $p_1^2=p_2^2=m^2$ (with or without the assumption
that the vertex function is being sandwiched between Dirac spinors);
\item 
the zero-momentum limit, when the gluon momentum vanishes ($p_3=0$).
\end{itemize}
Since in all these cases we can put $p_1^2=p_2^2\equiv p^2$,
we start by considering this as the first step towards all these
limits.

In the case $p_1^2=p_2^2\equiv p^2$ some of the tensor structures
(\ref{L_i}) and (\ref{T_i})
in the quark-gluon vertex become linearly dependent,
namely: $L_{2,\mu}$ and $T_{2,\mu}$, $L_{3,\mu}$ and $T_{1,\mu}$,
$T_{4,\mu}$ and $T_{7,\mu}$.
Moreover, according to Eq.~(\ref{l4t6}), $\lambda_4$ and
$\tau_6$ vanish.
Therefore, the quark-gluon vertex
in this limit can be written as [cf. Eqs.~(\ref{tau7t})
and (\ref{lambda23t})]
\begin{equation}
\label{Gamma12}
\left. \Gamma_{\mu}\right|_{p_1^2=p_2^2\equiv p^2}
= L_{1,\mu} \lambda_1
+ L_{2,\mu} \widetilde{\lambda}_2
+ L_{3,\mu} \widetilde{\lambda}_3
+ T_{3,\mu} \tau_3  + T_{5,\mu} \tau_5
+ T_{7,\mu} \widetilde{\tau}_7
+ T_{8,\mu} \tau_8 .
\end{equation}
In fact, only the $L_{1,\mu}$ contribution
remains ``non-transverse'' in this limit.

\subsection{Symmetric case}

In Ref.~\cite{PT} (see also in \cite{CG}) the ``symmetric'' limit
of the quark-gluon vertex, $p_1^2=p_2^2=p_3^2\equiv p^2=-\mu^2$,
has been examined (for the case of massless quarks, $m=0$).
The decomposition of the quark-gluon vertex is given in Eq.~(2.34)
of \cite{PT}. 
It basically corresponds to the naive decomposition presented
in Eq.~(\ref{decomp_1}) of this paper.
In Ref.~\cite{PT} explicit results are given only for two scalar
functions (out of twelve),
\begin{equation}
\Gamma_1 \leftrightarrow h_1 
\hspace{10mm}
\mbox{and} \hspace{10mm}
\Gamma_{12} \leftrightarrow -h_{12} \; .
\end{equation}
Taking into account Eq.~(\ref{l4t6}),
in terms of $\lambda$'s and $\tau$'s we get (see also in Appendix~A):
\begin{equation}
\label{Gammas}
\Gamma_1 = \lambda_1 + p_3^2 \tau_3 
+ (p_1 p_2) \tau_8 \; ,
\hspace{10mm}
\Gamma_{12} = - \tau_8 \; .
\end{equation}
Putting
$p_1^2=p_2^2=p_3^2\equiv p^2$ (implying $(p_1 p_2)=-\half p^2$) and $m=0$,
we arrive at the following one-loop results:
\begin{eqnarray}
\Gamma_1^{(1)} &=& -\frac{g^2\eta \left(C_F-\half C_A\right)}{(4\pi)^{n/2}}
\bigg\{ {\textstyle{1\over6}} \left[ 4(n-4)-n\xi\right] p^2 \varphi_0
        + (3-n+\xi) \kappa_{0,3} \bigg\}
\nn \\ && 
+ \frac{g^2\eta C_A}{(4\pi)^{n/2}}
\bigg \{ {\textstyle{1\over8}}
       \left[ 12 - 2(2n-5)\xi+(n-4)\xi^2 \right] \kappa_{0,3}
        - {\textstyle{1\over6}} (8-n\xi) p^2 \varphi_0 \bigg\} \; ,
\\
\Gamma_{12}^{(1)} &=& -\frac{g^2\eta \left(C_F-\half C_A\right)}{(4\pi)^{n/2}}
\; {\textstyle{1\over3}} 
\left[ (n-6)-(n-4)\xi \right] \varphi_0
\nn \\ && 
- \frac{g^2\eta C_A}{(4\pi)^{n/2}}
\left[ 1 + {\textstyle{1\over6}} (n-7) \xi
- {\textstyle{1\over24}} (n-6) \xi^2 \right] \varphi_0 \; .
\end{eqnarray}
Taking into account that the constant $R(1)$ used in \cite{PT,CG} can
be identified as
\begin{equation}
\label{R(1)}
R(1)= p^2 \left.\varphi_0(p^2,p^2,p^2)\right|_{n=4} 
= \frac{4}{\sqrt{3}}\Cl{2}{\frac{\pi}{3}} \; ,
\end{equation}
and expanding in $\ep$ (keeping the divergent and finite in $\ep$ terms), 
we arrive at the same result
as given in Eqs.~(2.36) of \cite{PT}\footnote{There is a 
misprint in Eq.~(2.35) of \cite{PT}: $\Gamma_{12}$ should read
$p^2\Gamma_{12}=-\mu^2 \Gamma_{12}$.
Note that the $\ep$ used in \cite{PT} has different sign, 
as compared to ours.}.
In Eq.~(\ref{R(1)}) 
$\mbox{Cl}_2(\theta)=\Im\left[\mbox{Li}_2(e^{{\rm i}\theta})\right]$
is the Clausen function.

%=================================================================

For the case of massive quarks ($m\neq 0$), a similar ``symmetric'' limit 
has been considered in Ref.~\cite{DTP}, where the renormalization factor
$Z_{1F}$ was calculated at the one-loop order in the MOM scheme.
For their calculation, only one of the scalar functions 
(namely, the one accompanying the $\gamma_{\mu}$ matrix)
was needed, $\Gamma_1\leftrightarrow h_1$,
which is related to $\lambda$'s and $\tau$'s via Eq.~(\ref{Gammas}). 
Putting $p_1^2=p_2^2=p_3^2\equiv p^2=-\mu^2$ in our expressions,
we arrive at the same result\footnote{Their $a$ corresponds to our 
$(1-\xi)$, whereas their $\lambda$ denotes the ratio $m^2/\mu^2=-m^2/p^2$.
Note that there is a misprint in the {\em journal} version of the result
for the last contribution in Eq.~(14), $C_b$: in the very last term,
$\ln\frac{1}{1+\lambda}$ should read $\ln\frac{\lambda}{1+\lambda}$.
In the {\em preprint} version of \cite{DTP} this equation is correct.}
as given in Eqs.~(13)--(14) of \cite{DTP}. In particular, we have 
taken into account that the functions $H$ and $M$ introduced
in Eq.~(15) of \cite{DTP} are related to our functions $\varphi_i$ 
through\footnote{We also note a misprint in the large-mass ($1/\lambda$)
expansion of the $M$ function presented in Appendix~C of Ref.~\cite{DTP}:
the last available term, $-2672/(11025\lambda^5)$,
should read $-1523/(6300\lambda^5)$. We are grateful to O.V.~Tarasov
for confirming this misprint.}
\begin{equation}
H = - p^2 \left.\varphi_1(p^2,p^2,p^2)\right|_{n=4}, \hspace{10mm}
M = - p^2 \left.\varphi_2(p^2,p^2,p^2)\right|_{n=4} .
\end{equation}
More details on these functions can be found in Appendix~C.

%=================================================================

\subsection{On-shell quarks}

The tensor structure of (\ref{Gamma12}) does not change
when we put $p^2=m^2$.
Let us introduce the notation
\be
\varphi_{1,2}^{\rm os} \equiv \varphi_{1,2}(m^2,m^2,p_3^2)
\ee
for the $\varphi_i$ functions in the on-shell limit.
According to Eq.~(\ref{3->2}), in this limit the function  
$\varphi_2$ (corresponding to diagram $a$) 
reduces to a two-point function and a tadpole,
\be
\varphi_2^{\rm os} = \frac{1}{(n-4)(4m^2-p_3^2)}
\left[ 2 (n-3)\kappa_{2,3}-(n-2)\widetilde{\kappa} \right] \; .
\ee
This is the reason why $\varphi_2$ does not appear 
in the on-shell results for diagram $a$.

The results for the relevant longitudinal and transverse
functions in the limit
$p_1^2=p_2^2=m^2$ are presented below.
 
\subsubsection{Diagram $a$}

\begin{eqnarray}
\label{Eq:lam-1a-onshell}
\lambda_1^{(1a)}(m^2,m^2,p_3^2) &=&
\frac{g^2\eta\left(C_F-\half C_A\right)}{(4\pi)^{n/2}}\;
\frac{(n-2)(1-\xi)}{2(n-3)}\; \widetilde{\kappa} \; ,
\\
\widetilde{\lambda}_2^{(1a)}(m^2,m^2,p_3^2) &=&
-\frac{g^2\eta\left(C_F-\half C_A\right)}{(4\pi)^{n/2}}\;
\frac{1}{(n-3)(n-4)(4m^2-p_3^2)^2}\;
\nn \\ && 
\times \bigg\{ 
(n-3) \left[ (2-\xi) (4m^2-p_3^2)+(n-5) \xi p_3^2 \right] 
\kappa_{2,3}
\nn \\ && 
- (n-2)[4m^2-p_3^2+2 (n-5) \xi m^2 ]\widetilde{\kappa} 
\bigg\} \; ,
\\
\label{Eq:lam-3a-onshell}
\widetilde{\lambda}_3^{(1a)}(m^2,m^2,p_3^2) &=&
\frac{g^2\eta\left(C_F\!-\!\half C_A\right)}{(4\pi)^{n/2}}\;
\frac{(n-\xi) m}{(n\!-\!3)(n\!-\!4)(4m^2\!-\!p_3^2)} 
\left[ (n\!-\!2) \widetilde{\kappa}
- 2(n\!-\!3) \kappa_{2,3} \right] \; ,
\\
\tau_3^{(1a)}(m^2,m^2,p_3^2) &=&
-\frac{g^2\eta\left(C_F-\half C_A\right)}{(4\pi)^{n/2}}\;
\frac{1}{2 (n-3) (n-4) p_3^2 (4m^2-p_3^2)} \;
\nn \\ && 
\times \bigg\{ 
\left[ (n-2)(n-3)-(n-4)\xi \right] p_3^2 
\left[ 2(n-3) \kappa_{2,3} - (n-2) \widetilde{\kappa} \right]
\nn \\ && 
+(n-3)(n-4)(4m^2-p_3^2) 
\left[ 2 \kappa_{2,3} - (n-2) \widetilde{\kappa} \right]
\bigg\} \; ,
\\
\tau_5^{(1a)}(m^2,m^2,p_3^2) &=&
-\frac{g^2\eta\left(C_F-\half C_A\right)}{(4\pi)^{n/2}}\;
\frac{1}{4(n-3) m (4 m^2-p_3^2)}\;
\bigg\{
(n-2)\xi (4 m^2-p_3^2) \widetilde{\kappa}
\nn \\ &&
+4 (n-3-\xi) m^2 
\left[ 2(n-3) \kappa_{2,3} - (n-2) \widetilde{\kappa} \right]
\bigg\} \; ,
\\
\widetilde{\tau}_7^{(1a)}(m^2,m^2,p_3^2) &=&
-\frac{g^2\eta\left(C_F-\half C_A\right)}{(4\pi)^{n/2}}\;
\frac{\xi}{2 (n\!-\!3) (n\!-\!4) m (4 m^2\!-\!p_3^2)^2}\;
\nn \\ &&
\times \bigg\{
4(n-5)m^2 
\left[ 2(n-3) \kappa_{2,3} - (n-2) \widetilde{\kappa} \right]
-(n-2)(n-4)(4 m^2-p_3^2) \widetilde{\kappa}
\bigg\} \; ,
\nn \\ &&
\hspace{5mm}
\\
\tau_8^{(1a)}(m^2,m^2,p_3^2) &=&
-\frac{g^2\eta\left(C_F-\half C_A\right)}{(4\pi)^{n/2}}\;
\frac{6-n+(n-4)\xi}{(n\!-\!3) (n\!-\!4) (4 m^2\!-\!p_3^2)}
\left[ 2(n\!-\!3)\kappa_{2,3} \!-\!
(n\!-\!2)\widetilde{\kappa} \right] \; .
\end{eqnarray}

\subsubsection{Diagram $b$}

\begin{eqnarray}
\label{Eq:lam-1b-onshell}
\lambda_1^{(1b)}(m^2,m^2,p_3^2) &=&
\frac{g^2\; \eta\; C_A}{(4\pi)^{n/2}}\;
\frac{2-\xi}{4 (n-3)}
\left[ (n-3) \kappa_{0,3} +(n-2) {\widetilde{\kappa}} \right] , 
\\
{\widetilde{\lambda}}_2^{(1b)}(m^2,m^2,p_3^2) &=&   
\frac{g^2\; \eta\; C_A}{(4\pi)^{n/2}}
\frac{1}{16 (n-3) (4 m^2-p_3^2)^2}
\bigg\{ 
8 (n-1)(n-3)(2-\xi)m^2 p_3^2 \varphi_1^{\rm os}
\nn \\ &&
-2(n\!-\!3)^2\xi^2 p_3^2 \big[2 (n\!-\!3) m^2 \!+\!p_3^2 \big] 
\varphi_1^{\rm os}
-2(n\!-\!3)^2\xi^2 \big[ 4(n\!-\!4)m^2 \!+\!3 p_3^2 \big] \kappa_{0,3}
\nn \\ &&
+4(n-3)(2-\xi) \big[ 4(n-2)m^2 +p_3^2 \big] \kappa_{0,3}
+ (n-1)(n-2)(n-3)\xi^2 p_3^2 {\widetilde{\kappa}}
\nn \\ &&
-2(n-2)(2-\xi) \big[ 8m^2+(n-3)p_3^2 \big] {\widetilde{\kappa}}
\bigg\} \; ,
\\
\label{Eq:lam-3b-onshell}
{\widetilde{\lambda}}_3^{(1b)}(m^2,m^2,p_3^2) &=&
\frac{g^2\; \eta\; C_A}{(4\pi)^{n/2}}\;
\frac{1}{32 (n-3) m (4 m^2-p_3^2)}
\bigg\{ 
2 (n-3)(n-4)\xi^2 p_3^2 (4 m^2-p_3^2) \varphi_1^{\rm os}
\nn \\ && 
+ \left[ 4 (n-1)+2 (n-4) \xi-(n-2) (n-3)\xi^2 \right]
\nn \\ &&
\times \left[ 4(n-3) m^2 p_3^2 \varphi_1^{\rm os} 
+8(n-3) m^2 \kappa_{0,3} - (n-2) p_3^2 {\widetilde{\kappa}} \right]
\nn \\ &&
+4(n-3)^2\xi^2 (4m^2-p_3^2) \kappa_{0,3}
-4(n-1)(n-2) (4m^2-p_3^2) {\widetilde{\kappa}}
\bigg\} \; ,
\\
\tau_3^{(1b)}(m^2,m^2,p_3^2) &=&
\frac{g^2\; \eta\; C_A}{(4\pi)^{n/2}}\;
\frac{1}{16 (n-3) p_3^2 (4 m^2- p_3^2)}
\nn \\ &&
\times \bigg\{ 
2 (n-3) \xi (4-\xi) (4 m^2-p_3^2) 
\left[ p_3^2 \varphi_1^{\rm os} - (n-4) \kappa_{0,3} \right]
\nn \\ && 
- \left[ 2 (2\!-\!\xi)-(n\!-\!3) \xi^2 \right] p_3^2
\left[ 4(n\!-\!3)m^2 \varphi_1^{\rm os} + 2(n\!-\!3) \kappa_{0,3}
       -(n\!-\!2) {\widetilde{\kappa}} \right]
\bigg\} \; ,
\\
\tau_5^{(1b)}(m^2,m^2,p_3^2) &=&
\frac{g^2\; \eta\; C_A}{(4\pi)^{n/2}}\;
\frac{1}{8 (n-3) m (4 m^2-p_3^2)}
\bigg\{ 
(n-2)(n-4)\xi (4 m^2-p_3^2) {\widetilde{\kappa}}
\nn \\ &&
+ \xi \big[ 2+(n-3) \xi \big] m^2 
\left[ 4(n-3) m^2 \varphi_1^{\rm os} + 2(n-3) \kappa_{0,3}
-(n-2) {\widetilde{\kappa}} \right]
\nn \\ &&
-2(n-3) \big[ 6 + 2(n-5)\xi + \xi^2 \big] 
m^2 (4m^2-p_3^2) \varphi_1^{\rm os}
\bigg\} \; ,
\\
{\widetilde{\tau}}_7^{(1b)}(m^2,m^2,p_3^2) &=&
-\frac{g^2\; \eta\; C_A}{(4\pi)^{n/2}} \;
\frac{\xi}{16 (n-3) m (4 m^2-p_3^2)^2}
\nn \\ &&
\times \bigg\{ 
2(n-3) (4 m^2-p_3^2) \left[ 4m^2 + 2(n-5)\xi m^2 
         - (n-4)\xi p_3^2 \right] \varphi_1^{\rm os}
\nn \\ &&
+(n\!-\!1)\big[2+(n-3)\xi\big] 
\left[ 4(n\!-\!3) m^2 p_3^2 \varphi_1^{\rm os}
+ 8(n\!-\!3)m^2 \kappa_{0,3} - (n\!-\!2) p_3^2 
{\widetilde{\kappa}}
\right]
\nn \\ &&
-4(n-3)^2 \xi (4 m^2-p_3^2) \kappa_{0,3}
-(n-2) \big[ 6 + (n-3)\xi \big] (4 m^2-p_3^2) 
{\widetilde{\kappa}}
\bigg \} \; ,
\\
\tau_8^{(1b)}(m^2,m^2,p_3^2) &=&
- \frac{g^2\; \eta\; C_A}{(4\pi)^{n/2}}\;
\frac{1}{8(n-3)(4 m^2-p_3^2)}
\bigg\{ 
2 (n-3)\xi(4-\xi)(4 m^2-p_3^2) \varphi_1^{\rm os}
\nn \\ &&
-\left[ 12+2(2n\!-\!7)\xi-(n\!-\!3)\xi^2 \right]
\left[ 4(n\!-\!3)m^2 \varphi_1^{\rm os} + 2(n\!-\!3)\kappa_{0,3}
-(n\!-\!2){\widetilde{\kappa}} \right]
\bigg\} \; .
\nn \\ &&
\hspace{5mm}
\label{Eq:tau-8b-onshell}
\end{eqnarray}

When $p_3^2\to 0$, the only problem which may arise
in Eqs.~(\ref{Eq:lam-1a-onshell})--(\ref{Eq:tau-8b-onshell}) 
concerns the functions $\tau_3^{(1a)}$ and $\tau_3^{(1b)}$ 
containing $p_3^2$ in the denominator.
However, the coefficients of $(p_3^2)^0$ in the numerators of 
$\tau_3^{(1a)}$ and $\tau_3^{(1b)}$ are proportional to 
$\left[ 2 \kappa_{2,3} -(n-2) \widetilde{\kappa} \right]$
and $\kappa_{0,3}$, respectively.
If we take into account that 
$\left.\kappa_{2,3}\right|_{p_3^2=0}=\half (n-2) \widetilde{\kappa}$
and $\left.\kappa_{0,3}\right|_{p_3^2=0}=0$ (massless tadpole),
we see that the limit $p_3^2\to 0$ is regular for $\tau_3^{(1)}$.

\subsubsection{Dirac and Pauli form factors}

Let us consider again the on-shell limit $p_1^2=p_2^2=m^2$,
without putting any further conditions on $p_3$.
If we recall that the ``physical'' quark-gluon vertex should be 
sandwiched between physical states obeying the Dirac equation,
\begin{equation}
\label{sandwich}
\overline{u}(-p_1) \Gamma_{\mu} u(p_2), \hspace{10mm}
\mbox{with} \hspace{5mm}
\overline{u}(-p_1)\pslash_1=-m \overline{u}(-p_1),
\hspace{5mm} \pslash_2 u(p_2) = m u(p_2) ,
\end{equation}
then (using the above Dirac conditions) we arrive at the standard 
decomposition 
\begin{eqnarray}
\overline{u}(-p_1) \Gamma_{\mu} u(p_2)\! &=& \!
F_1(p_3^2) \; \overline{u}(-p_1) \gamma_{\mu} u(p_2)
- \frac{1}{2m} F_2(p_3^2) \;
\overline{u}(-p_1) \sigma_{\mu \nu}\, p_3^\nu u(p_2)
\nn \\
\! &=& \! \left[F_1(p_3^2)+F_2(p_3^2)\right] \;  
\overline{u}(-p_1) \gamma_{\mu} u(p_2)
+ \frac{1}{2m} F_2(p_3^2)
\overline{u}(-p_1) (p_1-p_2)_{\mu} u(p_2) ,
\nn \\
&& \hspace{10mm}
\end{eqnarray}
where $F_1(p_3^2)$ and $F_2(p_3^2)$ are often associated with 
the Dirac and Pauli form factors, respectively.
In terms of the (modified) $\lambda$ and $\tau$ functions we get
\begin{eqnarray}
F_1+F_2 &=& \lambda_1 + p_3^2 \tau_3 -2m \tau_5 + 
\half ( p_3^2 - 4 m^2) \tau_8 ,
\\
\frac{1}{2m} F_2 &=& -2 m \widetilde{\lambda}_2
+ \widetilde{\lambda}_3 - \tau_5 + \half p_3^2 \widetilde{\tau}_7
- m \tau_8 .
\end{eqnarray}

Using our results we get
\begin{eqnarray}
(F_1+F_2)^{(1a)} &=&
-\frac{g^2\eta\left(C_F-\half C_A\right)}{(4\pi)^{n/2}}\;
\frac{1}{(n-3) (n-4) (4 m^2-p_3^2)}\;
\nn \\ && 
\times
\left\{ (n-3) \left[ 2 (n-3)p_3^2 -(n^2-9n+22)(4 m^2-p_3^2) \right]
\kappa_{2,3}
\right.
\nn \\
&& \left.
- 2(n-2) \left[ 2 (n-5) m^2 +p_3^2 \right] \widetilde{\kappa} \right\},
\\
(F_1+F_2)^{(1b)} &=&
- \frac{g^2\eta C_A}{(4\pi)^{n/2}}\;
\frac{1}{8 (n-3) (4 m^2-p_3^2)}
\bigg\{ 4 (n-3) p_3^2 \big( 2 m^2 \varphi_1^{\rm os}
+ \kappa_{0,3} \big)
\nn \\ &&  
+ (n-3) \left[ 8+2(4n-13)\xi-(n-4)\xi^2\right] (4 m^2-p_3^2) 
\kappa_{0,3}
\nn \\ && 
- 8 (n-2) (5 m^2-p_3^2) \widetilde{\kappa} \bigg\} \; ,
\end{eqnarray}
\begin{eqnarray}
\frac{1}{2m} F_2^{(1a)} &=&
\frac{g^2\eta\left(C_F-\half C_A\right)}{(4\pi)^{n/2}}\;
\frac{(n-5)m}{(n-3)(4 m^2-p_3^2)}\;
\left[ 2 (n-3)\kappa_{2,3}-(n-2)\widetilde{\kappa} \right] ,
\\
\frac{1}{2m} F_2^{(1b)} &=&
- \frac{g^2\eta C_A}{(4\pi)^{n/2}}\;
\frac{m}{2(n-3)(4 m^2-p_3^2)^2}
\nn \\ && 
\times \bigg\{ 
(n-3) p_3^2 \left[ 12 m^2+(n-4) p_3^2\right] \varphi_1^{\rm os}
+2 (n-3) \left[ 8 m^2+(n-3) p_3^2\right] \kappa_{0,3}
\nn \\ && 
+(n-2) \left[(n-6) (4 m^2-p_3^2) -(n-1) p_3^2\right] \widetilde{\kappa} 
\bigg\} \; .
\end{eqnarray}
The results for the (QED-like) diagram $a$
agree with Eqs.~(5.69) and (5.70) of \cite{Pokorski}.

In the massless limit, $m\to0$, and in Feynman gauge, $\xi=0$,
our results can be compared with those of \cite{NPS},
where renormalized results for the quark self energy,
$\Sigma$, and the quark-gluon vertex functions, 
$\Lambda_1^\mu$ and $\Lambda_2^\mu$ (their notation),
are collected in Tables~B.I
and B.II, respectively. The vertex functions are given
for off-shell gluon ($p_1^2=p_2^2=0$, $p_3^2\ne0$) and the out-going 
quark being off-shell ($p_1^2\ne0$, $p_2^2=p_3^2=0$).
We confirm these results.

\subsection{Zero-momentum limit}

Let us consider the 
zero-momentum limit $p_3=0$, with $p_2=-p_1\equiv p$
being off shell.
We can proceed in two steps.
The first step, putting $p_1^2=p_2^2=p^2$ (without putting $p^2=m^2$), 
has been already done in Eq.~(\ref{Gamma12}).
As the second step, we should put $p_3^2=0$ (which implies $p_3=0$). 
In this limit,
the quark-gluon vertex becomes
\begin{equation}
\label{qqg_zerom}
\left. \Gamma_{\mu}\right|_{p_2=-p_1\equiv p,\; p_3=0}
= L_{1,\mu} \lambda_1
+ L_{2,\mu} \widetilde{\lambda}_2
+ L_{3,\mu} \widetilde{\lambda}_3
= \gamma_{\mu} \lambda_1
+ 4p_{\mu}\pslash \widetilde{\lambda}_2
- 2p_{\mu}  \widetilde{\lambda}_3 .
\end{equation}
One can check that the corresponding scalar functions are regular.

The substitutions for the relevant integrals are\footnote{In general,
one should also make sure that the next term of the expansion
in $p_3^2$ does not contribute, which may happen when $p_3^2$
appears in the denominators. In this calculation, we did not
need such $p_3^2$ terms.}
\begin{eqnarray}
\left. J_2(1,1,1)\right|_{p_2=-p_1\equiv p,\; p_3=0}
&=& -\frac{1}{p^2-m^2}\left[ (n-3) J_2(0,1,1)-J_2(1,1,0) \right] \; ,
\\
\left. J_2(1,1,0)\right|_{p_2=-p_1\equiv p,\; p_3=0}
&=& \frac{n-2}{2 m^2}\; J_2(0,1,0) \; ,
\\
\left. J_1(1,1,1)\right|_{p_2=-p_1\equiv p,\; p_3=0}
&=& \frac{1}{(p^2-m^2)^2}
\left\{ (n-2) J_1(0,0,1) \right. \nn \\
& & \left.
- (n-3) (p^2+m^2) J_1(0,1,1) \right\} \; ,
\\
\left. J_1(1,1,0)\right|_{p_2=-p_1\equiv p,\; p_3=0}
&=& 0 \; .
\end{eqnarray}

The scalar functions from (\ref{qqg_zerom}) in this limit are
\begin{eqnarray}
\lambda_1^{(1a)}(p^2,p^2,0) &=& 
\frac{g^2\eta\left(C_F-\half C_A\right)}{(4\pi)^{n/2}}\;
\frac{(n-2)(1-\xi)}{2 p^2} \left[ (p^2+m^2) \kappa_1(p^2) 
- m^2 \widetilde{\kappa} \right] \; ,
\\
\widetilde{\lambda}_2^{(1a)}(p^2,p^2,0) &=&
-\frac{g^2\eta\left(C_F-\half C_A\right)}{(4\pi)^{n/2}}\;
\frac{(n-2)(1-\xi)}{8(p^2)^2 (p^2-m^2)}
\nn \\
&& \times
\left\{ \left[ (p^2-m^2) (p^2+3 m^2)-(n-3) (p^2+m^2)^2 \right]
\kappa_1(p^2)
\right.
\nn \\
&& \left.
+ \left[ 2 (n-2) p^2-n (p^2-m^2) \right] m^2 \widetilde{\kappa} 
\right\} \; ,
\\
\widetilde{\lambda}_3^{(1a)}(p^2,p^2,0) &=&
\frac{g^2\eta\left(C_F-\half C_A\right)}{(4\pi)^{n/2}}\;
\frac{(n-\xi)m}{2p^2(p^2-m^2)}\;
\nn \\
&& \times \left\{
\left[ (n-4)p^2 +(n-2) m^2 \right] \kappa_1(p^2)
- (n-2) m^2 \widetilde{\kappa} \right\} \; ,
\\
\lambda_1^{(1b)}(p^2,p^2,0)
&=& \frac{g^2\eta C_A}{(4\pi)^{n/2}}\;
\frac{2-\xi}{8 p^2 (p^2-m^2)}
\left\{ (n-2)(m^2-3p^2) m^2 {\widetilde{\kappa}}
\right.
\nn \\
&& \left.
+ \left[ 4 (n-3) m^2 p^2 + (n-2) (p^2-m^2) (3 p^2+m^2) \right] 
\kappa_1(p^2)
\right\} \; ,
\\
{\widetilde{\lambda}}_2^{(1b)}(p^2,p^2,0)
&=& - \frac{g^2\eta C_A}{(4\pi)^{n/2}}\;
\frac{(n-2) (2-\xi)}{32 (p^2)^2 (p^2-m^2)}
\left\{ \left[ (n-4) p^2+n m^2 \right] m^2 {\widetilde{\kappa}}
\right.
\nn \\
&& \left.
- \left[ 4(n-3)m^4+(n-4)(p^2-m^2)(p^2+3 m^2) \right]
\kappa_1(p^2)
\right\} \; ,
\\
{\widetilde{\lambda}}_3^{(1b)}(p^2,p^2,0)
&=& \frac{g^2\eta C_A}{(4\pi)^{n/2}}\;
\frac{m(n-1)}{4 p^2 (p^2-m^2)}
\left\{ \left[ (n-2)m^2+(n-4)p^2\right] \kappa_1(p^2) \right. \nn \\
&& \left.
-(n-2) m^2 {\widetilde{\kappa}} \right\} \; . 
\end{eqnarray}

If we now consider the on-shell case, i.e.\ put $p^2=m^2$, we need 
to be careful, since the above expressions
contain $(p^2-m^2)$ in their denominators.
Using two terms of the expansion of $\kappa_1(p^2)$ in 
$\delta\equiv(m^2-p^2)/m^2$,
\begin{equation}
\kappa_1(p^2)=\frac{n-2}{2 (n-3)} \widetilde{\kappa}(m^2) 
\left[ 1+\half \delta
+ {\cal{O}}(\delta^2) \right] \; ,
\end{equation}
we see that the pole at $p^2=m^2$ ($\delta=0$) is canceled. 
In this way, we arrive at
\begin{eqnarray}
\label{oszm1}
\lambda_1^{(1)}(m^2,m^2,0) &=&
\frac{g^2\eta}{(4\pi)^{n/2}}\;
\frac{(n-2)\widetilde{\kappa}}{2(n-3)} 
\left[\left(C_F -\half C_A\right)(1-\xi) +\half C_A (2-\xi) \right] ,
\\
\label{oszm2}
\widetilde{\lambda}_2^{(1)}(m^2,m^2,0) &=&
-\frac{g^2\eta}{(4\pi)^{n/2}}\;
\frac{(n-2)\widetilde{\kappa}}{4(n-3)m^2}
\left[\left(C_F-\half C_A\right)(1-\xi) +\quarter C_A (2-\xi) \right] ,
\hspace{5mm}
\\
\label{oszm3}
\widetilde{\lambda}_3^{(1)}(m^2,m^2,0) &=&
-\frac{g^2\eta}{(4\pi)^{n/2}}\;
\frac{(n-2)\widetilde{\kappa}}{4(n-3)m}
\left[\left(C_F-\half C_A\right)(n-\xi) +\half C_A \,(n-1) \right] .
\end{eqnarray} 
These results coincide with those obtained by first
taking the on-shell limit $p_1^2=p_2^2=m^2$ 
(see Eqs.~(\ref{Eq:lam-1a-onshell})--(\ref{Eq:lam-3a-onshell}) 
and (\ref{Eq:lam-1b-onshell})--(\ref{Eq:lam-3b-onshell}) above)
and then putting $p_3=0$.
If we assume that the vertex (\ref{qqg_zerom}) is sandwiched between
physical states (\ref{sandwich}), we see that both structures
containing $p_{\mu}$ can be transformed into $\gamma_{\mu}$,
so that the ``effective'' vertex becomes
\be
\label{sandwich2}
\gamma_{\mu} \big[ \lambda_1(m^2,m^2,0) 
+ 4 m^2 {\widetilde{\lambda}}_2(m^2,m^2,0)
- 2 m {\widetilde{\lambda}}_3(m^2,m^2,0) \big].
\ee
Substituting our results (\ref{oszm1})--(\ref{oszm3}) we obtain
for the one-loop contribution to Eq.~(\ref{sandwich2})
\be
\label{sandwich3}
\gamma_{\mu}\; \frac{g^2\eta}{(4\pi)^{n/2}}
        \frac{(n-1)(n-2)}{2(n-3)} C_F {\widetilde{\kappa}}
= \gamma_{\mu}\; \frac{g^2}{(4\pi)^{2-\ep}} \Gamma(\ep)
        \frac{(3-2\ep)}{(1-2\ep)} C_F (m^2)^{-\ep} ,
\ee
where we have taken into account the definitions of $\eta$
(\ref{eta}) and ${\widetilde{\kappa}}$ (\ref{kappa_tilde}).
We see that the result (\ref{sandwich3}) is gauge independent,
and it does not contain $C_A$. For the QED case,
it coincides with Eq.~(22) of \cite{ALV}.

In the massless ($m=0$) case, we can compare our results
for the zero-momentum limit with those given in \cite{BL}.
The definition of the zero-gluon-momentum vertex is given in
their Eq.~(A4) (the lower equation). It is proportional to
(their $q \leftrightarrow$ our $p$) 
\begin{equation}
\left\{ \left[ 1 + \Gamma_3(q^2)\right]\gamma_{\mu}
+ \Gamma_4(q^2)\gamma^{\nu} \left( g_{\mu\nu} 
-\frac{q_{\mu}q_{\nu}}{q^2} \right) \right\} .
\end{equation}
If we consider our results in the off-shell 
zero-momentum limit, we see that on putting $m=0$ the function 
$\widetilde{\lambda}_3$ vanishes. 
The remaining two functions can be mapped into $\Gamma_{3,4}$ as
\begin{equation}
\lambda_1^{\rm (ren)} \leftrightarrow \Gamma_3 + \Gamma_4, \hspace{10mm}
-4p^2 \widetilde{\lambda}_2 \leftrightarrow \Gamma_4 ,
\end{equation}
or, vice versa,
\begin{equation}
\Gamma_3 \leftrightarrow \lambda_1^{\rm (ren)} 
+ 4p^2 \widetilde{\lambda}_2, \hspace{10mm}
\Gamma_4  \leftrightarrow -4p^2 \widetilde{\lambda}_2 ,
\end{equation}
where the superscript ``(ren)'' means ``renormalized''.

In the massless case, taking into account that the massless tadpole
function $\left.{\widetilde{\kappa}}\right|_{m=0}$ vanishes, we get 
\begin{eqnarray}
\lambda_1^{(1)} &=& 
\frac{g^2\; \eta}{(4\pi)^{n/2}}\; \frac{n-2}{2} \kappa_0(p^2)
\left[ \left(C_F - \half C_A \right) (1-\xi) 
      +{\textstyle{3\over4}} C_A (2-\xi) \right] \; ,
\\
4p^2{\widetilde{\lambda}}_2^{(1)} &=& 
\frac{g^2\; \eta}{(4\pi)^{n/2}}\; \frac{(n-2)(n-4)}{2} \kappa_0(p^2)
\left[ \left(C_F - \half C_A \right) (1-\xi)  
      +{\textstyle{1\over4}} C_A (2-\xi) \right] \; .
\end{eqnarray}

Putting $p^2=\mu^2$ (as in \cite{BL}) means that the logarithms
$\ln(p^2/\mu^2)$ (in our case, $\ln(p^2)$, since we imply $\mu=1$) 
vanish. Therefore, we should substitute
$\kappa_0 \Rightarrow 1/\ep+2$ (see, for example, Eq.~(2.15) of \cite{DOT1}).
Omitting the ${\cal{O}}(\ep)$ terms we see that
\begin{eqnarray}
\lambda_1^{(1)} &\Rightarrow&  
\frac{g^2}{16\pi^2} \left[ \left(C_F - \half C_A \right)(1-\xi)
+ {\textstyle{3\over4}} C_A (2-\xi) \right] 
\left( \frac{1}{\ep} + 1 \right) , 
\\
\label{lt2_bl}
4p^2{\widetilde{\lambda}}_2^{(1)} &\Rightarrow& 
-\frac{g^2}{16\pi^2} 
\left[ 2 \left(C_F - \half C_A \right) (1-\xi) 
       + \half C_A (2-\xi) \right] \; ,
\end{eqnarray}
where we have explicitly separated the contributions of diagrams
$a$ and $b$. 
To perform the renormalization (in the $\overline{\mbox{MS}}$-scheme),
we need just to subtract the $1/\ep$ term, i.e.
\begin{equation}
\label{l1r_bl}
\lambda_1^{(1,{\rm ren})} \Rightarrow
\frac{g^2}{16\pi^2} \left[ \left(C_F - \half C_A \right)(1-\xi)
+ {\textstyle{3\over4}} C_A (2-\xi) \right].
\end{equation}

Now we can compare our Eqs.~(\ref{lt2_bl})--(\ref{l1r_bl}) with one-loop 
results for $\Gamma_{3,4}$ presented in Eq.~(A12) of \cite{BL}. 
Taking into
account that their $\xi$ corresponds to our $(1-\xi)$, 
and also that their $\alpha=g^2/(4\pi)$, we see that we
are in agreement with \cite{BL}.
Moreover, their result for the quark self-energy function $A(q^2)$
(see Eq.~(A9) of \cite{BL}) is in agreement with our $\alpha(p^2)$.
Note that $\beta(p^2)=0$ in the massless case.

%%%%%%%%%%%%%%%%%%%%%%%%%%%%%%%%%%%%%%%%%%%%%%%%%%%%%%%%%%%%%%%%%%%%%%%
\section{Conclusions}
\setcounter{equation}{0}
%%%%%%%%%%%%%%%%%%%%%%%%%%%%%%%%%%%%%%%%%%%%%%%%%%%%%%%%%%%%%%%%%%%%%%%

In this article we have given results for the one-loop quark-gluon
vertex in an arbitrary covariant gauge and in an arbitrary space-time
dimension. The calculation was carried out with massive quarks.

To calculate the quark-gluon vertex, we have decomposed it into
longitudinal (\ref{GammaL}) and transverse (\ref{GammaT}) parts,
$\Gamma^{(\rm L)}_{\mu}$ and $\Gamma^{(\rm T)}_{\mu}$ 
(like the decomposition in QED \cite{BC1}). 
Altogether twelve scalar functions (four $\lambda$'s and eight
$\tau$'s) are needed to define the quark-gluon vertex. We found that
the function $\lambda_4$, the coefficient of 
$\sigma_{\mu \nu}(p_1-p_2)^{\nu}$, which is absent in QED,
does not vanish in QCD (see Eq.~(\ref{lam4(1b)})) 
and contributes to the non-Abelian part of
the Ward--Slavnov--Taylor identity (\ref{WST2}). 
The general results for the longitudinal parts ($\lambda$'s) are given 
in Eqs.~(\ref{lam1(1a)})--(\ref{lam4(1b)}) (arbitrary gauge), 
and results for the transverse
parts ($\tau$'s) are given in Eqs.~(\ref{tau1(1a)})--(\ref{tau8(1b)}) 
(Feynman gauge) and
in Appendix~E (arbitrary covariant gauge).
Using recurrence relations (see Appendix~B),
all results have been expressed in terms of integrals with powers of
propagators equal to zero or one. 
Only two non-trivial scalar functions are required, 
$\varphi_i(p_1^2,p_2^2,p_3^2)$, where $i=1$ or 2 counts the number of 
{\it massive} propagators involved (see Section~2.1 and Appendix~C).

Starting from the general off-shell expressions ($p_1^2\ne m^2$,
$p_2^2\ne m^2$, $p_3^2\ne 0$) in an arbitrary space-time 
dimension, $n$, for the longitudinal and transverse
parts of the vertex, we have derived results for the on-shell
limit ($p_1^2=p_2^2=m^2$) which are also valid for an arbitrary $n$
(Section~4). 
Note that if we started from the off-shell results expanded around
$n=4$ (similar to the results of \cite{BC1,KRP} for the QED case),
we would get infrared
divergences from the on-shell-divergent logarithms.
Keeping the arbitrary space-time dimension, we see that the corresponding
infrared divergences result in extra poles in $\ep=(4-n)/2$.

Various special cases of the general results were compared with those
of Refs.~\cite{PT,DTP,NPS,BL,BC1,KRP,BKP,ALV} 
(for details, see Sections~3 and 4).

At the one-loop level, the Ward--Slavnov--Taylor identity for 
the quark-gluon vertex can be split in an Abelian and a non-Abelian part,
Eqs.~(\ref{WST1}) and (\ref{WST2}).
The Abelian part is similar to the Ward--Fradkin--Takahashi identity
in QED \cite{WFT}, whereas the non-Abelian part has
a nontrivial contribution involving the
quark-quark-ghost-ghost vertex (\ref{H}), 
which can be described by scalar 
functions $\chi_i$ ($i=0,...,3$). One-loop results for these functions are 
presented in Appendix~D.
Using the results for these $\chi$ functions,
and those for quark and ghost self energies, we 
have checked that our results for the longitudinal parts of the
vertex ($\lambda_1$, $\lambda_2$, $\lambda_3$, and $\lambda_4$)
satisfy the WST identity for arbitrary $n$ and $\xi$, as they should. 

In principle, some techniques which can be used for the calculation
of the two-loop off-shell quark-gluon vertex, at least in the $m=0$ case,
are already available \cite{UD,Tar}, although the problem of
higher powers of irreducible numerators is still difficult
for algorithmization. For special limits, the calculation is
very similar to the three-gluon vertex, which was calculated
at two loops in \cite{DOT2} (the zero-momentum limit) and in
\cite{DO1} (the on-shell case).

\vspace{3mm}

{\bf Note.} 
The results of this paper are available in {\sf REDUCE} format at \\
{\tt http://wwwthep.physik.uni-mainz.de/Publications/progdata/mzth9963/ }

\vspace{3mm}

{\bf Acknowledgements.}
It is a great pleasure to thank Oleg Tarasov for his contributions
in the early stages of this work.
We are grateful to the authors of \cite{KRP,BKP}, especially to 
Ayse K{\i}z{\i}lers{\"u}, for confirming our statements
about misprints.
A.~D. is grateful to A.G.~Grozin and J.G.~K\"orner for useful discussions.
This research has been supported by the Research Council of Norway.
Research by L.~S. was supported by the Norwegian State Educational 
Loan Fund.
A.~D.'s research was supported by the Alexander von Humboldt Foundation 
(before November 1999), and then by the DFG.
Partial support from the grants RFBR No.~98--02--16981
and Volkswagen No.~I/73611 is acknowledged.

%%%%%%%%%%%%%%%%%%%%%%%%%%%%%%%%%%%%%%%%%%%%%%%%%%%%%%%%%%%%%%%%%%%%%%%
\section*{Appendix A: Decomposition of the vertex}
\setcounter{equation}{0}
\renewcommand{\thesection}{A}
%%%%%%%%%%%%%%%%%%%%%%%%%%%%%%%%%%%%%%%%%%%%%%%%%%%%%%%%%%%%%%%%%%%%%%%

The general quark-gluon vertex can be expressed as
(see \cite{PT-QCD,BC1,KRP})
\begin{eqnarray}
\label{decomp_1}
\Gamma_{\mu}(p_1,p_2,p_3)
&=&\gamma_{\mu}h_1+p_{2\mu}h_2+p_{1\mu}h_3 +\gamma_{\mu}\pslash_2 h_4
+\gamma_{\mu}\pslash_1 h_5+p_{2\mu}\pslash_2 h_6
+p_{2\mu}\pslash_1 h_7 \nn \\
&&+ p_{1\mu}\pslash_2 h_8+p_{1\mu}\pslash_1 h_9
+p_{2\mu}\pslash_1 \pslash_2 h_{10}
+p_{1\mu}\pslash_1 \pslash_2 h_{11}
+\gamma_{\mu} \pslash_1 \pslash_2 h_{12},
\end{eqnarray}  
where $h_i\equiv h_i(p_1^2,p_2^2,p_3^2)$.

The longitudinal and transverse functions of Eqs.~(\ref{GammaL}) 
and (\ref{GammaT}) 
are related to this naive basis as follows:
\begin{eqnarray}
\lambda_1 
&=& h_1- \half(p_2 p_3)(h_6 + h_7)
- \half(p_1 p_3)(h_8 + h_9) + \half(p_3^2-2p_2^2)h_{12}, 
\nn \\
\lambda_2 
&=& \frac{1}{2(p_1^2 - p_2^2)} 
\left[ (p_2 p_3)(h_6-h_7)
+(p_1 p_3)(h_8-h_9)-p_3^2h_{12} \right], 
\nn \\
\lambda_3 
&=& -\frac{1}{p_1^2 - p_2^2}
\Bigl\{
(p_2 p_3)\left[h_2+h_4+(p_1 p_2)h_{10}\right]
+(p_1 p_3) \left[h_3+h_5 +(p_1 p_2)h_{11}\right] \Bigr\}, 
\nn \\
\lambda_4 
&=& \half\left[-h_4 + h_5 + (p_2 p_3)h_{10} 
+(p_1 p_3)h_{11}\right], \\[8pt]
\tau_1 &=& \frac{1}{p_1^2 - p_2^2} \left[h_2 + h_3 + h_4 + h_5+(p_1 
p_2)(h_{10}+h_{11})\right], 
\nn \\
\tau_2 &=& \frac{1}{2(p_1^2 - p_2^2)}(h_6- h_7+h_8 - h_9 +2h_{12} ), 
\nn \\
\tau_3 &=& - \quarter( h_6+ h_7 +h_8 + h_9), 
\nn \\
\tau_4 &=& \frac{1}{p_1^2 - p_2^2}(h_{10} + h_{11}), 
\nn \\
\tau_5 &=& - \half(h_4 + h_5), 
\nn \\
\tau_6 &=& \quarter(h_6 + h_7 -h_8 - h_9 -2h_{12}), 
\nn \\
\tau_7 &=& \frac{1}{p_1^2 - p_2^2} \left[(p_2 p_3)h_{10}
+(p_1 p_3)h_{11}\right], 
\nn \\
\tau_8 &=& -h_{12},
\end{eqnarray}
where $\lambda_i\equiv \lambda_i(p_1^2,p_2^2,p_3^2)$, 
$\tau_i\equiv \tau_i(p_1^2,p_2^2,p_3^2)$, and
$h_i\equiv h_i(p_1^2,p_2^2,p_3^2)$.

Inverting these relations, one finds:
\begin{eqnarray}
h_1&=&\lambda_1+\tau_3p_3^2+\tau_6(p_1^2-p_2^2)+\tau_8(p_1 p_2), 
\nn \\
h_2&=&-\lambda_3+\lambda_4
-\tau_1(p_1 p_3)+\tau_4(p_1 p_2)(p_1 p_3)+\tau_5
 +\tau_7(p_2^2-\half p_3^2), 
\nn \\ 
h_3&=&\lambda_3-\lambda_4
+\left[\tau_1-\tau_4(p_1 p_2)\right](p_2 p_3)+\tau_5
 -\tau_7(p_2^2-\half p_3^2) , 
\nn \\ 
h_4&=&-\lambda_4-\tau_5+\half\tau_7(p_1^2-p_2^2), 
\nn \\
h_5&=&\lambda_4-\tau_5-\half\tau_7(p_1^2-p_2^2), 
\nn \\
h_6&=&\lambda_2-\tau_2(p_1 p_3)-\tau_3+\tau_6, 
\nn \\ 
h_7&=&-\lambda_2+\tau_2(p_1 p_3)-\tau_3+\tau_6-\tau_8, 
\nn \\
h_8&=&-\lambda_2+\tau_2(p_2 p_3)-\tau_3-\tau_6+\tau_8, 
\nn \\
h_9&=&\lambda_2-\tau_2(p_2 p_3)-\tau_3-\tau_6, 
\nn \\
h_{10}&=&-\tau_4(p_1 p_3)+\tau_7, 
\nn \\
h_{11}&=&\tau_4(p_2 p_3)-\tau_7, 
\nn \\
h_{12}&=&-\tau_8.
\end{eqnarray} 

%%%%%%%%%%%%%%%%%%%%%%%%%%%%%%%%%%%%%%%%%%%%%%%%%%%%%%%%%%%%%%%%%%%%%%%
\section*{Appendix B: Recurrence relations for scalar integrals}
\setcounter{equation}{0}
\renewcommand{\thesection}{B}
%%%%%%%%%%%%%%%%%%%%%%%%%%%%%%%%%%%%%%%%%%%%%%%%%%%%%%%%%%%%%%%%%%%%%%%

To calculate scalar integrals with higher (integer) powers 
of denominators, a recurrence procedure based on the
integration-by-parts technique \cite{ibp} has been used.
For the three-point integrals, one can follow an approach
described in \cite{JPA} (see, in particular, Section~4
of \cite{JPA} where the massive case is discussed).
Using the recurrence relations, we can reduce all integrals
to the master integrals $J_i(1,1,1)$ ($i=1,2$) and
a few two-point integrals. All basic integrals are 
discussed in Appendix~C.

\subsection{Two-point integrals}

All two-point integrals occurring in this paper can be associated 
with certain special cases of
three-point integrals $J_2$ and $J_1$ (defined in 
Eqs.~(\ref{defJ2}) and (\ref{defJ1}),
respectively), when one of the indices
$\nu_i$ vanishes. The masses of internal particles ($m_1, m_2$)
can be equal to ($m,m$), ($m,0$), ($0,m$), or ($0,0$).

For two-point integrals with arbitrary masses,
\bea
\label{two-point-massive-integral_def}
J^{(2)}(\nu_1, \nu_2 | m_1,m_2)&
\equiv &
\int {\mbox{d}}^n q 
\frac{1}{
\left[(p-q)^2-m_1^2\right]^{\nu_1}
\left(q^2-m_2^2\right)^{\nu_2}
} \; ,
\eea
such procedure has been described in detail in
Appendix~A of \cite{BDS}. For positive $\nu_i$,
it is enough to apply
\bea
\label{two-point_int_eq_sys}
J^{(2)}(\nu_1+1,\nu_2| m_1,m_2)&=& 
\frac{1}{\nu_1 \Delta}
\bigg\{
[(n \!-\!\nu_1 \!-\!2 \nu_2 ) (p^2\!-\!m_1^2) 
     + (n \!-\! 3 \nu_1) m_2^2]
J^{(2)}(\nu_1,\nu_2| m_1,m_2)
\nn \\ &&
- \nu_1 (p^2 - m_1^2 -m_2^2) J^{(2)}(\nu_1 + 1,\nu_2 - 1| m_1,m_2)
\nn \\ &&
- 2 \nu_2 m_2^2 J^{(2)}(\nu_1 - 1,\nu_2 + 1| m_1,m_2)
\bigg\},
\eea
together with a similar equation with 
$(\nu_1, m_1)\leftrightarrow (\nu_2, m_2)$. 
In Eq.~(\ref{two-point_int_eq_sys}),
\be
\Delta \equiv
\Delta(m_1^2,m_2^2,p^2) = 4 m_1^2 m_2^2 - (p^2-m_1^2-m_2^2)^2 \; .
\ee
In our calculation, $\Delta$ may be equal to $p^2(4m^2-p^2)$,
$-(p^2-m^2)^2$ or $-(p^2)^2$.

Note that the sum of indices $\nu_i$ for any of the integrals
on the r.h.s. of Eq.~(\ref{two-point_int_eq_sys})
is less (by one) than such a sum for the integral on the l.h.s.
Therefore, $J^{(2)}$ with any (positive) integers $\nu_1$ and $\nu_2$
can be reduced to $J^{(2)}$ with $\nu_1=\nu_2=1$ and tadpole
integrals.

The reduction of tadpole integrals is trivial:
\be
J_1(0,0,N) = 
\frac{1}{(N-1)!} 
\left(1-{\textstyle{n\over2}}\right)_{N-1}
\; (-m^2)^{1-N} \; J_1(0,0,1) \; 
\ee
(and similarly for $J_2(N,0,0)=J_2(0,N,0)$),
where $(a)_j\equiv \Gamma(a+j)/\Gamma(a)$ is the Pochhammer
symbol.

\subsection{Three-point integrals $J_2$}

For the integrals $J_2$ with positive $\nu_i$ (with
at least one of them being greater than one),
the following solution of recurrence relations can be used:
\bea
\label{J2_nu3+}
J_2(\nu_1,\nu_2,\nu_3+1)&=&
\frac{1}{2 \nu_3 {\cal{M}}_2}
\Biggl\{
\Biggl[
(\nu_1\!-\!\nu_3) p_1^2 p_3^2
+(\nu_2\!-\!\nu_3) p_2^2 p_3^2
+2 (n\!-\!\nu_1\!-\!\nu_2\!-\!2 \nu_3) p_3^2 (p_1 p_2)
\nn \\ &&
+2 (\nu_2-\nu_1) m^2 (p_1^2 -p_2^2) 
-(2 n-\nu_1-\nu_2-6 \nu_3) m^2 p_3^2
\Biggr] J_2(\nu_1,\nu_2,\nu_3)
\nn \\ &&
- p_3^2 (p_3^2 -4 m^2  )
\bigl[\nu_1
J_2(\nu_1\!+\!1,\nu_2,\nu_3\!-\!1)
+\nu_2
J_2(\nu_1,\nu_2\!+\!1,\nu_3\!-\!1)
\bigr]
\nn \\ &&
+\bigl[(p_2^2 -m^2) p_3^2  + 2 (p_1^2  - p_2^2 ) m^2\bigr]
\nn \\ &&
\times \bigl[\nu_1 
J_2(\nu_1+1,\nu_2-1,\nu_3)
+\nu_3
J_2(\nu_1,\nu_2-1,\nu_3+1)
\bigr]
\nn \\ &&
+\bigl[(p_1^2 -m^2) p_3^2  - 2 (p_1^2  - p_2^2 ) m^2\bigr]
\nn \\ &&
\times 
\bigl[\nu_2
J_2(\nu_1-1,\nu_2+1,\nu_3)
+\nu_3
J_2(\nu_1-1,\nu_2,\nu_3+1)
\bigr]\Biggr\},
\eea
\bea
\label{J2_nu1+}
J_2(\nu_1+1,\nu_2,\nu_3) &=&
\frac{1}{2 \nu_1 {\cal{M}}_2}
 \bigg\{ 
2 (n-\nu_1-\nu_2-2\nu_3)
\left[ p_1^2 (p_2 p_3) -m^2 (p_1 p_3) \right]
J_2(\nu_1,\nu_2,\nu_3)
\nn \\ &&
+(p_1^2-m^2)
\left[ (\nu_3-\nu_1) p_1^2+(\nu_2-\nu_3) p_2^2
+(\nu_1-\nu_2) m^2 \right]
J_2(\nu_1,\nu_2,\nu_3)
\nn \\ &&
+\Big[p_3^2 (p_1^2-m^2) -2 m^2 (p_1^2 - p_2^2)\Big]
\nn \\ && \times
[\nu_1 J_2(\nu_1 + 1,\nu_2,\nu_3 - 1)
+ \nu_2 J_2(\nu_1,\nu_2 + 1,\nu_3 - 1)]
\nn \\ &&
+(p_1^2\!-\!m^2 ) (p_2^2\!-\!m^2 )
[\nu_1 J_2(\nu_1 \!+\! 1,\nu_2 \!-\! 1,\nu_3)
+ \nu_3 J_2(\nu_1,\nu_2 \!-\! 1,\nu_3 \!+\! 1)]
\nn \\ &&
- (p_1^2\!-\!m^2 )^2 [\nu_2 J_2(\nu_1 \!-\! 1,\nu_2 \!+\! 1,\nu_3)
+ \nu_3 J_2(\nu_1 \!-\! 1,\nu_2,\nu_3 \!+\! 1)]
\bigg\}\; ,
\eea
and also an equation for $J_2(\nu_1,\nu_2+1,\nu_3)$
which can be obtained from
(\ref{J2_nu1+}) via $(\nu_1,p_1^2)\leftrightarrow (\nu_2,p_2^2)$.
The quantity ${\cal{M}}_2$ is defined in Eq.~(\ref{calM2}).

In fact, for our calculation we needed only the
$\nu_1=\nu_2=\nu_3=1$ case of Eq.~(\ref{J2_nu3+}),
\bea
J_2(1,1,2)&=&
\frac{1}{{2\cal{M}}_2}
\bigg\{
2 (n - 4) p_3^2 \left[(p_1  p_2)-m^2\right] J_2(1,1,1)
- 2 p_3^2 (p_3^2-4 m^2) J_2(2,1,0)
\nn \\ &&
+\left[(p_2^2-m^2) p_3^2+
 2(p_1^2-p_2^2) m^2\right]\left[J_2(2,0,1)+J_2(1,0,2)\right]
\nn \\ &&
+\left[(p_1^2-m^2) p_3^2-2
(p_1^2-p_2^2) m^2 \right]\left[J_2(0,2,1)+J_2(0,1,2)\right]
\bigg\} \; ,
\eea
where we have taken into account the obvious symmetry 
$J_2(\nu_1,\nu_2,0)=J_2(\nu_2,\nu_1,0)$.
We note a remarkable fact that the coefficient of $J_2(1,1,1)$
is proportional to $(n-4)$. Since, in the off-shell case, 
$J_2(1,1,1)$ is finite in four dimensions, this means that
we do not get any nontrivial function in finite parts of 
all triangle integrals with higher $\nu_i$. 
In fact, this property is valid for arbitrary internal masses
\cite{geom} (a special case has been discussed in \cite{KW}).

\subsection{Three-point integrals $J_1$}

For the integrals $J_1$ with positive $\nu_i$ (with
at least one of them being greater than one), 
the following solution of recurrence relations can
be used:
\bea
\label{J1_nu3+}
J_1(\nu_1,\nu_2,\nu_3+1) &=&
\frac{1}{2\nu_3 {\cal{M}}_1}
\bigg\{
\bigg[
2 (n-\nu_1-\nu_2-\nu_3)(p_1 p_2) 
+\nu_1 p_1^2 +\nu_2 p_2^2 -\nu_3 p_3^2 
\nn \\ &&
+\left(2 n -3\nu_1 -3\nu_2 - 2\nu_3 \right) m^2 \bigg]
J_1(\nu_1,\nu_2,\nu_3)
\nn \\ &&
+(p_2^2-m^2) \bigl[
\nu_1 J_1(\nu_1+1,\nu_2-1,\nu_3)
+\nu_3  J_1(\nu_1,\nu_2-1,\nu_3+1)
\bigr]
\nn \\ &&
+(p_1^2 -m^2) \bigl[\nu_2 J_1(\nu_1-1,\nu_2+1,\nu_3)
+ \nu_3  J_1(\nu_1-1,\nu_2,\nu_3+1)\bigr]
\nn \\ &&
-p_3^2 \bigl[\nu_1 J_1(\nu_1+1,\nu_2,\nu_3-1)
+\nu_2 J_1(\nu_1,\nu_2+1,\nu_3-1)\bigr]
\bigg\} \; ,
\eea
\bea
\label{J1_nu1+}
J_1(\nu_1+1,\nu_2,\nu_3)&=&
\frac{1}{2 \nu_1 p_3^2 {\cal{M}}_1}
\bigg\{
2(n-\nu_1-\nu_2-\nu_3)
\left[p_1^2(p_2 p_3)+m^2(p_1 p_3)\right] 
J_1(\nu_1,\nu_2,\nu_3)
\nn \\ &&
-(\nu_1 p_1^2  - \nu_2 p_2^2  - \nu_3 p_3^2 ) p_1^2
J_1(\nu_1,\nu_2,\nu_3)
\nn \\ &&
+ m^2 \left[ 2\nu_1 p_1^2  + 2 \nu_2 (p_1 p_2) + (\nu_2-\nu_3) p_3^2
-(\nu_1-\nu_2)m^2 \right]
J_1(\nu_1,\nu_2,\nu_3)
\nn \\ &&
+\bigl( {\cal{M}}_1  +  p_3^2m^2 \bigr)
\bigl[\nu_1 J_1(\nu_1 + 1,\nu_2-1,\nu_3)
+\nu_3 J_1(\nu_1,\nu_2-1,\nu_3+1)\bigr]
\nn \\ &&
+ p_3^2 (p_1^2-m^2) \bigl[\nu_1 J_1(\nu_1+1,\nu_2,\nu_3-1)
+\nu_2 J_1(\nu_1,\nu_2+1,\nu_3-1)\bigr]
\nn \\ &&
-(p_1^2\!-\!m^2)^2 \bigl[\nu_2 J_1(\nu_1\!-\!1,\nu_2\!+\!1,\nu_3)
+\nu_3 J_1(\nu_1\!-\!1,\nu_2,\nu_3\!+\!1)\bigr]
\bigg\} \; , \hspace*{4mm}
\eea
and also an equation for $J_1(\nu_1,\nu_2+1,\nu_3)$
which can be obtained from
(\ref{J1_nu1+}) via $(\nu_1,p_1^2)\leftrightarrow (\nu_2,p_2^2)$.
The quantity ${\cal{M}}_1$ is defined in Eq.~(\ref{calM1}).

The ``highest'' integral $J_1$ which occurred in our calculation 
was $J_1(2,2,1)$. Using Eq.~(\ref{J1_nu1+}) and the 
$(\nu_1,p_1^2)\leftrightarrow (\nu_2,p_2^2)$ symmetry, we get
\bea
J_1(2,2,1) &=&
\frac{1}{2 p_3^2 {\cal M}_1}
\bigg\{ \left( {\cal M}_1 + m^2 p_3^2 \right) 
\left[ J_1(2,1,1) + J_1(1,2,1) + J_1(1,1,2) \right]
\nn \\ &&
+ (n\!-\!5) \left[ p_2^2 (p_1 p_3)\!+\!m^2 (p_2 p_3) \right] 
J_1(2,1,1)
+ (n\!-\!5) \left[ p_1^2 (p_2 p_3)\!+\!m^2 (p_1 p_3) \right] 
J_1(1,2,1)
\nn \\ &&
-(p_1^2\!-\!m^2)^2 \left[ J_1(0,3,1)\!+\!\half J_1(0,2,2) \right]
-(p_2^2\!-\!m^2)^2 \left[ J_1(3,0,1)\!+\!\half J_1(2,0,2) \right]
\nn \\ &&
+p_3^2 \left( p_1^2+p_2^2-2m^2 \right)
\left[ J_1(3,1,0)+\half J_1(2,2,0) \right]
\bigg\} \; .
\eea

For the integrals $J_1(1,1,2)$, $J_1(2,1,1)$ and $J_1(1,2,1)$,
direct application of Eqs.~(\ref{J1_nu3+})--(\ref{J1_nu1+}) 
yields:
\bea
J_1(1,1,2) &=& \frac{1}{2 {\cal{M}}_1} 
\bigg\{  
2 (n-4) \left[(p_1 p_2) + m^2\right] J_1(1,1,1)
-2 p_3^2 J_1(2,1,0)
\nn \\ &&
+ (p_2^2 \!-\! m^2) [J_1(2,0,1) \!+\! J_1(1,0,2)]
\!+\! (p_1^2 \!-\! m^2) [J_1(0,2,1) \!+\! J_1(0,1,2)]
\bigg\} , \hspace*{7mm}
\\
J_1(2,1,1)  &=& \frac{1}{2 {\cal{M}}_1 p_3^2}
\bigg\{
2 (n - 4) \left[p_1^2 (p_2 p_3)+m^2 (p_1 p_3)\right] J_1(1,1,1)
+2 p_3^2 (p_1^2-m^2) J_1(2,1,0)
\nn \\ &&
+ ({\cal{M}}_1\!+\!m^2 p_3^2) \left[J_1(2,0,1) \!+\! J_1(1,0,2)\right]
   - (p_1^2\!-\!m^2)^2 \left[J_1(0,2,1) \!+\! J_1(0,1,2)\right]
\bigg\}
\nn \\ && {}
\eea
and similarly for $J_1(1,2,1)$.

\subsection{Integrals with numerators}

We also need some integrals with negative powers of
denominators (i.e., when the corresponding denominator
is in the numerator). Such formulae can be obtained
in a standard way, via representing the numerators in
terms of the external invariants. For the integrals
$J_2$, we only needed the cases when one of the
$\nu_i$ was equal to $(-1)$,
\begin{eqnarray*}
J_2(\nu_1,\nu_2,-1) &=& 
\left[m^2-(p_1  p_2)\right]J_2(\nu_1,\nu_2,0)+J_2(\nu_1,\nu_2-1,0)
,\\ \nn \\
J_2(\nu_1,-1,\nu_3) &=&  -(p_2^2)^{-1}(p_2
p_3)\left[J_2(\nu_1,0,\nu_3-1)
-m^2J_2(\nu_1,0,\nu_3) \right]-(p_1 
p_3)J_2(\nu_1,0,\nu_3) 
,\\ \nn \\
 J_2(-1,\nu_2,\nu_3) &=&-(p_1^2)^{-1}(p_1  p_3)
\left[J_2(0,\nu_2,\nu_3-1)-m^2J_2(0,\nu_2,\nu_3) \right]
-(p_2  p_3)J_2(0,\nu_2,\nu_3) .
\end{eqnarray*}
We also list analogous results for $J_1$: 
\begin{eqnarray*}
J_1(\nu_1,\nu_2,-1)&=& -(p_3^2)^{-1}
\bigg\{ p_3^2 \bigl[(p_1 p_2)+m^2\bigr] J_1(\nu_1,\nu_2,0)
                   +(p_1 p_3) J_1(\nu_1-1,\nu_2,0)
\\ &&
                   +(p_2 p_3) J_1(\nu_1,\nu_2-1,0) \bigg\} \; ,
\\
J_1(\nu_1,-1,\nu_3)&=& -(p_2^2)^{-1}
\bigg\{
\bigl[ p_2^2(p_1 p_3)+m^2(p_2 p_3) \bigr]
J_1(\nu_1,0,\nu_3)
\\ &&
+(p_2 p_3) J_1(\nu_1,0,\nu_3-1)+(p_1 p_2)
J_1(\nu_1-1,0,\nu_3)
\bigg\},
\\ 
J_1(-1,\nu_2,\nu_3)&=& -(p_1^2)^{-1} 
\bigg\{ 
\bigl[ p_1^2 (p_2 p_3) + m^2 (p_1 p_3)\bigr] 
J_1(0,\nu_2,\nu_3)
\\ &&
+(p_1 p_3) J_1(0,\nu_2,\nu_3-1)
+(p_1 p_2) J_1(0,\nu_2-1,\nu_3) \bigg\}.
\end{eqnarray*}
Moreover, an integral with $\nu_3=-2$  has occurred,
\begin{eqnarray*}
J_1(\nu_1,\nu_2,-2) &=&
\bigg\{ \bigl[(p_1 p_2)+m^2\bigr]^2 - (n-1)^{-1} {\cal K} \bigg\}
 J_1(\nu_1,\nu_2,0)
\\ &&
+ 2(p_3^2)^{-1} 
\bigg\{ (p_1 p_3) \bigl[(p_1 p_2)+m^2\bigr]
+ (n-1)^{-1} {\cal K} \bigg\} J_1(\nu_1-1,\nu_2,0)
\\ &&
+ 2(p_3^2)^{-1}
\bigg\{ (p_2 p_3) \bigl[(p_1 p_2)+m^2\bigr]
+ (n-1)^{-1} {\cal K} \bigg\} J_1(\nu_1,\nu_2-1,0)
\\ &&
+ (p_3^2)^{-2}
\bigg[ (p_1 p_3)^2- (n-1)^{-1} {\cal K} \bigg]
J_1(\nu_1-2,\nu_2,0)
\\ &&
+ (p_3^2)^{-2}
\bigg[ (p_2 p_3)^2- (n-1)^{-1} {\cal K} \bigg] 
J_1(\nu_1,\nu_2-2,0)
\\ &&
+ 2 (p_3^2)^{-2} \bigg[ (p_1 p_3) (p_2 p_3)
+ (n-1)^{-1} {\cal K} \bigg] J_1(\nu_1-1,\nu_2-1,0) \; ,
\end{eqnarray*}
with ${\cal K}$ defined in Eq.~(\ref{Eq:calK}).

%%%%%%%%%%%%%%%%%%%%%%%%%%%%%%%%%%%%%%%%%%%%%%%%%%%%%%%%%%%%%%%%%%%%%%%
\section*{Appendix C: Basic scalar integrals}
\setcounter{equation}{0}
\setcounter{subsection}{0}
\renewcommand{\thesection}{C}
%%%%%%%%%%%%%%%%%%%%%%%%%%%%%%%%%%%%%%%%%%%%%%%%%%%%%%%%%%%%%%%%%%%%%%%

As discussed in Section~2.1 (see also Appendix~B), all results for the scalar 
functions ($\lambda_i$, $\tau_i$, etc.) can be expressed in terms of
three-point integrals $J_1(1,1,1)$ and $J_2(1,1,1)$, two-point
integrals (\ref{J1_110})--(\ref{J2_110}) and a tadpole integral (\ref{J1_001}).
Remember that we are interested in results which are 
valid for arbitrary values of the space-time dimension $n$.

\subsection{Two-point integrals}

The massless two-point integral is trivial,
\begin{equation}
J_1(1,1,0)=J_0(1,1,0)={\mbox{i}}\; \pi^{n/2}\; 
(-p_3^2)^{n/2-2}\;
\frac{\Gamma^2\left({\textstyle{n\over2}}-1\right) 
      \Gamma\left(2-{\textstyle{n\over2}}\right)}
     {\Gamma(n-2)} \; .
\end{equation}
Two-point integrals with one or two massive lines can be expressed
in terms of the Gauss hypergeometric function (see, e.g., in \cite{BD-TMF}):
\begin{eqnarray}
\label{2ptm0}
J_1(0,1,1)=J_2(0,1,1)=2{\mbox{i}}\; \pi^{n/2}\; (m^2)^{n/2-2}\;
\frac{\Gamma\left(2-\frac{n}{2}\right)}{n-2}\;
\left. _2F_1\left(
\begin{array}{c} 1, \; 2-n/2 \\ n/2 \end{array}
\right| \frac{p_1^2}{m^2} \right) \; ,
\hspace*{8mm}
\\
\label{2ptmm}
J_2(1,1,0) = \mbox{i} \pi^{n/2} \; (m^2)^{n/2-2}\;
\Gamma\left(2-{\textstyle{n\over2}}\right)\;
\left. _2F_1\left(
\begin{array}{c} 1, \; 2-n/2 \\ 3/2 \end{array}
\right| \frac{p_3^2}{4m^2} \right) .
\hspace*{30mm}
\end{eqnarray}
The results for the integrals $J_1(1,0,1)=J_2(1,0,1)$ can be obtained
from (\ref{2ptm0}) by substituting $p_1^2\to p_2^2$.

In three dimensions, Eqs.~(\ref{2ptm0}) and (\ref{2ptmm}) yield
(cf., e.g., in \cite{Nickel})
\begin{eqnarray}
\left. J_1(0,1,1)\right|_{n=3} &=& \left. J_2(0,1,1)\right|_{n=3}
=\frac{2\mbox{i}\pi^2}{m}\,
f\left(\frac{p_1^2}{m^2}\right) \; ,
\nn \\
\left. J_2(1,1,0)\right|_{n=3} &=& \frac{\mbox{i}\pi^2}{m}\,
f\left(\frac{p_3^2}{4 m^2}\right) \; ,
\end{eqnarray}
with
\begin{eqnarray}
\label{f}
f(z) = \left\{ 
\begin{array}{ll} \frac{1}{2\sqrt{z}} 
                 \ln\frac{1+\sqrt{z}}{1-\sqrt{z}} \; , &
                 \;\;\; z>0 \; , \\
                 \frac{1}{\sqrt{-z}} \arctan\sqrt{-z} \; , &
                 \;\;\; z<0 \; .
\end{array} \right.
\end{eqnarray}

Around four dimensions ($n=4-2\varepsilon$), they are singular
(see, e.g., \cite{tHV'79}): 
\begin{equation}
J_1(0,1,1)=J_2(0,1,1)
=\mbox{i}\pi^{2-\varepsilon}m^{-2\varepsilon}
\Gamma(1+\varepsilon)
\left\{ \frac{1}{\varepsilon} + 2 
+ \frac{m^2\!-\!p_1^2}{p_1^2} \ln\frac{m^2\!-\!p_1^2}{m^2} 
+{\cal O}(\varepsilon)
\right\} ,
\end{equation}
\begin{equation}
J_2(1,1,0) = \mbox{i}\pi^{2-\varepsilon}m^{-2\varepsilon}\,
\Gamma(1+\varepsilon)
\left\{ \frac{1}{\varepsilon} +2 
- 2 f\left(\frac{p_3^2}{p_3^2-4m^2}\right) 
+{\cal O}(\varepsilon)
\right\} ,
\end{equation}
with the same function $f$ as in Eq.~(\ref{f}).
Above the corresponding threshold ($p_1^2>m^2$ or $p_3^2>4m^2$),
these functions acquire imaginary part, whose sign is
defined by the causal prescription 
$p_i^2\leftrightarrow p_i^2+\mbox{i}0$
(for details, see Appendix~A of \cite{BDS}).
%================================================================

In the limit $p_1^2=p_2^2=p^2$ (in particular, in the on-shell case), 
the combinations
\[
\frac{J_2(1,0,1)-J_2(0,1,1)}{p_2^2-p_1^2} 
\]
should be treated in the following way:
\begin{equation}  
\left. \left\{\frac{J_2(1,0,1)
-J_2(0,1,1)}{p_2^2-p_1^2}\right\}\right|_{p_1^2=p_2^2=p^2}
= \frac{\mbox{d}}{\mbox{d}p^2}
\left\{ \left. J_2(0,1,1)\right|_{p_1^2=p^2} \right\}\; .
\end{equation}
Using Eq.~(\ref{2ptm0}), and also the fact that
\begin{equation}
\frac{\mbox{d}}{\mbox{d}z}
\left. _2F_1\left(
\begin{array}{c} a, \; b \\ c \end{array}   
\right| z\right)  
= \frac{ab}{c}\;
\left. _2F_1\left(
\begin{array}{c} a+1, \; b+1 \\ c+1 \end{array}
\right| z\right) ,
\end{equation}
we obtain
\begin{equation}
\left. \left\{\frac{J_2(1,0,1)
-J_2(0,1,1)}{p_2^2-p_1^2}\right\}\right|_{p_1^2=p_2^2=p^2}
= -\frac{1}{2m^2} \; \left. J_2(0,1,1)\right|_{p_1^2=p^2} .
\end{equation}

In the on-shell case, we get
\begin{equation}
\left. J_2(0,1,1)\right|_{p_1^2=m^2}
=\frac{n-2}{2 m^2 (n-3)} J_2(0,1,0) .
\end{equation}
Moreover, for some functions we need the expansion of
this integral in $\delta_1=(m^2-p_1^2)/m^2$ up to the linear term,
\begin{equation}
J_2(0,1,1)=\frac{n-2}{2m^2(n-3)} J_2(0,1,0)
\left[ 1 + \half\delta_1 
+ {\cal{O}}(\delta_1^2)
\right] \, .
\end{equation}

\subsection{Three-point integrals}

For the triangle integrals $J_1$ and $J_2$,
results in arbitrary dimension, and also for any powers
of the propagators, can be presented in terms of
multiple hypergeometric functions \cite{BD-TMF}.
For our purposes, we need integer powers of
propagators. Using recurrence relations \cite{JPA} based on 
the integration-by-parts technique \cite{ibp},
all scalar integrals can be reduced to $J_i(1,1,1)$
and two-point functions.

\subsubsection{General off-shell case}

Transforming Feynman parametric integrals (for example, using
the Cheng--Wu theorem \cite{CW}), we can present
$J_1(1,1,1)$ and $J_2(1,1,1)$ as 
\be
\label{2fold1}
J_1(1,1,1) 
= -{\mbox{i}}\pi^{n/2} \Gamma(3-{\textstyle{n\over2}})
\int\limits_0^{\infty} \int\limits_0^{\infty}
\frac{{\mbox{d}}\xi\;\; {\mbox{d}}\eta}
{(1\!+\!\xi\!+\!\eta)^{n-3}
\left[m^2\eta(1\!+\!\xi\!+\!\eta) \!-\!\eta p_1^2
      \!-\!\xi \eta p_2^2\!-\!\xi p_3^2\right]^{3-n/2}} \; ,
\ee
\be
\label{2fold2}
J_2(1,1,1) = -{\mbox{i}}\pi^{n/2} \Gamma(3-{\textstyle{n\over2}})
\int\limits_0^{\infty} \int\limits_0^{\infty}
\frac{{\mbox{d}}\xi\;\; {\mbox{d}}\eta}{(1+\xi+\eta)^{n-3}
\left[m^2(1\!+\!\xi\!+\!\eta)(1\!+\!\eta)\!-\!\xi p_1^2 \!-\!\xi\eta p_2^2 
      \!-\!\eta p_3^2\right]^{3-n/2}} \; .
\ee

In the three-dimensional case ($n=3$) 
the denominator $(1+\xi+\eta)$ disappears in both integrals,
and one can easily integrate over $\xi$, and then over $\eta$. This yields
\be
\left. J_1(1,1,1)\right|_{n=3} =
-\frac{{\mbox{i}}\pi^2}{\sqrt{p_3^2{\cal{M}}_1}}
\ln\left[\frac{m+\sqrt{{\cal{M}}_1/p_3^2}}
              {m-\sqrt{{\cal{M}}_1/p_3^2}}\right] \; ,
\ee
\be
\left. J_2(1,1,1)\right|_{n=3} =
-\frac{{\mbox{i}}\pi^2}{\sqrt{{\cal{M}}_2}}
\ln\left[\frac{m(2m^2-p_1^2-p_2^2)+\sqrt{{\cal{M}}_2}}
              {m(2m^2-p_1^2-p_2^2)-\sqrt{{\cal{M}}_2}}\right] \; ,
\ee
where ${\cal M}_1$ and ${\cal M}_2$ are defined in
Eqs.~(\ref{calM1}) and (\ref{calM2}), respectively.
If ${\cal{M}}_1/p_3^2$ or ${\cal{M}}_2$ is negative (in $J_1$
and $J_2$, respectively), the logarithms should be substituted
by arctan functions (see Eq.~(\ref{f})),
which correspond to the limiting cases of the result obtained 
in \cite{Nickel}
(see also in \cite{geom}) for three-dimensional three-point function
with arbitrary masses. Note that in the massless case we arrive 
at the well-known result \cite{uniq3} (see also in \cite{BKP})
\be
\left. J_{1,2}(1,1,1)\right|_{n=3,\; m=0} =
\left. J_0(1,1,1)\right|_{n=3} =
-\frac{{\mbox{i}}\pi^3}{\sqrt{-p_1^2 p_2^2 p_3^2}} \; .
\ee

In four dimensions ($n=4$), we can also integrate over $\xi$. 
Then, performing the remaining $\eta$ integral, we arrive at the
known results in terms of dilogarithms \cite{tHV'79}.

\subsubsection{Symmetric case}

The ``symmetric'' case is of a certain interest (see e.g.\ in 
\cite{DTP}), when all external invariants are equal,
$p_1^2=p_2^2=p_3^2=-\mu^2$. 
Then we obtain
\be
\left. J_1(1,1,1)\right|_{n=4,\; p_i^2=-\mu^2}
= -\frac{{\mbox{i}}\pi^2}{\mu^2}
\int\limits_0^{\infty}
\frac{{\mbox{d}}\eta}{1+\eta+\eta^2}
\ln\left[\frac{(1+\eta)(m^2+\mu^2+\mu^2\eta)}
              {m^2+\mu^2+m^2\eta}\right] \; ,
\ee
\be
\left. J_2(1,1,1)\right|_{n=4,\; p_i^2=-\mu^2}
= -\frac{{\mbox{i}}\pi^2}{\mu^2}
\int\limits_0^{\infty}
\frac{{\mbox{d}}\eta}{1+\eta+\eta^2}
\ln\left[\frac{(m^2+\mu^2)(1+\eta)^2}{m^2(1+\eta)^2+\mu^2\eta}\right] \; .
\ee
These parametric representations are equivalent to those for
the $H$ and $M$ functions given in \cite{DTP}. Note that
$\mu^2J_1\leftrightarrow{\mbox{i}}\pi^2{}H$ and
$\mu^2J_2\leftrightarrow{\mbox{i}}\pi^2{}M$.
These integrals can be evaluated in terms of Clausen functions as
\begin{eqnarray}
\left. J_1(1,1,1)\right|_{n=4,\; p_i^2=-\mu^2}
&=& -\frac{{\mbox{i}}\pi^2}{\mu^2\sqrt{3}}
\bigg\{ 2\Cl{2}{\frac{\pi}{3}} + 2 \Cl{2}{\frac{\pi}{3}+2\theta_{{\rm s}1}}
+ \Cl{2}{\frac{\pi}{3}-2\theta_{{\rm s}1}}
\nonumber \\
&&\hspace*{10mm}
+ \Cl{2}{\pi-2\theta_{{\rm s}1}} \bigg\} \; ,
\end{eqnarray}
\be
\left. J_2(1,1,1)\right|_{n=4,\; p_i^2=-\mu^2}
= -\frac{2{\mbox{i}}\pi^2}{\mu^2\sqrt{3}}
\left\{ 2 \Cl{2}{\frac{2\pi}{3}} +  \Cl{2}{\frac{\pi}{3}+2\theta_{{\rm s}2}}
+ \Cl{2}{\frac{\pi}{3}-2\theta_{{\rm s}2}} \right\} \; ,
\ee
where 
\be
\tan\theta_{{\rm s}1}=\frac{\mu^2+2m^2}{\mu^2\sqrt{3}}, \hspace{10mm}
\tan\theta_{{\rm s}2}=\sqrt{\frac{\mu^2+4m^2}{3\mu^2}} \; .
\ee
This gives analytical results for the $H$ and $M$ functions from \cite{DTP}.
In the massless limit ($m\to 0$), $\theta_{{\rm s}1}=\theta_{{\rm s}2}
=\pi/6$, and, remembering that
$\Cl{2}{\frac{2\pi}{3}}={\textstyle{2\over3}}\Cl{2}{\frac{\pi}{3}}$,
we reproduce the well-known result \cite{CG}
\be
\left. J_0(1,1,1)\right|_{n=4,\; p_i^2=-\mu^2}
= -\frac{4{\mbox{i}}\pi^2}{\mu^2\sqrt{3}} \Cl{2}{\frac{\pi}{3}} \; .
\ee

\subsubsection{On-shell limit}

Now let us consider the on-shell limit $p_1^2=p_2^2=m^2$.
For $J_1(1,1,1)$, the two-fold parametric integral (\ref{2fold1}) 
yields
\be
\label{osJ1}
\left. J_1(1,1,1)\right|_{p_1^2=p_2^2=m^2} =
-{\mbox{i}}\pi^{n/2} \Gamma\left(3-{\textstyle{n\over2}}\right)
\int\limits_0^{\infty} \int\limits_0^{\infty}
\frac{{\mbox{d}}\xi \; {\mbox{d}}\eta}
{(1+\xi+\eta)^{n-3} \left[ m^2 \eta^2 - \xi p_3^2 \right]^{3-n/2}} .
\ee
Using Mellin--Barnes contour integral for the second denominator,
we find
\bea
\left. J_1(1,1,1)\right|_{p_1^2=p_2^2=m^2} &=&
-{\mbox{i}}\pi^{n/2} (m^2)^{n/2-3} \frac{1}{\Gamma(n-3)}\;
\frac{1}{2\pi{\mbox{i}}}
\int\limits_{-{\rm{i}}\infty}^{{\rm{i}}\infty}
{\mbox{d}}s \left( -\frac{p_3^2}{m^2} \right)^s
\nn \\
&& \times \Gamma(-s)\; \Gamma(n-5-2s)\;
\Gamma^2(1+s)\; \Gamma\left(3-{\textstyle{n\over2}}+s\right)\; .
\eea
Closing the contour to the right, we get the result in terms 
of $_2F_1$ functions of the argument $p_3^2/(4m^2)$.
Note that the integral $J_1(1,1,1)$ is not divergent
in the on-shell limit, so that we can put $n=4$ in Eq.~(\ref{osJ1}). 
For instance,
the following simple representation in terms of $\mbox{Cl}_2$
can be mentioned
\be
\left. J_1(1,1,1)\right|_{n=4,\; p_1^2=p_2^2=m^2} =
\frac{2{\mbox{i}}\pi^2}{\sqrt{p_3^2 (4m^2-p_3^2)}}
\left\{ \Cl{2}{\pi-2\psi_{\rm os}} - {\mbox{i}}\pi \psi_{\rm os} 
\right\} \; ,
\ee
with
\[
\psi_{\rm os} = \arctan\sqrt{(4m^2-p_3^2)/p_3^2}, \hspace{10mm}
0<p_3^2<4m^2 \; .
\]

Let us also consider the integral $J_2(1,1,1)$ in the on-shell limit.
Starting from the representation
(\ref{2fold2}) in arbitrary dimension $n$ and integrating 
over $\xi$, we get
\be
\left. J_2(1,1,1)\right|_{p_1^2=p_2^2=m^2} =
{\textstyle{1\over2}} {\mbox{i}}\pi^{n/2} 
\Gamma\left(2-{\textstyle{n\over2}}\right)
\int\limits_0^{\infty}
\frac{{\mbox{d}}\eta}{(1+\eta)^{n-4}
\left[ m^2(1+\eta)^2-\eta p_3^2 \right]^{3-n/2}} \; .
\ee
It is easy to show that this three-point function 
(in the on-shell limit)  
reduces to a two-point function (\ref{2ptmm})
with a shifted space-time dimension $n\to n-2$,
\begin{equation}
\label{3to2}
\left. J_2(1,1,1)\right|_{p_1^2=p_2^2=m^2}
=\left.\frac{\pi}{4-n} J_2(1,1,0)
\right|_{n\to n-2}.
\end{equation}
Using Eq.~(\ref{2ptmm}) we get
\begin{equation}
\label{ons111}
\left. J_2(1,1,1)\right|_{p_1^2=p_2^2=m^2}
= {\textstyle{1\over2}} \; \mbox{i} \pi^{n/2} \; (m^2)^{n/2-3}\;
\Gamma\left(2-{\textstyle{n\over2}}\right)\;
\left. _2F_1\left(
\begin{array}{c} 1, \; 3-n/2 \\ 3/2 \end{array}   
\right| \frac{p_3^2}{4m^2} \right) .
\end{equation}
In the limit $n\to 4$, because of the singular factor in front
of the r.h.s. of Eq.~(\ref{3to2}), 
we need to expand the two-point function up to the $\ep$ term.
This is how dilogarithms (or Clausen functions) arise
in the finite part of the three-point function.

The same result (\ref{ons111}) can be obtained from Eqs.~(32) and (34)
of \cite{BD-TMF}, taking into account that the first two arguments of 
the $\Phi_2$ function are $z_1=z_2\equiv z=1$.
Therefore, the sum over $j$ in Eq.~(34) of \cite{BD-TMF}
(corresponding to an $_2F_1$ function of unit argument)
can be performed in terms of $\Gamma$ functions.
As a result, we arrive at Eq.~(\ref{ons111}).

Moreover, using Kummer relations for contiguous
$_2F_1$ functions, we get 
\begin{equation}
(n-3)\; \left. _2F_1\left(
\begin{array}{c} 1, \; 2-n/2 \\ 3/2 \end{array}
\right| z\right) = (n-4)(1-z)\; 
\left. _2F_1\left(
\begin{array}{c} 1, \; 3-n/2 \\ 3/2 \end{array}
\right| z\right) + 1 .
\end{equation}
Therefore,
\begin{equation}
\label{3->2}
\left. J_2(1,1,1)\right|_{p_1^2=p_2^2=m^2}
=\frac{2(n-3)}{(n-4)(4m^2-p_3^2)}\;
\left[ J_2(1,1,0) - \left. J_2(0,1,1)\right|_{p_1^2=m^2} \right] .
\end{equation}

%%%%%%%%%%%%%%%%%%%%%%%%%%%%%%%%%%%%%%%%%%%%%%%%%%%%%%%%%%%%%%%%%%%%%%%
\section*{Appendix D: Results for the $\chi$ functions}
\setcounter{equation}{0}
\setcounter{subsection}{0}
\renewcommand{\thesection}{D}
%%%%%%%%%%%%%%%%%%%%%%%%%%%%%%%%%%%%%%%%%%%%%%%%%%%%%%%%%%%%%%%%%%%%%%%

The one-loop results for $\chi_i$ functions from
Eq.~(\ref{H}) are collected below:
\begin{eqnarray}
\label{chi0}
\chi_0^{(1)}(p_1^2,p_2^2,p_3^2) &=& 
-\frac{g^2\eta}{(4\pi)^{n/2}}\;
\frac{C_A}{8 {\cal{M}}_1}
\bigg\{ 
\big[2+(n-3)\xi\big] {\cal M}_1 (p_1^2-m^2) \varphi_1
-(n-2)\xi p_1^2 p_2^2 p_3^2 \varphi_1
\nn \\ &&
+(n-2)\xi {\cal M}_1 \big[ (p_1 p_2) +m^2 \big] \varphi_1
-\xi m^2 p_3^2 \big[ n (p_1 p_2) + 2 m^2 \big] \varphi_1
\nn \\ && 
+(n-3)\xi \left[ m^2 p_3^2 \kappa_{0,3}
+ p_1^2 (p_2^2-m^2) \kappa_{1,1}
+ m^2 (p_1^2-p_3^2-m^2) \kappa_{1,2} \right]
\nn \\ &&
-(2-\xi) {\cal M}_1 \big( \kappa_{0,3} + \kappa_{1,2} \big)
+ \xi {\cal M}_1 \kappa_{1,1}
+(n-2) \xi m^2 \big[ (p_1 p_2)+m^2 \big] {\widetilde{\kappa}}
\bigg\} \; ,
\\ 
\label{chi1}
\chi_1^{(1)}(p_1^2,p_2^2,p_3^2) &=&
\frac{g^2\eta}{(4\pi)^{n/2}}\;
\frac{m \; C_A}{8 {\cal K} {\cal M}_1}
\bigg\{ 
(2-\xi) {\cal M}_1 (p_2 p_3) \big[(p_1 p_2)+m^2\big]
\varphi_1   
-\xi m^2 {\cal K} p_3^2 \varphi_1
\nn \\ && 
+ \big[2+(n-3)\xi\big] {\cal K} {\cal M}_1 \varphi_1
+(n-3) \xi p_3^2 \big[ (p_1 p_2)+m^2\big] 
\left[ p_1^2 p_2^2+m^2 (p_1 p_2) \right] \varphi_1
\nn \\ && 
+ (2-\xi) {\cal M}_1 
\left[ (p_1 p_2) \kappa_{1,1} + p_2^2 \kappa_{1,2}
       + (p_2 p_3) \kappa_{0,3} \right] 
\nn \\ &&
+(n-3) \xi p_3^2 \left[ p_1^2 p_2^2+m^2 (p_1 p_2)
\right] \kappa_{0,3}
+(n-3) \xi p_1^2 \left[ p_2^2 (p_1 p_3)+m^2 (p_2 p_3) 
\right] \kappa_{1,1}
\nn \\ && 
+(n-3) \xi p_2^2 \left[ p_1^2 (p_2 p_3)+m^2 (p_1 p_3) 
\right] \kappa_{1,2}
+(n-2) \xi m^2 {\cal{K}} \; {\widetilde{\kappa}} 
\bigg\} \; ,
\\
\label{chi2}
\chi_2^{(1)}(p_1^2,p_2^2,p_3^2) &=& 
\frac{g^2\eta}{(4\pi)^{n/2}}\; 
\frac{m\; C_A}{8 {\cal K} {\cal M}_1 }
\bigg\{ 
\big[ 2+(n-3)\xi \big] {\cal K} {\cal M}_1 \varphi_1
-(2-\xi) {\cal M}_1 (p_1 p_3)  \big[ (p_1 p_2)+m^2 \big] 
\varphi_1
\nn \\ &&
+\xi  
\left[ (n-4) {\cal K} -(n-3) {\cal M}_1 \right] p_1^2 p_3^2 \varphi_1
- (n-3) \xi p_1^2 p_3^2 \big[ (p_1 p_2)+m^2 \big] \kappa_{0,3}
\nn \\ && 
+ (n-3) \xi p_1^2 
\left[ p_3^2 (p_1 p_2) + (p_1 p_3) (p_2^2-m^2) \right] \kappa_{1,1}
\nn \\ && 
- (n-3) \xi p_1^2 (p_2 p_3) \big[ (p_1 p_2)+m^2 \big] \kappa_{1,2}
+ \xi p_1^2 {\cal K} \kappa_{1,2}
\nn \\ &&
- (2-\xi) {\cal M}_1
\left[ p_1^2 \kappa_{1,1} + (p_1 p_2) \kappa_{1,2}
      +(p_1 p_3) \kappa_{0,3} \right]
\nn \\ && 
+(n-2) \xi m^2 {\cal{K}} (p_2^2)^{-1} \big[ 2 (p_1 p_2)+m^2 \big]
\big( \kappa_{1,2} - {\widetilde{\kappa}} \big) 
\bigg\} \; ,
\\
\chi_3^{(1)}(p_1^2,p_2^2,p_3^2) &=& 
- \frac{g^2\eta}{(4\pi)^{n/2}}\;
\frac{C_A}{8 {\cal K} {\cal M}_1 }
\bigg\{
(n-4) \xi m^2  {\cal K} p_3^2 \varphi_1
- m^2 {\cal M}_1 \big[ 2 (p_1 p_3)+\xi (p_2 p_3) \big] \varphi_1
\nn \\ && 
- \big[ 2+(n-3)\xi \big] {\cal M}_1 p_1^2 (p_2 p_3) \varphi_1 
- (n-2) \xi {\cal M}_1 (p_1 p_2) (p_2 p_3) \varphi_1
\nn \\ &&
- {\cal M}_1 \big[ 2 (p_1 p_3)\!+\!\xi (p_2 p_3) \big] \kappa_{0,3}
- {\cal M}_1 \big[ 2 p_1^2\!+\!\xi (p_1 p_2) \big] \kappa_{1,1}
- {\cal M}_1 \big[ 2 (p_1 p_2)\!+\!\xi p_2^2 \big] \kappa_{1,2}
\nn \\ && 
+ (n-3) \xi p_3^2 \left[ p_1^2 p_2^2+m^2 (p_1 p_2) \right]
\kappa_{0,3}
+ (n-3) \xi p_1^2 \left[ p_2^2 (p_1 p_3)+m^2 (p_2 p_3) \right]
\kappa_{1,1}
\nn \\ && 
+ (n-3) \xi p_2^2 \left[ p_1^2 (p_2 p_3)+m^2 (p_1 p_3) \right]
\kappa_{1,2}
+(n-2) \xi m^2 {\cal{K}} \; {\widetilde{\kappa}} 
\bigg\} \; .
\end{eqnarray}

%%%%%%%%%%%%%%%%%%%%%%%%%%%%%%%%%%%%%%%%%%%%%%%%%%%%%%%%%%%%%%%%%%%%%%%
\section*{Appendix E: Transverse functions in an arbitrary gauge}
\setcounter{equation}{0}
\setcounter{subsection}{0}
\renewcommand{\thesection}{E}
%%%%%%%%%%%%%%%%%%%%%%%%%%%%%%%%%%%%%%%%%%%%%%%%%%%%%%%%%%%%%%%%%%%%%%%

We here collect results for the transverse parts of the vertex,
$\tau_i$, valid for arbitrary covariant gauge and dimension.

%%%%%%%%%%%%%%%%%%%%%%%%%%%%%%%%%%%%%%%%%%%%%%%%%%%%%%%%%%%%%%%%%%%%%%%%

\subsection{Transverse functions of diagram $a$}

Using the decomposition (\ref{combs}), all $\tau$'s
of diagram $a$ can be presented as\footnote{As mentioned in
Section~3.2, the combinations of $\kappa$'s 
in Eqs.~(\ref{tau_dec_a}) and (\ref{tau_dec_b})
are linearly dependent. As a result, we can (simultaneously)
shift $t_{i,1}\to t_{i,1}+2c_i$, 
$t_{i,2}\to t_{i,2}+p_3^2c_i$ and
$t_{i,5}\to t_{i,5}+(p_1^2-p_2^2)^2c_i$
(where $c_i$ are arbitrary functions of the momenta,
which can be chosen separately for each $\tau_i$),
without change of the value of the corresponding $\tau_i$.
This possibility allows us to write some of the $t$'s in 
a more compact form.}
\bea
\label{tau_dec_a}
\tau_i^{(1a)}(p_1^2,p_2^2,p_3^2)
&=&\frac{g^2\eta \left(C_F-\half C_A\right)}{(4\pi)^{n/2}}  
\bigg\{ t_{i,0}^{(1a)} \varphi_2
+ t_{i,1}^{(1a)} 
\left[(p_1 p_3) \kappa_{1,1}
      +(p_2 p_3) \kappa_{1,2}+p_3^2 \kappa_{2,3}\right]
\nn \\
&& + t_{i,2}^{(1a)} 
\left( \kappa_{1,1} + \kappa_{1,2}-2 \kappa_{2,3} \right)
+ t_{i,3}^{(1a)} 
\left( \kappa_{1,1} + \kappa_{1,2}-2 \widetilde{\kappa} \right)
\nn \\
&& + t_{i,4}^{(1a)} \left( \kappa_{1,1} + \kappa_{1,2} \right)
+ t_{i,5}^{(1a)} \frac{\kappa_{1,1} - \kappa_{1,2}}{p_1^2-p_2^2} 
\bigg\} \; .
\eea
The results for the scalar functions $t_i^{(1a)}$ 
(which depend on the invariants $p_1^2, p_2^2, p_3^2$) are listed below,
for all eight $\tau$'s:
\begin{eqnarray*}
t_{1,0}^{(1a)} &=& - (n-\xi) m {\cal{K}}^{-1} 
[ (p_1 p_2)-m^2 ]\; ,
\nn \\
t_{1,1}^{(1a)} &=& t_{1,3}^{(1a)} = t_{1,4}^{(1a)} = 0 \; ,
\nn \\
t_{1,2}^{(1a)} &=& \half (n-\xi) m {\cal{K}}^{-1} \; ,
\nn \\
t_{1,5}^{(1a)} &=& \half (n-\xi) m {\cal{K}}^{-1} (p_1-p_2)^2 \; , \\[8pt]
%%%%%%%%%%%%%%%%%%%%%%%%%%%%%%%%%%%%%%%%%%%%%%%%%%%%%%%%%%%%%%%%%%%%%%
t_{2,0}^{(1a)} &=&
-\quarter
{\cal{K}}^{-1}
\{ 2  (n-4) \xi {\cal{M}}_2^{-1} p_3^2
(p_1^2-m^2) (p_2^2-m^2) [ (p_1 p_2)-m^2 ]
+(1+\xi) p_3^2-4m^2
\nn \\ &&
+ (n-1) {\cal{K}}^{-1} p_3^2 [ (p_1 p_2)-m^2 ]
   [ (1+\xi) (p_1 p_2) -(1-\xi) m^2]
+4\xi (p_1 p_2)
\} ,
\nn \\
%%%%%%%%%%%%%%%%%%%%%%%%%%%%%%%
t_{2,1}^{(1a)} &=& -\quarter (n-1) {\cal K}^{-2} 
[(1+\xi ) (p_1 p_2)-(1-\xi ) m^2] \; ,
\nn \\
t_{2,2}^{(1a)} &=& \quarter \xi  {\cal K}^{-1} 
[(n-2)-(n-3) m^2{\cal M}_2^{-1} (p_1^2-p_2^2)^2] \; ,
\nn \\
t_{2,3}^{(1a)} &=& -{\textstyle{1\over8}}(n-2) m^2{\cal {K}}^{-1}
 \{(1-\xi ) (p_1^2 p_2^2)^{-1}(p_1 p_2) 
 + 2 \xi {\cal M}_2^{-1} p_3^2 [(p_1 p_2)-m^2]
+ 4 \xi {\cal K} {\cal M}_2^{-1} \} \; ,
\nn \\
t_{2,4}^{(1a)} &=& -{\textstyle{1\over8}}(n-4){\cal K}^{-1} {\cal M}_2^{-1}
 \{(1-\xi ) {\cal M}_2+2 \xi  p_3^2 [p_1^2 p_2^2-m^2(p_1 p_2) ]
-4 \xi m^2 {\cal K} \} \; ,
\nn \\
t_{2,5}^{(1a)} &=& -{\textstyle{1\over8}}{\cal K}^{-1} (p_1^2 p_2^2)^{-1}
 \{(n-2)(1-\xi) (p_1-p_2)^2 
[ p_1^2 p_2^2-m^2 (p_1 p_2) ] 
+ 2 (n-2)(1-\xi) m^2 {\cal K}  \nn \\
&&
+2 (n-3) \xi m^2 {\cal M}_2^{-1}
p_1^2 p_2^2 
(p_1-p_2)^2 (p_1^2-p_2^2)^2\} \; , \\[8pt]
%%%%%%%%%%%%%%%%%%%%%%%%%%%%%%%%%%%%%%%%%%%%%%%%%%%%%%%%%%%%%%%%%%%%%%
t_{3,0}^{(1a)} &=&   
{\textstyle{1\over8}}{\cal{K}}^{-1}
\{
(p_1^2+p_2^2-2m^2)
[ (1+\xi) (p_1^2+p_2^2)-2(1-\xi)m^2 ]
-4(n-2) {\cal{K}}
\nn \\ &&
+(n-1){\cal{K}}^{-1} (p_1^2-p_2^2)^2
[ (p_1 p_2)-m^2 ]
[ (1+\xi) (p_1 p_2) - (1-\xi) m^2 ]
\nn \\ &&
- 2(n-4)\xi m^2 {\cal{M}}_2^{-1}
(p_1^2-p_2^2)^2
[ (p_1 p_2)-m^2 ]
(p_1^2+p_2^2-2m^2) \} ,
\nn \\
%%%%%%%%%%%%%%%%%%%%%%%%%%%%%%%
t_{3,1}^{(1a)} &=& 0 \; ,
\nn \\
t_{3,2}^{(1a)} &=& {\textstyle{1\over16}}{\cal K}^{-2}
 \{ 
2 (n-3) \xi m^2 {\cal K} {\cal M}_2^{-1} (p_1^2-p_2^2)^2
(p_1^2+p_2^2-2 m^2)
\nn \\ &&
- [(n-1) (p_1^2-p_2^2)^2+4 {\cal K}] 
[(1+\xi ) (p_1 p_2)-(1-\xi ) m^2]
\} \; ,
\nn \\
t_{3,3}^{(1a)} &=& {\textstyle{1\over16}}(n-2) m^2{\cal K}^{-1}
 \{(1-\xi ) (p_1^2 p_2^2)^{-1}[(p_1-p_2)^2 (p_1 p_2)- 2 {\cal K}]
 +2 \xi {\cal M}_2^{-1} (p_1^2-p_2^2)^2 [(p_1 p_2)-m^2] \} \; ,
\nn \\
t_{3,4}^{(1a)} &=& {\textstyle{1\over16}}(n-4){\cal K}^{-1}
 \{(1+\xi ) (p_1-p_2)^2
     + 2 \xi m^2 {\cal M}_2^{-1} (p_1^2-p_2^2)^2 [(p_1 p_2)-m^2] 
\} \; ,
\nn \\
t_{3,5}^{(1a)} &=& {\textstyle{1\over16}}{\cal K}^{-2}
 (p_1^2-p_2^2)^2
  \{
   (n-1+2 \xi ) [2{\cal K} -(p_1 p_2)(p_1-p_2)^2] 
- (n-n\xi+6\xi) {\cal K} 
\nn \\ &&
   -(1-\xi ) m^2 [(n-2) {\cal K} (p_1^2 p_2^2)^{-1}(p_1 p_2)
-(n-1) (p_1-p_2)^2]
\nn \\
&&
+(n-3) \xi {\cal M}_2^{-1} (p_1-p_2)^2
[ 2m^2 p_1^2 p_2^2 p_3^2 
- p_3^2 (p_1p_2)(p_1^2 p_2^2 +m^4)
-4 m^4 {\cal K} ] \} \; ,
\\[8pt]
%%%%%%%%%%%%%%%%%%%%%%%%%%%%%%%%%%%%%%%%%%%%%%%%%%%%%%%%%%%%%%%%%%%%%%
t_{4,0}^{(1a)} &=& -\half \xi m{\cal K}^{-1}
 \{ 
(n-1) {\cal K}^{-1} p_3^2 [(p_1p_2)-m^2] 
-2 (n-5)
\nn \\ &&
  + 2 (n-4) {\cal K} {\cal{M}}_2^{-1} (p_1^2+p_2^2-2 m^2)
- (n-4) {\cal{M}}_2^{-1} (p_1^2-p_2^2)^2 
[ (p_1p_2)-m^2 ] \} \; ,
\nn \\
t_{4,1}^{(1a)} &=& 
\half \xi m {\cal{K}}^{-1} [ 
(n-3){\cal{M}}_2^{-1} (p_1^2+p_2^2-2m^2)
- (n-1){\cal{K}}^{-1} 
] \; ,
\nn \\
t_{4,2}^{(1a)} &=& 0 \; ,
\nn \\
t_{4,3}^{(1a)} &=& \quarter (n-2)\xi m^3 
 {\cal K}^{-1} {\cal M}_2^{-1} \, (p_1^2 p_2^2)^{-1}
[ 2 {\cal K} (p_1^2+p_2^2) -p_1^2 p_2^2 p_3^2
+m^2 p_3^2 (p_1p_2) ] \; ,
\nn \\
t_{4,4}^{(1a)} &=& -\quarter (n-4)\xi m {\cal K}^{-1} {\cal M}_2^{-1}
  p_3^2 [(p_1p_2)-m^2] \; ,
\nn \\
t_{4,5}^{(1a)} &=& - \quarter \xi m
 \{ (n-2){\cal K}^{-1} 
[(p_1^2 p_2^2)^{-1}(p_1p_2)(p_1^2+p_2^2) -2]  
  +4(n-3){\cal M}_2^{-1} (p_1^2+p_2^2-2 m^2)\} \; , 
\\[8pt]
%%%%%%%%%%%%%%%%%%%%%%%%%%%%%%%%%%%%%%%%%%%%%%%%%%%%%%%%%%%%%%%%%%%%%%
t_{5,0}^{(1a)} &=& 
-\quarter m {\cal K}^{-1}
\{
4 (n-4-\xi) {\cal K} 
+ (n-4)\xi {\cal K} {\cal M}_2^{-1} (p_1^2-p_2^2)^2 (p_3^2-4 m^2)
\nn \\ &&
-(n-3)\xi (p_1^2-p_2^2)^2 - 2\xi p_3^2 [ (p_1p_2)-m^2 ]
\} \; ,
\nn \\
t_{5,1}^{(1a)} &=& \quarter \xi m {\cal K}^{-1}
 \{(n-3) {\cal M}_2^{-1}
(p_1^2-p_2^2)^2 [(p_1p_2)-m^2] +2 \} \; ,
\nn \\
t_{5,2}^{(1a)} &=& 0 \; ,
\nn \\
t_{5,3}^{(1a)} &=& {\textstyle{1\over8}}(n-2) \xi m^3 {\cal M}_2^{-1} 
(p_1^2 p_2^2)^{-1}
 [2 (p_1^2-p_2^2)^2 (p_1p_2)
 +2 p_1^2 p_2^2 p_3^2-m^2 p_3^2 (p_1^2+p_2^2) ] \; ,
\nn \\
t_{5,4}^{(1a)} &=& {\textstyle{1\over8}}(n-4) \xi m {\cal M}_2^{-1}
 p_3^2 (p_1^2+p_2^2-2 m^2) \; ,
\nn \\
t_{5,5}^{(1a)} &=& {\textstyle{1\over8}} \xi m (p_1^2-p_2^2)^2
   \{ (n-2)(p_1^2 p_2^2)^{-1} - 4 (n-3) {\cal M}_2^{-1} 
      [(p_1p_2)-m^2] \} \; , 
\\[8pt]
%%%%%%%%%%%%%%%%%%%%%%%%%%%%%%%%%%%%%%%%%%%%%%%%%%%%%%%%%%%%%%%%%%%%%%
t_{6,0}^{(1a)} &=& -{\textstyle{1\over8}}{\cal K}^{-1} 
{\cal M}_2^{-1}(p_1^2-p_2^2)
 \{{\cal M}_2 [(1+\xi) p_3^2-4 m^2] 
+6 \xi m^2 p_3^2 [(p_1p_2)-m^2] (p_1^2+p_2^2-2 m^2) \nn \\
&&
+(n-1) {\cal K}^{-1} p_3^2 [(p_1p_2)-m^2]^2 
[{\cal M}_2+\xi p_3^2 (p_1^2 p_2^2-m^4)]\} \; ,
\nn \\
t_{6,1}^{(1a)} &=& {\textstyle{1\over8}}{\cal K}^{-1}(p_1^2-p_2^2)
 \{ 
2 (n-3)\xi m^2 {\cal M}_2^{-1} (p_1^2+p_2^2-2 m^2)
-(n-1){\cal K}^{-1} [(1+\xi) (p_1p_2)-(1-\xi) m^2] 
\} \; ,
\nn \\
t_{6,2}^{(1a)} &=& 0 \; ,
\nn \\
t_{6,3}^{(1a)} &=& -{\textstyle{1\over16}}(n-2) m^2 
{\cal K}^{-1} (p_1^2-p_2^2)
 \{(1-\xi) (p_1^2 p_2^2)^{-1}(p_1p_2)
+2 \xi  {\cal M}_2^{-1} p_3^2 [(p_1p_2)-m^2]\} \; ,
\nn \\
t_{6,4}^{(1a)} &=& -{\textstyle{1\over16}}(n-4) 
{\cal K}^{-1} {\cal M}_2^{-1}(p_1^2-p_2^2)
 \{(1-\xi) {\cal M}_2+2 \xi p_1^2 p_2^2 p_3^2
 -2 \xi m^2 [4 {\cal K}+p_3^2 (p_1p_2)]\} \; ,
\nn \\
t_{6,5}^{(1a)} &=&  {\textstyle{1\over16}}{\cal K}^{-1}(p_1^2-p_2^2)
 \{ (n-2) (1-\xi) m^2 
[(p_1^2 p_2^2)^{-1}(p_1 p_2)(p_1^2+p_2^2) -2 ]
-(n-2) (1+\xi) (p_1-p_2)^2
\nn \\ &&
-8 (n-3) \xi m^2 {\cal K}{\cal M}_2^{-1} (p_1^2+p_2^2-2 m^2)
\} \; , 
\\[8pt]
%%%%%%%%%%%%%%%%%%%%%%%%%%%%%%%%%%%%%%%%%%%%%%%%%%%%%%%%%%%%%%%%%%%%%%
t_{7,0}^{(1a)} &=& \half \xi m p_3^2
 [(n-4){\cal M}_2^{-1} (p_1^2+p_2^2-2 m^2) 
- (n-3){\cal K}^{-1}
] \; ,
\nn \\
t_{7,1}^{(1a)} &=& -\half (n-3)\xi m {\cal M}_2^{-1}
\{ {\cal K}^{-1} p_3^2[(p_1p_2)-m^2] +2\} \; ,
\nn \\
t_{7,2}^{(1a)} &=& 0 \; ,
\nn \\
t_{7,3}^{(1a)} &=& 
-\quarter(n-2)\xi m^3 {\cal{M}}_2^{-1} (p_1^2 p_2^2)^{-1}
      \{ p_3^2[ 2(p_1 p_2)-m^2] + 4{\cal{K}}\} \; ,
\nn \\
t_{7,4}^{(1a)} &=& -\quarter (n-4) \xi m {\cal M}_2^{-1} p_3^2 \; ,
\nn \\
t_{7,5}^{(1a)} &=& \quarter \xi m
 \{ 4 (n-3) {\cal M}_2^{-1} 
[ 2 {\cal K} + p_3^2 (p_1p_2)-m^2 p_3^2 ]
-(n-2) (p_1^2 p_2^2)^{-1}(p_1^2+p_2^2)  \} \; ,
\\[8pt]
%%%%%%%%%%%%%%%%%%%%%%%%%%%%%%%%%%%%%%%%%%%%%%%%%%%%%%%%%%%%%%%%%%%%%%
t_{8,0}^{(1a)} &=& -\half[n-6-(n-4) \xi]{\cal K}^{-1} p_3^2 
[(p_1p_2)-m^2] \; ,
\nn \\
t_{8,1}^{(1a)} &=& -\half [n-6-(n-4) \xi]{\cal K}^{-1} \; ,
\nn \\
t_{8,2}^{(1a)} &=& t_{8,3}^{(1a)} = t_{8,4}^{(1a)} = t_{8,5}^{(1a)} = 0 \; .
\end{eqnarray*}

\subsection{Transverse functions of diagram $b$}

By analogy with diagram $a$, (\ref{tau_dec_a}), all $\tau$'s
of diagram $b$ can be presented as 
\bea
\label{tau_dec_b}
\tau_i^{(1b)}(p_1^2,p_2^2,p_3^2)
&=&\frac{g^2\eta C_A}{(4\pi)^{n/2}}
\bigg\{ t_{i,0}^{(1b)} \varphi_1
+ t_{i,1}^{(1b)}
[(p_1 p_3) \kappa_{1,1}
      +(p_2 p_3) \kappa_{1,2}+p_3^2 \kappa_{0,3}]
\nn \\
&& + t_{i,2}^{(1b)}
( \kappa_{1,1} + \kappa_{1,2}-2 \kappa_{0,3} )
+ t_{i,3}^{(1b)}
( \kappa_{1,1} + \kappa_{1,2}-2 \widetilde{\kappa} )
\nn \\
&& + t_{i,4}^{(1b)} ( \kappa_{1,1} + \kappa_{1,2} )
+ t_{i,5}^{(1b)} \frac{\kappa_{1,1} - \kappa_{1,2}}{p_1^2-p_2^2}
\bigg\} \; .
\eea
The results for the scalar functions $t_i^{(1b)}$
(which depend on the invariants $p_1^2, p_2^2, p_3^2$) are listed below,
for all eight $\tau$'s:
\begin{eqnarray*}
t_{1,0}^{(1b)} &=& {\textstyle{1\over8}}m {\cal K}^{-1} {\cal M}_1^{-1}
 \{(n-4) (n-6) \xi^2 m^2 
           {\cal K}{\cal M}_1^{-1} p_3^2 [(p_1 p_2)+m^2]  
+(n-4)\xi {\cal K} p_3^2
\nn \\ &&
+(n-4)\xi^2 [ (p_1 p_2)+m^2 ]
[ (n-2){\cal K} + (n-3) p_3^2 (p_1 p_2) ]
\nn \\ &&
-(n-3)\xi(1-\xi){\cal M}_1 p_3^2
-2 (2n-2-\xi) {\cal M}_1 [ (p_1 p_2)+m^2 ]
\} \; ,
\nn \\
t_{1,1}^{(1b)} &=& -{\textstyle{1\over8}}(n-3) \xi m {\cal K}^{-1} 
{\cal M}_1^{-1}
  \{[1- (n -3) \xi] (p_1 p_2)
+[1-\xi - (n-6) \xi {\cal K}{\cal M}_1^{-1}] m^2\} \; ,
\nn \\
t_{1,2}^{(1b)} &=& {\textstyle{1\over16}} m
[ 
2 (2n-2 -\xi){\cal K}^{-1}
-(n-2) (n-3) \xi^2{\cal M}_1^{-1}] \; ,
\nn \\
t_{1,3}^{(1b)} &=& {\textstyle{1\over16}} (n-2)\xi m
 \{(p_1^2 p_2^2)^{-1}-{\cal M}_1^{-1}
               -(n-6) \xi m^2 {\cal M}_1^{-2}[(p_1 p_2)+m^2]\} \; ,
\nn \\
t_{1,4}^{(1b)} &=& {\textstyle{1\over16}} (n-4)\xi m {\cal{M}}_1^{-1}
\{3 \xi -1  
+ (n-6) \xi {\cal M}_1^{-1} [p_1^2 p_2^2 + m^2 (p_1 p_2)  ]  \} \; ,
\nn \\
t_{1,5}^{(1b)} &=& {\textstyle{1\over16}} m {\cal K}^{-1}
 [2 (2 n - 2 - \xi) (p_1-p_2)^2 
-(n-2) \xi {\cal K} (p_1^2 p_2^2)^{-1} (p_1^2+p_2^2) ] \; ,
\\[8pt]
%%%%%%%%%%%%%%%%%%%%%%%%%%%%%%%%%%%%%%%%%%%%%%%%%%%%%%%%%%%%%%%%%%%%%%
t_{2,0}^{(1b)} &=&   
{\textstyle{1\over16}}{\cal{K}}^{-1}
\{ [ 2-\xi -(n-3) \xi^2 ] p_3^2
- (n-3) (n-4)\xi^2 m^2 {\cal{M}}_1^{-1} p_3^2 
 [ (p_1 p_2)+m^2 ]
\nn \\ &&
+ (n-1)(2-\xi) {\cal{K}}^{-1} p_3^2  
 [ (p_1 p_2)+m^2 ]^2
- (n-1)(n-3)\xi^2 {\cal{K}}^{-1} p_3^2 (p_1 p_2)
 [ (p_1 p_2)+m^2 ]
\nn \\ &&
+ [ 4(n-3) - 2(2n-5)\xi-(n-2)(n-3)\xi^2 ]
 [ (p_1 p_2)+m^2 ]
\} ,
\nn \\   
%%%%%%%%%%%%%%%%%%%%%%%%%%%%%%%
t_{2,1}^{(1b)} &=&
{\textstyle{1\over16}}{\cal{K}}^{-2}
\{ (n-1)(2-\xi) [(p_1 p_2)+m^2]
-(n-3)\xi^2[(n-1)(p_1 p_2)+(n-3)m^2{\cal{K}}{\cal{M}}_1^{-1}
] \} ,
\nn \\ 
%%%%%%%%%%%%%%%%%%%%%%%%%%%%%%%
t_{2,2}^{(1b)} &=& -{\textstyle{1\over32}}{\cal K}^{-1}
 [ 4 (n-3)- 2(2n-5)\xi -(n-2)(n-3)\xi^2 ] \; ,
\nn \\
%%%%%%%%%%%%%%%%%%%%%%%%%%%%%%%
t_{2,3}^{(1b)} &=&
{\textstyle{1\over32}}(n-2)m^2 {\cal{K}}^{-1}
\{ (n-3)\xi^2 {\cal{M}}_1^{-1}
        [(p_1 p_2)+m^2]
-(2-\xi) (p_1^2 p_2^2)^{-1}(p_1 p_2) 
\} ,
\nn \\
%%%%%%%%%%%%%%%%%%%%%%%%%%%%%%%
t_{2,4}^{(1b)} &=& {\textstyle{1\over32}}(n-4){\cal K}^{-1} {\cal M}_1^{-1}
  \{(2-\xi) {\cal M}_1 
- (n-3) \xi^2 [p_1^2 p_2^2 +  m^2 (p_1 p_2)]\} \; ,
\nn \\
t_{2,5}^{(1b)} &=& {\textstyle{1\over32}}{\cal K}^{-1}
 \{(n-2) (2-\xi) m^2 
[(p_1^2 p_2^2)^{-1}(p_1 p_2)(p_1^2+p_2^2) -2] \nn \\
&&
  -[2(n-4)-(3n-8)\xi] (p_1-p_2)^2\} \; ,
\\[8pt]
%%%%%%%%%%%%%%%%%%%%%%%%%%%%%%%%%%%%%%%%%%%%%%%%%%%%%%%%%%%%%%%%%%%%%%
t_{3,0}^{(1b)} &=&
{\textstyle{1\over32}}{\cal{K}}^{-1}
\{ (n-4)\xi m^2{\cal{M}}_1^{-1}
        [(p_1 p_2)+m^2]
        [ (n-3)\xi (p_1^2-p_2^2)^2-4(4-\xi){\cal{K}} ]
\nn \\ &&
+\xi [ 4+(n-4)\xi] [ (p_1 p_2)+m^2]
     [ (n-3) (p_1-p_2)^2 -4(n-4){\cal{K}}(p_3^2)^{-1} ]
\nn \\ &&
+(2-\xi) [ (n-2) (p_1^2-p_2^2)^2 
              ( 1-{\cal{K}}^{-1}{\cal{M}}_1 )
- {\cal{K}}^{-1}{\cal{M}}_1  p_3^2 (p_1-p_2)^2 ]
+8 (n\!-\!3) \xi {\cal{K}} 
\nn \\ &&
+(n\!-\!3)\xi^2 (p_1^2\!-\!p_2^2)^2 
+(n\!-\!3)\xi^2 (p_1 p_2)\!
[ (p_1 p_2)\!+\!m^2 ] 
[ 4 +  (n\!-\!1){\cal{K}}^{-1} (p_1^2\!-\!p_2^2)^2] 
\} ,
\nn \\
%%%%%%%%%%%%%%%%%%%%%%%%%%%%%%%
t_{3,1}^{(1b)} &=&
{\textstyle{1\over32}}{\cal{K}}^{-1}(p_3^2)^{-1}
\{ 
(n-3)\xi m^2 {\cal{M}}_1^{-1}
[ (n-3)\xi (p_1^2-p_2^2)^2 -4(4-\xi) {\cal{K}} ]
\nn \\ &&
+ [ (n-1) {\cal{K}}^{-1} (p_1^2-p_2^2)^2 +4 ]
  [ (n-3)\xi^2 (p_1 p_2)-(2-\xi) (p_1 p_2)-(2-\xi)m^2 ]
\} ,
\nn \\
%%%%%%%%%%%%%%%%%%%%%%%%%%%%%%%
t_{3,2}^{(1b)} &=& -{\textstyle{1\over64}}\xi {\cal K}^{-1} (p_3^2)^{-1}
 [4+(n -4) \xi] [4 {\cal K}+(n-3) (p_1^2-p_2^2)^2] \; ,
\nn \\
t_{3,3}^{(1b)} &=& {\textstyle{1\over64}}(n-2)m^2{\cal K}^{-1} (p_3^2)^{-1} 
 \{\xi {\cal M}_1^{-1} [(p_1 p_2)+m^2]
[4 (4-\xi) {\cal K} -(n-3) \xi (p_1^2-p_2^2)^2 ] 
\nn \\
&&
+(2-\xi) (p_1^2 p_2^2)^{-1}p_3^2 
[(p_1 p_2) (p_1-p_2)^2-2 {\cal K}]\} \; ,
\nn \\
t_{3,4}^{(1b)} &=& {\textstyle{1\over64}}(n-4){\cal K}^{-1} (p_3^2)^{-1}
\{
(n-3) \xi^2{\cal M}_1^{-1} (p_1^2-p_2^2)^2 
           [ p_1^2 p_2^2+m^2(p_1 p_2)]
\nn \\ &&
-(2-\xi) p_3^2 (p_1-p_2)^2
-4\xi(4-\xi) {\cal K} {\cal M}_1^{-1}
           [ p_1^2 p_2^2+m^2(p_1 p_2)]
\} \; ,
\nn \\
t_{3,5}^{(1b)} &=& {\textstyle{1\over64}}{\cal K}^{-1} (p_3^2)^{-1}
 (2-\xi) (p_1^2-p_2^2)^2
 \{4 (n-2)  m^2{\cal K}(p_1^2 p_2^2)^{-1}
 -2 (n-3)\xi (p_1-p_2)^2
\nn \\ &&
-(n-2) (p_1-p_2)^2 
[ m^2(p_1^2 p_2^2)^{-1}(p_1 p_2)  + 1 ]
\} \; ,
\\[8pt]
%%%%%%%%%%%%%%%%%%%%%%%%%%%%%%%%%%%%%%%%%%%%%%%%%%%%%%%%%%%%%%%%%%%%%%
t_{4,0}^{(1b)} &=&
{\textstyle{1\over8}}\xi m {\cal{K}}^{-1}{\cal{M}}_1^{-1}
\{ (n-4)(n-6)\xi m^2
{\cal{K}}{\cal{M}}_1^{-1}  p_3^2
+(n-2)(n-4) \xi{\cal{K}}
\nn \\ &&       
+[ 1+(n-3) \xi] p_3^2(p_1 p_2)
 [ n-4-(n-1) {\cal{K}}^{-1}{\cal{M}}_1  ]
-(n-1)m^2{\cal{K}}^{-1} {\cal{M}}_1   p_3^2
\nn \\ &&
+(n-4) [ 1-(n-3)\xi] m^2 p_3^2
-(n-2) (n-3) \xi {\cal{M}}_1
\} ,
\nn \\
%%%%%%%%%%%%%%%%%%%%%%%%%%%%%%%
t_{4,1}^{(1b)} &=& {\textstyle{1\over8}}\xi m {\cal K}^{-1} {\cal M}_1^{-1}
 \{ (n-3)\xi [ p_1^2 p_2^2+m^2 (p_1 p_2) ]
[ (n-6) {\cal M}_1^{-1} - (n-1) {\cal K}^{-1} ]
\nn \\ &&
-(n-1) {\cal K}^{-1}{\cal M}_1 + (n-3) (1+3\xi) 
\} \; ,
\nn \\
t_{4,2}^{(1b)} &=& {\textstyle{1\over16}}(n-2) (n-3)
\xi^2 m {\cal K}^{-1} {\cal M}_1^{-1}
  [(p_1 p_2)+m^2] \; ,
\nn \\
t_{4,3}^{(1b)} &=& {\textstyle{1\over16}}(n-2)\xi m {\cal K}^{-1} 
{\cal M}_1^{-1}
 [ {\cal M}_1 (p_1 p_2) (p_1^2 p_2^2)^{-1} - (p_1 p_2)
-m^2 + (n-3)\xi m^2
-(n-6) \xi m^2 {\cal K} {\cal M}_1^{-1} 
] \; ,
\nn \\
t_{4,4}^{(1b)} &=& -{\textstyle{1\over16}}(n-4) \xi m {\cal K}^{-1} 
{\cal M}_1^{-1}
 \{ (p_1 p_2)+m^2 +(n-3)\xi(p_1 p_2)
+(n-6) \xi m^2 {\cal K}{\cal M}_1^{-1} \} \; ,
\nn \\
t_{4,5}^{(1b)} &=& -{\textstyle{1\over16}}(n-2)\xi m {\cal K}^{-1}
  [(p_1^2 p_2^2)^{-1}(p_1 p_2)(p_1^2+p_2^2) -2] \; ,
\\[8pt]
%%%%%%%%%%%%%%%%%%%%%%%%%%%%%%%%%%%%%%%%%%%%%%%%%%%%%%%%%%%%%%%%%%%%%%
t_{5,0}^{(1b)} &=&
-{\textstyle{1\over8}} m {\cal{K}}^{-1}
\{ \xi p_3^2 
[ p_1^2+p_2^2+(2-\xi) m^2 ]
[ (n-4){\cal{K}}{\cal{M}}_1^{-1} -(n-3) ]
\nn \\ &&
+[(n-3)(2-\xi)-1] \xi p_3^2 
[ (p_1 p_2)+m^2 ]
+ 
[ 12-2\xi+(n-4)\xi(4-\xi) ] {\cal{K}} 
\} ,
\nn \\
%%%%%%%%%%%%%%%%%%%%%%%%%%%%%%%
t_{5,1}^{(1b)} &=& {\textstyle{1\over8}}\xi m  {\cal K}^{-1} \; ,
\nn \\
%%%%%%%%%%%%%%%%%%%%%%%%%%%%%%%
t_{5,2}^{(1b)} &=&
{\textstyle{1\over16}}(n\!-\!3) \xi m {\cal{K}}^{-1}{\cal{M}}_1^{-1}
\{ 2{\cal{K}}( p_1^2\!+\!p_2^2\!-\!2m^2 )
-(p_1^2\!-\!p_2^2)^2 [ (p_1 p_2)\!+\!m^2 ]
-\xi p_3^2 [ p_1^2 p_2^2 \!+\! m^2 (p_1 p_2) ]\!
\} ,
\nn \\
%%%%%%%%%%%%%%%%%%%%%%%%%%%%%%%
t_{5,3}^{(1b)} &=& {\textstyle{1\over16}}(n-2)\xi m {\cal M}_1^{-1}
\{ (2-\xi) m^2 
- [ {\cal M}_1(p_1^2 p_2^2)^{-1}-1 ] 
(p_1^2+p_2^2) 
\} \; ,
\nn \\
t_{5,4}^{(1b)} &=& {\textstyle{1\over16}}(n-4)\xi m {\cal M}_1^{-1}
  [p_1^2+p_2^2+(2-\xi) m^2] \; ,
\nn \\
t_{5,5}^{(1b)} &=& {\textstyle{1\over16}}\xi m {\cal K}^{-1} {\cal M}_1^{-1}
 (p_1^2-p_2^2)^2
 \{
(n-3)(2-\xi) [p_1^2 p_2^2+m^2(p_1 p_2)]
\nn \\
&&
+(n-2) {\cal K} {\cal M}_1(p_1^2 p_2^2)^{-1}
-(n-3) (p_1^2+p_2^2) [(p_1 p_2) + m^2 ]
\} \; ,
\\[8pt]
%%%%%%%%%%%%%%%%%%%%%%%%%%%%%%%%%%%%%%%%%%%%%%%%%%%%%%%%%%%%%%%%%%%%%%
t_{6,0}^{(1b)} &=&
-{\textstyle{1\over32}}{\cal{K}}^{-1}(p_1^2-p_2^2)
\{ (n-3) (n-4) 
\xi^2 ( m^2 {\cal{M}}_1^{-1} p_3^2+1)[ (p_1 p_2)+m^2 ]
\nn \\ &&
+(n-1) (n-3)\xi^2 {\cal{K}}^{-1} p_3^2 (p_1 p_2)
[ (p_1 p_2)+m^2 ]
+ [ (n-2)(2-\xi)+(n-3)\xi^2] p_3^2
\nn \\ &&
-2[ 2-(n-3)\xi ]
[ (p_1 p_2)+m^2 ]
- (n-1) (2-\xi) {\cal{K}}^{-1} {\cal{M}}_1 p_3^2 
\} ,
\nn \\
%%%%%%%%%%%%%%%%%%%%%%%%%%%%%%%
t_{6,1}^{(1b)} &=&
{\textstyle{1\over32}}
{\cal{K}}^{-2}(p_1^2\!-\!p_2^2)
\{ \!
(n\!-\!1) (2\!-\!\xi) [ (p_1 p_2)\!+\!m^2 ] 
\!-\!(n\!-\!1) (n\!-\!3) \xi^2 (p_1 p_2)
\!-\!(n\!-\!3)^2 \xi^2 m^2 {\cal{K}}{\cal{M}}_1^{-1} \!
\} ,
\nn \\
%%%%%%%%%%%%%%%%%%%%%%%%%%%%%%%
t_{6,2}^{(1b)} &=&
-{\textstyle{1\over64}}{\cal{K}}^{-1}(p_1^2-p_2^2)
[ 4 - 2(n-3)\xi - (n-3)(n-4)\xi^2 ] , 
\nn \\
%%%%%%%%%%%%%%%%%%%%%%%%%%%%%%%
t_{6,3}^{(1b)} &=&
{\textstyle{1\over64}}(n-2){\cal{K}}^{-1} (p_1^2-p_2^2)
\{ (n-3)\xi^2 m^2{\cal{M}}_1^{-1} 
[ (p_1 p_2) + m^2 ]
-(2-\xi) m^2 (p_1^2 p_2^2)^{-1} (p_1 p_2)\} ,
\nn \\
%%%%%%%%%%%%%%%%%%%%%%%%%%%%%%%
t_{6,4}^{(1b)} &=& 
{\textstyle{1\over64}}(n-4){\cal K}^{-1} {\cal M}_1^{-1}(p_1^2-p_2^2)
  \{(2-\xi) {\cal M}_1
-(n-3) \xi^2 [p_1^2 p_2^2+m^2(p_1 p_2) ]
\} \; ,
\nn \\
t_{6,5}^{(1b)} &=& {\textstyle{1\over64}}{\cal K}^{-1}(p_1^2-p_2^2)
 \{(n-2) (2-\xi) m^2 
[(p_1^2 p_2^2)^{-1}(p_1 p_2)(p_1^2+p_2^2) -2] 
\nn \\ &&
  +(p_1-p_2)^2 [(n-4) (2+\xi)-2(n-3) \xi^2]\} \; ,
\\[8pt]
%%%%%%%%%%%%%%%%%%%%%%%%%%%%%%%%%%%%%%%%%%%%%%%%%%%%%%%%%%%%%%%%%%%%%%
t_{7,0}^{(1b)} &=& {\textstyle{1\over8}} \xi m p_3^2 
[(n-4){\cal M}_1^{-1} - (n-3) {\cal K}^{-1}
] \; ,
\nn \\
t_{7,1}^{(1b)} &=& -{\textstyle{1\over8}}(n-3) \xi m{\cal K}^{-1}
{\cal M}_1^{-1}
 [(p_1 p_2)+m^2] \; ,
\nn \\
t_{7,2}^{(1b)} &=& 0 \; ,
\nn \\
t_{7,3}^{(1b)} &=& 
{\textstyle{1\over16}} (n-2) \xi m
[(p_1^2 p_2^2)^{-1}-{\cal M}_1^{-1}] \; ,
\nn \\
t_{7,4}^{(1b)} &=& -{\textstyle{1\over16}}
(n-4) \xi m {\cal M}_1^{-1} \; , 
\nn \\
t_{7,5}^{(1b)} &=& -{\textstyle{1\over16}} (n-2)\xi m 
(p_1^2 p_2^2)^{-1}(p_1^2+p_2^2)  \; , 
\\[8pt]
%%%%%%%%%%%%%%%%%%%%%%%%%%%%%%%%%%%%%%%%%%%%%%%%%%%%%%%%%%%%%%%%%%%%%%
t_{8,0}^{(1b)} &=&
{\textstyle{1\over8}}{\cal{K}}^{-1}
\{ (n-4)\xi(4-\xi) {\cal{K}} 
        ( m^2  {\cal{M}}_1^{-1} p_3^2 + 1 )
+(n-3)\xi (4-\xi) p_3^2 (p_1 p_2)
\nn \\ &&
+(6-\xi) [ p_3^2 (p_1 p_2)+m^2 p_3^2 +2{\cal{K}} ] 
\},
\nn \\
%%%%%%%%%%%%%%%%%%%%%%%%%%%%%%%
t_{8,1}^{(1b)} &=& {\textstyle{1\over8}}{\cal K}^{-1} {\cal M}_1^{-1}
 \{(n-3) \xi (4-\xi)  [p_1^2 p_2^2+m^2 (p_1 p_2)]
+(6-\xi) {\cal M}_1 \} \; ,
\nn \\
t_{8,2}^{(1b)} &=& t_{8,5}^{(1b)} = 0 \; ,
\nn \\
t_{8,3}^{(1b)} &=& 
- {\textstyle{1\over16}} (n-2) \xi (4-\xi) m^2 {\cal M}_1^{-1} \; ,
\nn \\
t_{8,4}^{(1b)} &=&
- {\textstyle{1\over16}} (n-4) \xi (4-\xi) m^2 {\cal M}_1^{-1} \; .
\end{eqnarray*}

In the limit $m\to 0$, $\tau_1^{(1)}$, $\tau_4^{(1)}$, $\tau_5^{(1)}$, 
and $\tau_7^{(1)}$ vanish. We also note that all $t_{i,4}^{(1a)}$
and $t_{i,4}^{(1b)}$ are proportional to $(n-4)$, as they should,
since the transverse part cannot contain UV-poles in $\ep$ at one loop.
%%%%%%%%%%%%%%%%%%%%%%%%%%%%%%%%%%

%\newpage

\end{document}